\documentclass[12pt]{utarticle}
\usepackage[osf]{mathpazo}
\usepackage[T1]{fontenc}
\usepackage{newpxtext,newpxmath}
\usepackage{amsmath}
\usepackage{amsfonts}
\usepackage{amssymb}
\usepackage{mathtools}
\usepackage{longtable}
\usepackage{graphicx}
\usepackage[dvipsnames]{xcolor}
\usepackage{color}
\usepackage{ucs}
\usepackage[utf8]{inputenc}
\usepackage{xparse}
\usepackage{tikz}
\usetikzlibrary{arrows}
\usetikzlibrary{calc}
\usepackage{cite}
\usepackage{booktabs}
\usepackage{blindtext}
\usepackage{enumitem}
\usetikzlibrary{arrows.meta}
\usepackage[citecolor=OliveGreen,colorlinks=true,linkcolor=black,urlcolor=MidnightBlue,pdfborder={0 0 0}]{hyperref} 
\usepackage{framed}
\numberwithin{equation}{section}

\newtagform{conditions}[]()

\newcommand{\gt}{>}

\preprint{
UTTG--22--18\\
}

\title{Masses, Sheets and Rigid SCFTs}
\begin{document}
\author{Aswin Balasubramanian \address{ NHETC and Department of Physics and Astronomy, Rutgers University,  Piscataway, NJ 08885, USA \\ DESY (Theory) and Department of Mathematics, University of Hamburg, 20146 Hamburg, Germany \\ \email{aswinb.phys@gmail.com} } and Jacques Distler \address{ Theory Group, Department of Physics,  University of Texas at Austin, Austin, TX 78712, USA \\  \email{distler@golem.ph.utexas.edu}  } }

\Abstract{
We study mass deformations of certain three dimensional $\mathcal{N}=4$ Superconformal Field Theories (SCFTs) that have come to be called $T^\rho[G]$ theories. These are associated to tame defects of the six dimensional $(0,2)$ SCFT $X[\mathfrak{j}]$ for $\mathfrak{j}=A,D,E$.  We describe these deformations using a refined version of the theory of sheets, a subject of interest in Geometric Representation Theory. In mathematical terms, we parameterize local mass-like deformations of the tamely ramified Hitchin integrable system and identify the subset of the deformations that do admit an interpretation as a mass deformation for the theories under consideration. We point out the existence of non-trivial \textit{Rigid SCFTs} among these theories. We classify the Rigid theories within this set of SCFTs and give a description of their Higgs and Coulomb branches. We then study the implications for the endpoints of RG flows triggered by mass deformations in these 3d $\mathcal{N}=4$ theories. Finally, we discuss connections with the recently proposed idea of Symplectic Duality and describe some conjectures about its action.}
\maketitle
\newpage

\tableofcontents

\section{Introduction}
 
  In their pioneering work, Seiberg and Witten determined the low energy effective theory on the Coulomb branch for pure $SU(2)$ $\mathcal{N}=2$ gauge theory and for $SU(2)$ $\mathcal{N}=2$ SQCD with $N_f \leq 4$  using constraints from supersymmetry and electric magnetic duality of the abelian theory at a generic point of the Coulomb branch \cite{Seiberg:1994rs, Seiberg:1994aj}. Subsequent studies into the structure of Coulomb branches of other $\mathcal{N}=2$ theories have used various different strategies. One approach involves obtaining 4d $\mathcal{N}=2$ theories from higher dimensional field theories. This has the advantage that it ``geometrizes" many subtle aspects of the 4d gauge theory and allows one  to construct a large family of 4d $\mathcal{N}=2$ theories in a uniform way. 
 
 An important feature of the theory on the Coulomb branch of a 4d $\mathcal{N}=2$ theory is that it is governed by an associated complex integrable system \cite{Martinec:1995by,Donagi:1995cf}. Complex Integrable systems that arise in this fashion are called Seiberg-Witten integrable systems (see \cite{Donagi:1997sr} for a review). The base manifold of a complex integrable system is naturally a special K\"{a}hler manifold \cite{Freed:1997dp} and in the case of a Seiberg-Witten integrable system, it is identified with the Coulomb branch of the four dimensional theory. The total space of the integrable system, on the other hand, is identified with the Coulomb branch of the three dimensional theory obtained by compactifying the 4d theory on a circle \cite{Seiberg:1996nz}.

An example of a large family of 4d $\mathcal{N}=2$ theories that can be constructed by dimensional reduction are the theories of Class $\mathcal{S}$. These are theories that are obtained by twisted compactifications of the six dimensional $(0,2)$ SCFTs $X[j]$ of type $\mathfrak{j}\in A,D,E$ together with its codimension two (or four dimensional) defects on a punctured Riemann surface $C_{g,n}$ where the defects are inserted at the punctures \cite{Gaiotto:2009we,Gaiotto:2009hg}.  This list of theories includes several familiar Lagrangian theories and numerous theories for which a UV Lagrangian is not known. The range of theories that can be obtained using such a construction can be explored in a systematic manner and it has been in \cite{Chacaltana:2010ks} and subsequent works in the Tinkertoys program for Class $\mathcal{S}$ theories.
 
A key virtue of these theories is that the associated Seiberg-Witten integrable system is Hitchin's integrable system corresponding to the Lie algebra $\mathfrak{j}$ and the Riemann surface $C_{g,n}$. This is clearly an instance where a construction from higher dimensions greatly aids in understanding the geometry underlying the four dimensional theory. In the presence of codimension two defects, one actually studies the ramified Hitchin system.  As we recall below, the codimension two defects of theory $X[j]$ fall into two major categories, \textit{tame} and \textit{wild}. They manifest themselves as tame or wild singularities in the Hitchin system. The present paper is concerning the behaviour of the tame defects of the 6d theory under mass deformations.

 Let us now recall some basic properties of a large class of tame codimension two defects of the six dimensional $(0,2)$ SCFT $X[j]$ that have been studied by \cite{Chacaltana:2012zy}. In general, these are to be understood as modifications of the six dimensional theory along a four dimensional submanifold. Ideally, one would like to understand these defects directly in their native home, namely the six dimensional theory. But, the lack of a Lagrangian description for the non-abelian phase of the six dimensional theory makes such a direct approach difficult. The approach that we take here, following what has become standard practice, is to study the behaviour of these defects under dimensional reductions. At least three different dimensional reductions from six dimensions are directly relevant for the current paper. The first reduction can be used to associate a Nahm label to the defect (as in \cite{Chacaltana:2012zy}, see \S\ref{nahmlabel} below). The second reduction can be used to associate a Hitchin label to the defect \cite{Gaiotto:2009hg} (see \S\ref{hitchinlabel} below). In both cases, a gauge theory appears in five dimensions. We will denote the gauge group on the Nahm side to be $G$ and the gauge group on the Hitchin side to be $G^\vee$ in keeping with the notation used in \cite{Chacaltana:2012zy}. Here, we use $G$ to denote the compact Lie group and $\mathfrak{g}$ to denote the underlying complex Lie algebra.
 
  If the dimensional reduction on the circle does not involve an outer-automorphism twist, then $G$ and $G^\vee$ are both simply laced and have the same underlying complex Lie algebra $\mathfrak{g} = \mathfrak{j}$. We will denote the corresponding complex groups as $G_{\mathbb{C}}$ and $G_{\mathbb{C}}^\vee$. For the purposes of this paper, the precise global form of the gauge group will not matter. Our considerations will mostly be at the Lie algebra level. When the outer-automorphism twist $\mathfrak{o}$ is non-trivial, $\mathfrak{g}$ and $\mathfrak{j}$ are related as below \cite{Chacaltana:2012zy},
 
 \begin{center}
 \begin{tabular}{|c|c|c|c|}
 \hline 
 $\mathfrak{j}$ & $\mathfrak{o}$ & $\mathfrak{g}$ & $\mathfrak{g}^\vee$ \\ 
\toprule
 $A_{2n-1}$ & $\mathbb{Z}_2$ & $C_n$ & $B_n$ \\ 
 \hline 
 $D_n$ & $\mathbb{Z}_2$ & $B_{n-1}$ & $C_{n-1}$ \\ 
 \hline 
 $D_4$ & $\mathbb{Z}_3$ & $G_2$ & $G_2$ \\ 
 \hline 
 $E_6$ & $\mathbb{Z}_2$ & $F_4$ & $F_4$ \\ 
 \hline 
 \end{tabular} 
 \end{center}
 
 The third dimensional reduction that is relevant for the paper is the one applicable to the construction of class $\mathcal{S}[\mathfrak{j}]$ theories using a (partially twisted) compactification of the six dimensional theory $X[\mathfrak{j}]$ on a Riemann surface $C_{g,n}$ of genus $g$ with $n$ punctures \cite{Gaiotto:2009we,Gaiotto:2009hg}. The third reduction is also the setting for the AGT correspondence \cite{Alday:2009aq} which, among other things, sets up a map between codimension two defects and certain primaries of $Toda[j]$ theories \cite{Kanno:2009ga,Balasubramanian:2014jca,Haouzi:2016ohr}.

 In the limit where mass parameters are turned off, a pair of nilpotent orbits $(\mathcal{O}_N, \mathcal{O}_H)$ offer an efficient description of the defect \cite{Chacaltana:2012zy,Balasubramanian:2014jca}. As we review in \S\ref{nahmlabel}, the nilpotent orbit $\mathcal{O}_N$, the Nahm label, is a nilpotent in the Lie algebra $\mathfrak{g}$ while $\mathcal{O}_H$, the Hitchin label, is a nilpotent orbit in $\mathfrak{g}^\vee$. For the known tame defects, the Hitchin label is restricted to be a special nilpotent orbit while the Nahm label can be any nilpotent orbit. For a complete description of the local 4d Coulomb branch data associated to a tame defect, one has to additionally identify the Sommers-Achar group $C(\mathcal{O}_H)$ associated to the defect. The map from $\mathcal{O}_N$ to $(\mathcal{O}_H,\mathcal{C}(\mathcal{O}_H))$ is known. For the untwisted defects, this is a refinement of Spaltenstein's duality map and more generally, it is Sommers' refinement \cite{sommers2001lusztig} of the duality map due to Barbasch-Vogan (see \cite{Chacaltana:2012zy,Balasubramanian:2014jca} for more detailed discussions and references).  
 
 One would like to extend this understanding further to the case of non-zero hypermultiplet masses for several reasons. The most important reason is perhaps the fact that this will help us determine Seiberg-Witten curves for class $\mathcal{S}$ theories with nonzero hypermultiplet masses. Starting with the mass deformed form of the curve, one can arrive at the Seiberg-Witten curve for several other theories by taking various limits of large masses. This set of theories includes, for example, asymptotically free theories. For defects in Type A, mass deformations are well understood \cite{Gaiotto:2009we,Gaiotto:2009hg,Chacaltana:2012zy}. The primary motivation for the current work is to understand mass deformations for Class $\mathcal{S}$ theories outside of type A. 
 
 In this present paper, we study only the \textit{local} aspects of this problem. By this, we mean the effect of a mass deformation on a single tame defect.  In fact, we will only present the solution to a corresponding problem about 3d $\mathcal{N}=4$ SCFTs. These SCFTs arise when the four dimensional defect is wrapped on a circle. The resulting three dimensional defect in 5d SYM can be described by a coupling to a 3d $\mathcal{N}=4$ SCFT living on the world volume of the defect. For the tame defects that we study, these SCFTs are the $T^\rho[G]$ theories introduced by Gaiotto-Witten in their study of S-duality for conformal 1/2 BPS boundary conditions in 4d $\mathcal{N}=4$ SYM \cite{Gaiotto:2008ak}. Our reasons for restricting to the problem in three dimensions are two fold. First, we find several interesting structures already in the three dimensional problem of mass deformations which are perhaps best elucidated separately. Second, there are additional subtleties that enter the description of the four dimensional Coulomb branch. We comment briefly on how to incorporate them in \S\ref{deformationsin4d} but we leave a fuller discussion for a later  work. 

Already in \cite{Chacaltana:2012zy}, a proposal for the study of mass deformations outside of type A was put forward and this involves using nilpotent orbit induction and the theory of sheets to studying mass deformations. The present work can be taken to be a precise realization of this proposal.  But, we encounter some surprising new results in the process. Specifically, we find that there is a richer variation in RG flows (triggered by mass deformations) between the 3d SCFTs associated to non-Type A defects. Elucidating the ways in which these RG flows (and their endpoints) can be different outside of type A is a major goal of this paper. We do this in \S\ref{taxonomy} where we observe that each defect belongs to one of three deformation classes : \textit{Smoothable}, \textit{Malleable} or \textit{Rigid} and then describe the end point of a mass deformation for each of the defects. 

For our purposes, we find it useful to first re-interpret known results about  local hyper-K\"{a}hler geometry of the tamely ramified Hitchin System using the theory of sheets and use this to describe all \textit{mass-like} deformations of tame Hitchin system. Then, we further specialize to those deformations that do admit an interpretation as a mass deformation for the codimension two defects of the six dimensional $X[j]$ SCFT. We give a more detailed outline of this strategy in the \S\ref{strategy}.

The paper is organized in the following way. In \S\ref{massless}, we recall  properties of the defects in the massless limit and give a summary of our strategy for approaching the problem of mass deformations in \S\ref{strategy}. In \S\ref{SymplecticSingularities}, we review aspects of the local geometry of the ramified Hitchin system and discuss the conditions that need to be obeyed by a mass-like deformation of the tame Hitchin system. In \S\ref{sheets}, we give a detailed review of the theory of sheets in complex Lie algebras. After the detailed review, we give an explicit parameterization of mass-like deformations of the tame Hitchin system using the theory of sheets. 

 In \S\ref{massdeformations}, we identify the subset of mass-like deformations that further admit an interpretation as a mass deformation for a codimension two defect of theory $X[j]$ and give a description of the geometric consequences of a mass deformation associated to a defect. In the process of giving this description, we introduce the idea of a \textit{Rigid SCFT} and explain its relation to a refinement of the theory of sheets. We also slot the defects under study into some natural groups (deformation classes). 
 
 In \S\ref{tables}, we describe mass deformed defects for various simply laced $\mathfrak{g}$ in explicit fashion. In \S\ref{futher}, we provide further discussion highlighting some important aspects of our results with the help of examples from \S\ref{tables}.  In \S\ref{FurtherComments}, we raise the question of when the 3d SCFTs under study have UV Lagrangians and make some comments towards answering this question. 
 
 In \S\ref{SymplecticDuality}, we explore connections to works centered on the idea of \textit{Symplectic Duality}. We argue that symplectic duality for the $T^\rho[G]$ theories leads to a duality between slices (to nilpotent orbits) in the Lie algebra $\mathfrak{g}$ and a certain refined version of (special) sheets in the Lie algebra $\mathfrak{g}^\vee$. We explain this argument in the form of two conjectures in \S\ref{slicessheets} and \S\ref{correctrigidity}.
 
 Finally, in Appendix \S\ref{OrbitsAppendix}, we have summarized some aspects of the structure theory of nilpotent orbits that we could not explain as part of the main text. We have also provided a quick guide in \S\ref{definitionsappendix} to several terms that appear in the body of the paper. And in Appendix \S\ref{MoreTaxnonomy}, we have described mass deformed twisted defects for certain low rank cases.

\section{Review and Outline of Strategy}
\label{massless}
\subsection{The Nahm label}
\label{nahmlabel}

To associate a Nahm label to the defect, we consider a dimensional reduction from six to five dimensions by formulating the theory on a circle transverse to the defect. One obtains $\mathcal{N}=2$ SYM with a gauge group that we take to be $G$, a compact simple Lie group. The resulting defect in five dimensions is now co-dimension one. More accurately, it is a special case of a codimension one defect that can be identified with a boundary condition in the five dimensional theory. When this setup is further compactified down to four dimensions, one obtains 4d SYM with gauge group $G$ and a conformal 1/2 BPS boundary condition. This boundary condition is the one that corresponds to a `pure Nahm pole' in the terminology of \cite{Gaiotto:2008sa}. Associated to this boundary condition is a $\mathfrak{sl}_2$ embedding $\rho : \mathfrak{sl}_2 \rightarrow \mathfrak{g}$. By the Jacobson-Morozov theorem, picking $\rho$ is equivalent to picking a nilpotent orbit in the Lie algebra $\mathfrak{g}$. Let us denote the nilpotent orbit associated to the embedding $\rho$ by $\mathcal{O}_N$. This nilpotent orbit constitutes the `Nahm label' for the defect.

The Nahm label is convenient to identity one of the components of the moduli space of vacua associated of the CFT to the defect. Once we have compactified the defect, one is describing a three dimensional defect which has 3d $\mathcal{N}=4$ SUSY on its world volume. In this compactified setting, one can attach a \textit{Higgs branch} to the defect. For a defect with a Nahm label $\mathcal{O}_N$, the Higgs branch is a particular stratum inside the nilpotent cone $\mathcal{N}_\mathfrak{g}$. This stratum can be described as the intersection $\mathcal{S}_{\mathcal{O}_N} \bigcap \mathcal{N}$, where $\mathcal{S}_{\mathcal{O}_N}$ is the Slodowy slice to the nilpotent orbit $\mathcal{O}_N$. We refer to \cite{Chacaltana:2012zy,Balasubramanian:2014jca} for more elaborate recollections of these matters.

The construction of Slodowy's slice proceeds as follows. Pick a standard triple $(e,f,h)$ for $\mathfrak{sl}_2$ and then consider the space 
\begin{equation}
S_{\mathcal{O}_N} = \rho(e) + Z_{\mathfrak{g}}(\rho(f)),
\end{equation}
where $\rho(e)$ is the image under $\rho$ of the nilpositive element and $Z_{\mathfrak{g}}(\rho(f))$ is the centralizer (in $\mathfrak{g}$) of the image of the nilnegative element $f$. Under the $\mathfrak{g}$-action, this space sweeps a stratum inside the Lie algebra $\mathfrak{g}$ and this is called the Slodowy slice to the orbit $\mathcal{O}_N$. The intersection $\mathcal{S}_{\mathcal{O}_N} \bigcap \mathcal{N}$ appears naturally as the moduli space of solutions to Nahm's equations on a semi-infinite line with the pure Nahm pole boundary condition at the boundary and fixed regular values at infinity. The problem of finding the solution space to Nahm's equations can be cast as a hyper-K\"{a}hler quotient construction. Thus, the resulting moduli space of solutions carries a natural hyper-K\"{a}hler structure. Taken together with the natural hyper-K\"{a}hler metric that is obtained in this construction, this moduli space constitutes the Higgs branch $\mathcal{M}_{Higgs}$ that one attaches to the defect. If considers the moduli space that is fibered over the boundary values, one obtains a different moduli space $\widehat{\mathcal{M}_{Higgs}}$. In a distinguished complex structure and for a specific pattern of boundary values, $\widehat{\mathcal{M}_{Higgs}}$ is a hyper-K\"{a}hler resolution of $\mathcal{M}_{Higgs}$ \cite{bielawski1996hyperkahler,bielawski1997hyperkahler}. We will not discuss this resolution of singularities in this paper. 

But, the following aspect of the Higgs Branch will play an important role for us. Let $\mathfrak{f}$ be the joint (Lie algebra) centralizer of the triple $(\rho(e),\rho(f),\rho(h))$ in the Lie algebra $\mathfrak{g}$. And let $F$ be a compact group group associated to the complex Lie algebra $\mathfrak{f}$. The group $F$ acts on the Higgs branch $\mathcal{M}_{Higgs}$ by hyper-K\"{a}hler isometries.\footnote{We regret the similarity in notation for the nilnegative element $f$ in $\mathfrak{sl}_2$ and the Lie algebra $\mathfrak{f} \subset \mathfrak{g}$ of the Flavour symmetry group $F$. We have continued to use what are standard labels in the literature. The context should clarify any potential confusion between the two.} It is natural to call this a flavour symmetry since  in Lagrangian theories, flavour symmetries appear precisely as hyper-K\"{a}hler isometries of Higgs branches. 

\subsection{The Hitchin Label}
\label{hitchinlabel}

To identify the Hitchin label associated to the defect, we take a compactification to five dimensions on a circle that the defect wraps. This leads to a three dimensional defect in $\mathcal{N}=2$ SYM with gauge group $G^\vee$, the GNO/Langlands dual group associated to $G$. This three dimensional defect involves coupling the 5d theory to a 3d $\mathcal{N}=4$  SCFT. For the defects that we study, these 3d $\mathcal{N}=4$ SCFTs turn out to be the $T^\rho[G]$ theories of Gaiotto-Witten \cite{Gaiotto:2008ak}. 

For studying the properties of a single defect, it is convenient to specialize to the case where the transverse space to the defect is topologically a punctured disc. Describing the behaviour of the field of the 5d theory near the defect turns out to be equivalent to describing a boundary condition for the ramified Hitchin system for the group $G^\vee$ near the puncture\cite{Gaiotto:2009hg}. We review some very basic facts here and give a more elaborate discussion about the Hitchin system in \S\ref{hitchinintegrable} and \S\ref{localhitchin}. 

 In the massless limit, the defect induces a singular behaviour in the Higgs field $\phi$ belonging to a Hitchin system (for the group $G^\vee$) that is of the following form,
\begin{equation}
\label{hitchintame}
\phi = \frac{a}{z} + (\ldots),
\end{equation}
where $a$ is a nilpotent element in the Lie algebra $\mathfrak{g}^\vee$, $z=0$ is the location of the defect on the Riemann surface $C$ and $(\ldots)$ contains terms that are strictly regular. There is a gauge equivalence that identifies any two defects whose residues $a, a'$  happen to be conjugates in $\mathfrak{g}^\vee$. So, the gauge invariant data fixing the singularity type of the Higgs field is the adjoint orbit to which the nilpotent element $a$ belongs. Let this nilpotent orbit be $\mathcal{O}_H$. This constitutes the `Hitchin label' for the defect. In type A, this is a complete description of the Hitchin label. Outside of type A, two further considerations enter the description of the Hitchin boundary condition. First, it becomes necessary to restrict to the case where $\mathcal{O}_H$ is a \textit{special} nilpotent orbit \cite{Chacaltana:2012zy,Balasubramanian:2014jca}.\footnote{The tame defects with Hitchin labels being a special nilpotent orbit are the ones for which the local Higgs branch and the associated flavour symmetry $F$ are known.} Second, specifying the contribution of a defect to the graded dimension of the four dimensional Coulomb branch requires some additional data. A way to specify this data is by the identification Sommers-Achar group associated the defect \cite{Chacaltana:2012zy}. 

\subsection{The Hitchin Integrable System}
\label{hitchinintegrable}

One of the defining properties of theories of the $\mathcal{N}=2$ theories of class $\mathcal{S}$ is that their Seiberg-Witten solution is encoded in the geometry of the Hitchin system. Here, we recall some basic aspects of the Hitchin system. For this discussion, we will restrict to the case of $\mathfrak{g}=\mathfrak{sl}_N$.  

The Hitchin system for a Lie algebra $\mathfrak{g}$  \cite{hitchin1987self} describes pairs $(A,\phi)$ that obeys the equations
\begin{equation}
\begin{split}
F_A + [\phi , \phi^\dagger] &= 0, \\
\bar{\partial}_A \phi &= 0.
\label{hitchinsystem}
\end{split}
\end{equation}
 The moduli space of solutions to the above equations is usually termed the Hitchin moduli space $\mathcal{M}_H$. This space is hyper-K\"{a}hler. The Hitchin moduli space has several possible presentations. One of them is as the space of Higgs bundles. These are pairs $(E,\phi)$ where $E$ is a holomorphic bundle and $\phi : E \rightarrow E \otimes K $ is an $End(E)$ valued one form.  A second presentation is as a moduli space of flat $G_{\mathbb{C}}$-connections on a vector bundle $V$ over the Riemann surface $C$. Consider a connection of the form,
 \begin{equation}
 \mathbf{\nabla} = \nabla_{A} + \phi + \phi^\dagger,
 \end{equation}
where $\phi+\phi^\dagger$ is an $End(V)$ valued one form. Now, demanding that $\nabla$ be flat \footnote{More generally, one could considered connections of the form $\nabla = \nabla_A + \zeta \phi + \zeta^{-1} \phi^\dagger$ for a $\zeta \in \mathbb{C}^\star$.} amounts to demanding that Hitchin's equations (\ref{hitchinsystem}) are obeyed. The hermitian adjoint $\phi^\dagger$ is taken with respect to a hermitian metric $h$ on $E$. The metric $h$ is completely specified by a solution to Hitchin's equation. In this way, one can use solutions to Hitchin's equations to go between the Higgs bundle picture and the flat connections picture. This translation between the Higgs bundle picture and the flat connection picture is called the non-abelian Hodge correspondence \cite{simpson1992higgs}. For the case of a Riemann surface without punctures, the result is obtained by combining the results of Hitchin\cite{hitchin1987self}-Simpson \cite{simpson1988constructing} relating Higgs bundles obeying a stability condition to harmonic maps and of Corlette \cite{corlette1988flat}-Donaldson \cite{donaldson1987twisted} which relates harmonic metrics and moduli spaces of flat connections. See \cite{wentworth2016higgs} for a recent review of this circle of ideas. For our present discussion, we actually need the version of the correspondence for a punctured Riemann surface $C_{g,n}$ with tame singularities at the punctures \cite{simpson1990harmonic}.  The tameness constraint translates to the condition that both $\nabla_{(0,1)}$ and $\phi$ have simple poles at the punctures as in Eq (\ref{hitchintame}) above. It also implies a particular behaviour for the hermitian metric $h$ on the bundle $E$ (of rank $r$, say) near the punctures \cite{simpson1992higgs,biquard1991fibres}. Let $\alpha_i$ be the residue of $\nabla_{(0,1)}$. Then, the metric $h$ locally takes the following form, 
\begin{equation}
h = \text{diag} (\mid z \mid^{-2\alpha_1}, \mid z \mid^{-2\alpha_2}, \ldots \mid z \mid^{-2\alpha_r} ).
\end{equation}
Given the metric $h$, the connection $\nabla_{0,1}$ can be recovered as the $(0,1)$ part of the Chern connection $d+h^{-1}\partial h$ associated to $h$. Away from the punctures, the metric is extended in a smooth way that is consistent with Hitchin's equations. The relation between the residues of $\nabla_{(0,1)}$ and $\phi$ is somewhat subtle. On the one hand, the nilpotent parts of $Res(\nabla_{(0,1)})$ and $Res(\phi)$ are identical but there is a non-canonical relationship between their semi-simple parts  \cite{simpson1990harmonic}. To keep matters simple, we only discuss $Res(\phi)$ in what follows with the understanding that one can always deduce $Res(\nabla_{(0,1)})$ using the table of formulas in  \cite{simpson1990harmonic}. In this context, see also the discussion in \S4 of \cite{nakajima1996hype}.

The precise map between the Higgs bundle picture and the flat connections picture necessarily involves a careful specification of stability conditions on both sides. In the tamely ramified case, this, in turn, depends on the assignment of parabolic weights to each of the punctures\cite{simpson1990harmonic}. In our discussions, which will remain purely local, we will not be explicit about stability conditions. But, they will enter the picture if one were to make statements globally.
 
When described as the space of Higgs bundles, the Hitchin moduli space has the structure of a complex integrable system \cite{hitchin1987stable}. This is seen by studying the Hitchin map
\begin{equation}
\mu_H : \mathcal{M}_H \rightarrow \mathcal{B}
\end{equation}
where $\mathcal{B} = \bigoplus_{i=2}^{k}H^0(\Sigma,K^{d_i})$. Locally, the map $\mu_H$ takes the pair $(E,\phi)$ to the coefficients $p_k$ of the characteristic polynomial of $\phi$,
\begin{equation}
\mu_H : (E,\phi) \mapsto (p_2(\phi),p_3(\phi) \ldots p_k(\phi)).
\end{equation}

If $b$ is a smooth point of the base $\mathcal{B}$, the fibers $f_b=\mu_H^{-1}(b)$ of the map $\mu_H$ are complex tori. In unramified case, the complex dimension of moduli space of stable Higgs bundles $\mathcal{M}_H(C_g, \mathfrak{g}), g \geq 2 $ is given by
\begin{equation}
\dim(\mathcal{M}_H) = \dim (\mathfrak{g}) (2g-2).
\end{equation}

The dimension of the Hitchin base $\mathcal{B}$ is half the dimension of $\mathcal{M}_H$,
\begin{equation}
\label{basedimension}
\dim (\mathcal{B}) =  \dim (\mathfrak{g}) (g-1)
\end{equation}

In the presence of co-dimension two defects, the appropriate integrable system to consider is the ramified Hitchin system. The integrable system structure of the ramified Hitchin system has been studied in \cite{markman1994spectral,bottacin1995symplectic,donagi1996spectral}. See \cite{dalakov2016meromorphic} for a recent survey. The present paper is concerned with the local geometry of the tamely ramified Hitchin system and we will delve into more of the details in \S\ref{localhitchin}. For the moment, we note that the presence of the tame defects modifies the dimension in the following way,\footnote{One can prove this using the Riemann-Roch theorem. The local contribution can also deduced by studying the tame Hitchin system on a punctured disk and noticing that the local moduli space is nothing but the adjoint orbit of the residue.} 
\begin{equation}
\dim(\mathcal{M}_H) = \dim (\mathfrak{g}) (2g-2) + \sum_i \dim(\mathcal{O}_H^i)
\label{defectsdimension}
\end{equation}
where $\mathcal{O}_H^i$ are the Hitchin labels associated to each tame defect. The dimension of the base $\mathcal{B}$ increases such that the relation $\dim(\mathcal{B})=\frac{1}{2}\dim(\mathcal{M}_H)$ still holds. 

The presence of tame defects (or equivalently, simple poles for the Higgs field) can be interpreted as the appearance of source terms in Hitchin's equations (\ref{hitchinsystem}) (see, for ex \cite{Donagi:1995cf}). 

With the addition of the source terms, the equations take the following form,
\begin{equation}
\begin{split}
F + [\phi , \phi^\dagger] &= \sum_i \mu_{\mathbb{R}} \delta(z-z_i) \\
\bar{\partial}_A \phi &=  \sum_i \mu_{\mathbb{C}} \delta(z-z_i)
\label{hitchinsystemsource}
\end{split}
\end{equation}
where the parameters $\mu_{\mathbb{R}},\mu_{\mathbb{C}}$ are certain linear combinations of $Res(\phi)$ and $Res(\nabla)$ and $z_i$ are the locations of the defects. Locally, such source terms can be interpreted as the result of coupling the Hitchin system without the singularity to the adjoint orbit (see discussion in \S3 of \cite{Gukov:2008sn} and \S3.1.6 of \cite{Gaiotto:2009hg} for this interpretation). 

We now clarify a point of terminology. When we speak of the tamely ramified Hitchin system, we mean the meromorphic Hitchin system in which the Higgs field has simple poles and the residues at these poles are \textit{fixed} adjoint orbits. When the residue data is not fixed, one obtains only a Poisson structure on the total space of the meromorphic Hitchin system \cite{markman1994spectral,bottacin1995symplectic}. But, upon fixing the residue, the total space is restricted to being a symplectic leaf of the larger Poisson manifold and we obtain a complex integrable system whose total space is now equipped with a holomorphic symplectic form. A closely related notion is that of a parabolic Hitchin system where a parabolic structure is fixed at each puncture and only residues that respect this parabolic structure are allowed. In our discussions, we do not need to fix a parabolic structure and hence, we avoid referring to the parabolic Hitchin system or parabolic Higgs bundles. We clarify this further in \S\ref{localhitchin} where we study the local geometry of the tame Hitchin system.

The Seiberg-Witten curve is the spectral curve associated to the Hitchin system and the Seiberg-Witten differential is the canonical 1-form on $T^\star C_{g,n}$, restricted to the SW curve. We denote it by $\lambda dz$, where $(\lambda, z)$ are co-ordinates for $T^\star C_{g,n}$. 

For the case $\mathfrak{g}=\mathfrak{sl}_n$, the spectral curve $S$ is a $n$-sheeted cover $S \rightarrow C$ defined by
\begin{equation}
det_{\underline{N}}(\phi(z) - \lambda(z) I) = \lambda^N + \lambda^{N-2} p_2 + \ldots + p_n =0 
\end{equation}
where $p_i$ for $i=2,\dots, n$ are the invariant symmetric polynomials of degree $i$. In particular, $p_2=\tfrac{1}{2}Tr(\phi^2)$ and $p_n = det(\phi)$.

When hypermultiplet masses are set to zero, the corresponding ramified Hitchin systems describes the geometry of the Coulomb of a super-\textit{conformal} field theory (SCFT). The dimension of the integrable system is a measure of the massless degrees of freedom of this SCFT. It is known, for example, that there is a direct relationship between the dimension of the integrable system and the trace anomalies $(a,c)$ \cite{Benini:2009gi}. 

\subsection{Our Strategy}
\label{strategy} 

In this section, we explain our strategy for studying mass deformations of tame defects. It entails breaking up the problem into two steps. The first involves classifying all mass-like deformations of the tame Hitchin system. The second step involves identifying those mass-like deformations that do correspond to actual mass deformations of $T^\rho[G]$ theories. We now explain both these steps in more detail. 

\subsubsection{Step 1 : Classifying mass-like deformations}

An intriguing feature of mass deformed Seiberg-Witten integrable systems is the fact that the cohomology class of the holomorphic symplectic form $[\Omega]$ varies linearly with the hypermultiplet mass parameters $m_i$. This was already noted in the study of the SU(2) $N_f=4$ theory in \cite{Seiberg:1994aj} (\S 17) and this property was subsequently studied in the dimensionally reduced theory in three dimensions in \cite{Seiberg:1996nz}. It turns out that \textit{demanding} that $[\Omega] \propto m_i$ is highly constraining and goes a long way in determining the form of the Seiberg-Witten solution of the mass deformed theory \cite{Seiberg:1994aj}. Following these works, we take our first requirement for a mass-like deformation to be
\usetagform{conditions}
\begin{equation}
\boxed{[\Omega]_{m \neq 0} \propto m_i}
\label{omega}
\end{equation}

It is known that the local geometry of the ramified Hitchin system provides \text{families} of holomorphic symplectic manifolds in which $[\Omega]$ has precisely the required linear dependence and so, it follows that they are suitable candidates to describe mass deformations \cite{Donagi:1995cf}.
 
Now, the above condition constraints how the geometry of the Coulomb branch can vary once we have a non-zero mass-like deformation. But, we also need some way to relate it to the undeformed Coulomb branch. For this, we also impose the following natural condition :
 \begin{equation}
 \label{dimensioncondition}
\boxed{\dim(\mathcal{O}_{a_M}) = \dim(\mathcal{O}_{a_0})}.
 \end{equation}
It is clear that any mass deformation should obey this condition. This is because we expect hypermultiplet masses to parameterize a family of Coulomb branches that are of the same dimension as the Coulomb branch of the original theory with zero hypermultiplet masses. We will call (\ref{dimensioncondition}) the \textit{dimension condition}. 
 
 If the two conditions above in Eqs (\ref{omega}) (\ref{dimensioncondition}) are obeyed, then we call the corresponding deformation of the integrable system to be a mass-like deformation of the integrable system $(\mathcal{M}_H, \Omega)$. We classify all mass-like deformations of the tame Hitchin system (for Lie algebra $\mathfrak{g}^\vee$) in \S\ref{sheets}. These deformation will turn out to be labeled by a pair $(\mathfrak{l}^\vee,\mathcal{O})$ where $\mathfrak{l}^\vee$ is a Levi subalgebra of $\mathfrak{g}^\vee$ and $\mathcal{O}$ is a nilpotent orbit in $\mathfrak{l}^\vee$. The mass-like parameters themselves will turn to be valued in $Z(\mathfrak{l}^\vee)$, the center of the Levi subalgebra and they parameterize a \textit{sheet} of the Lie algebra $\mathfrak{g}^\vee$. Sheets offer a particular stratification of the Lie algebra $\mathfrak{g}^\vee$. They are defined in the following way. Let $\mathcal{U}_d$ be the union of all adjoint orbits of a fixed dimension $d$. Then, the irreducible components of $\mathcal{U}_d$ are called sheets. The Levi $\mathfrak{l}^\vee$ corresponding to a mass-like deformation of the integrable system is obtained as the centralizer of the semi-simple part of the non-nilpotent orbits that belong to a sheet. An account of how sheets arise in the study of the Hitchin system is provided in \S\ref{SymplecticSingularities}. A longer introduction to the theory of sheets is contained in \S\ref{sheets}.
 
 \subsubsection{Step 2 : The Flavour condition}
  
 There is one further condition for a mass-like deformation to be an actual mass deformation. Let us denote $m_i \in Z(\mathfrak{l}^\vee)$ to be the mass-like deformation parameters of the complex integrable system. For this to correspond to an actual mass deformation,\footnote{We confine ourselves to instances of a maximal mass deformation.} it should obey the additional condition that the vector space dual of the space of deformation can be identified with the Cartan subalgebra of the Flavour group $\mathfrak{h}(F)$. In other words, we require that 
 \begin{equation}
 \boxed{m_i^\star \in \mathfrak{h}(F)  }
 \label{flavour}
 \end{equation}
This is the third and final condition that must be obeyed and it follows from the fact that the masses and the flavour current Lie in the same multiplet of the superconformal algebra $\mathfrak{osp}(4 \mid 4)$ (see \S\ref{relevantrecollections}). We call (\ref{flavour}) the \textit{Flavour condition}. As we reviewed in \S\ref{nahmlabel}, the flavour symmetry is the group of  continuous hyper-K\"{a}hler isometries of the local Higgs branch and can be inferred directly from the Nahm label.  The Flavour Condition implies that the Levi $\mathfrak{l}^\vee_{\text{sheet}}$ associated to the deformation of the integrable systems is the Langlands dual to Bala-Carter Levi $\mathfrak{l}_{BC}$ of the Nahm nilpotent orbit. Let us see how this comes about. From the structure theory of nilpotent orbits, it follows that the Bala-Carter Levi $\mathfrak{l}_{BC}$ is the largest Levi-subalgebra that obeys property that $\mathfrak{h}(F) \cap \mathfrak{h}(L)= \{\emptyset \}$, where $L$ is the compact group corresponding to the complex Lie algebra $\mathfrak{l}$. 
 
Let $\mathfrak{l}^\vee_{\text{sheet}}$ be the Levi associated to a mass-like deformation of the integrable system. It follows that the mass-like parameters are valued in the center of the Levi subalgebra $Z(\mathfrak{l}^\vee_{\text{sheet}})$. First, let us assume $Z(\mathfrak{l}^\vee_{\text{sheet}}) \cap \mathfrak{h}(\mathfrak{l}_{BC}) \neq \{ \emptyset \}$. Then, using \ref{flavour}, we see that our assumption contradicts the statement that $\mathfrak{l}_{BC}$ is the Bala-Carter Levi for the Nahm orbit $\mathcal{O}_N$. So, this implies that we need to have $Z(\mathfrak{l}^\vee_{\text{sheet}}) \cap \mathfrak{h}(\mathfrak{l}_{BC}) = \{ \emptyset \}$ for any mass deformation. 

Second, if $\mathfrak{l}$ were a proper Levi subalgebra of $\mathfrak{l}_{BC}$, then this deformation would not be a maximal mass deformation. While this is entirely consistent, it is not interesting from the point of view of classification. To classify mass deformations, it makes sense to first classify the maximal ones. The partial mass deformations can then be obtained as further restrictions on the maximal mass deformations. So, we need to consider the largest deformation for which $Z(\mathfrak{l}^\vee_{\text{sheet}}) \cap \mathfrak{h}(\mathfrak{l}_{BC}) = \{ \emptyset \}$ holds. There is a unique solution (up to conjugacy), 

 \begin{equation}
\boxed{\mathfrak{l}^\vee_{\text{sheet}} = \mathfrak{l}_{BC}^{\vee}.}
 \label{sheetbclevisame}
 \end{equation}
 

 Since the condition (\ref{sheetbclevisame}) follows from (\ref{flavour}), we will interchangeably refer to (\ref{sheetbclevisame}) or (\ref{flavour}) as the Flavour Condition.
 
 It is important to note that turning on a maximal mass deformation does \textit{not} imply that the Higgs branch of the resulting theory in the IR is trivial. We will meet several examples in which a non-trivial Higgs branch remains after a maximal mass deformation. Such residual Higgs branches are always rigid in the sense that they can't be further lifted by another mass deformation. They are the Higgs branches of what we call \textit{Rigid SCFTs}. In the theories we are studying, we find that there are no ``dangerously irrelevant" operators which  become mass deformation operators in the IR along a RG flow triggered by a maximal mass deformation. If there had been such operators, then the theory after a maximal mass deformation would have admitted further mass deformation by the operator that became a relevant operator along the RG flow. It has been conjectured by the authors of \cite{Argyres:2015ffa} that such marginality crossings do not occur in 3d $\mathcal{N}=4$ theories. Our results are consistent with this conjecture being true.
 
 The main result of this paper is the identification of the unique solution to the above three conditions (Eqns \ref{omega}, \ref{dimensionD} and \ref{flavour}) for all known tame defects of the 6d theory. To identify these solutions, we find it convenient to first impose conditions (\ref{omega}) and (\ref{dimensioncondition}) and classify all possible mass-like deformations of the tamely ramified Hitchin system. Then, we impose (\ref{sheetbclevisame}) and restrict to those deformation that do admit an interpretation as mass deformations. Imposing (\ref{sheetbclevisame}) leads to a refinement of the usual theory of sheets. We explain this refinement in \S\ref{massdeformations}. 
 
 In several instances,  we also find that there exist mass-like deformations for the tame Hitchin system with special nilpotent residue that do not actually correspond to mass deformations of any $T^\rho[G]$ theory. In other words, there exist deformations that obey condition \ref{omega} and \ref{dimensioncondition} but do not obey condition \ref{sheetbclevisame} (or, equivalently, condition \ref{flavour}). Every such deformation corresponds to a \textit{non-special sheet} attached to a special nilpotent orbit. We believe the existence of such deformations of the tame Hitchin system is, by itself, an interesting observation.
 
 The existence of such deformations of the integrable system is revealed only through a study of the Hitchin system outside of type A since non-special sheets exist only outside of type A. We explain this notion and give many examples of such deformations in \S\ref{specialvsnonspecial}. We do not, at present, know if the deformations corresponding to non-special sheets could be given the interpretation of a mass deformation of a different set of 3d $\mathcal{N}=4$ SCFTs.

\section{Local geometry of the Hitchin system }
\label{SymplecticSingularities}
As reviewed in the earlier sections, our study of the properties of a single defect has the goal of obtaining a systematic understanding of mass deformations in an arbitrary Class $\mathcal{S}$ theory built out of these tame defects. In the presence of tame defects, the dimension of the 3d Coulomb Branch increases according to (\ref{defectsdimension}).  Mass deformations deform the geometry of the Coulomb Branch in a way that preserves its dimension. In this section, we will see that the condition that the dimension of the Coulomb Branch be preserved leads to an interesting constraint on how the residue of the Higgs field can vary. 

Important background references for this section include \cite{nakajima1996hype,biquard1996equations} where the local structure of the tame Hitchin system was studied and \cite{Gukov:2006jk} where these local moduli spaces were further studied in the context of of Surface operators in $\mathcal{N}=4$ SYM and the tamely ramified geometric Langlands program. The local moduli spaces are obtained by solving Hitchin's equations locally and they are endowed with a natural hyper-K\"{a}hler structure. As in the global case, the resulting setup is an infinite dimensional hyper-K\"{a}hler quotient construction.  In some special cases, it becomes possible to also supply certain finite dimensional  hyper-K\"{a}hler quotient constructions for these local moduli spaces (see, for example,\cite{nakajima1994instantons,kobak1996classical}). These finite dimensional quotients will not play a major role in this paper but we comment on some of them in \S\ref{FurtherComments}.

In the mathematical literature on Higgs bundles, the term \textit{strongly parabolic Higgs bundles} is used to describe  the cases where the residue at the punctures is  nilpotent and the term \textit{weakly parabolic Higgs bundles} is used for the case where the residue is non-nilpotent \cite{logares2010moduli}. See also \cite{biswas1994infinitesimal}, \cite{biswas1997parabolic} for useful background material on parabolic Higgs bundles. In this paper, we will uniformly use the term \textit{tamely ramified Hitchin System} to denote both situations (following \cite{Gukov:2006jk}). As we will see, studying mass deformations amounts to studying how the geometry changes as one varies this residue from nilpotent to non-nilpotent values. 

\subsection{The local moduli space}
\label{localhitchin}
First, we follow \cite{biquard1996equations,Gukov:2006jk} and relate the local geometry of the ramified Hitchin system at the point of ramification to  symplectic resolutions/deformations of nilpotent orbits. Consider the ramified Hitchin system for the group $G^\vee$ that controls the physics of Coulomb branch of a theory of class $\mathcal{S}$. We study the local geometry of this Hitchin system near the point of ramification. We choose polar co-ordinates $(r,\theta)$ near the point of ramification such that $(z-z_i)= r e^{i\theta}$, where $z_i$ is the location of the defect. The most general $\mathbb{S}^1_{\theta}$ invariant solution to Hitchin's system of equations would have following behaviour near the ramification point
\begin{equation}
\begin{split}
A &= a(r) d \theta + h(r) \frac{dr}{r}, \\
\phi &=  b(r) d \theta + c(r) \frac{dr}{r}.
\end{split}
\end{equation}
Now, by choosing $s=-log(r)$ and $\frac{D}{Ds}= \frac{d}{ds} + [h,\cdot]$, Hitchin's equations can be recast as Nahm's equations for the functions $a(r),b(r),c(r)$.
\begin{equation}
\begin{split}
\frac{Da}{Ds} &= [b,c] \\
\frac{Db}{Ds} &= [c,a] \\
\frac{Dc}{Ds} &= [a,b] 
\label{nahmtwo}
\end{split}
\end{equation}

By a gauge transformation, one could choose to set $h(r)$ to zero. The pole boundary condition for the Hitchin system at $r=0$ now becomes a condition at $s \rightarrow + \infty $ for the Nahm system (\ref{nahmtwo}). The reader must note that the Nahm system appearing here is \textit{different} from the one that is relevant for the Higgs branch associated to the defect in \S\ref{nahmlabel}. For one thing, this Nahm system is for the group $G^\vee$. Additionally, the boundary conditions imposed are different. Let us denote the boundary values of $a(r),b(r),c(r)$ by
\begin{equation}
\begin{split}
\alpha &= a(0) \\
\beta &= b (0) \\
\gamma &= c(0)
\label{nahmbc}
\end{split}
\end{equation}

We will denote by $\mathcal{M}^{loc}(\alpha,\beta,\gamma)$ the moduli space of solutions to the Nahm system (\ref{nahmtwo}) with the above boundary conditions (\ref{nahmbc}). The combination $\beta + i \gamma$ is the precisely the residue of the Higgs field $Res(\phi)$. Since we have denoted this residue by $a$ in (\ref{hitchintame}), we set $a=\beta+ i \gamma$. 

Nahm's equations can be recast as an infinite dimensional hyper-K\"{a}hler quotient \cite{kronheimer1990hyper,kronheimer1990instantons,kronheimer2004hyperkahler}. The resulting moduli space  $\mathcal{M}^{loc}$ is a finite dimensional hyper-K\"{a}hler manifold. This implies that there are three complex structures $I,J,K$ that obey the quaternionic identities $I^2=J^2=K^2=IJK=-1$. Correspondingly, there exist three K\"{a}hler form $\omega_I, \omega_J,\omega_K$ and a Riemannian metric $g$. This Riemannian metric is K\"{a}hler with respect to each of the pairs $(I,\omega_I), (J,\omega_J), (K,\omega_K) $. The complex structures $I,J,K$ can be used to parameterize the full $\mathbb{CP}^1$ worth of complex structures that exist on $\mathcal{M}^{loc}$. In any fixed complex structure (say $I$), one can also build a holomorphic symplectic form $\Omega_I = \omega_J + i \omega_K$ that is non-degenerate. Singular manifolds that admit a non-degenerate holomorphic symplectic form are known as Symplectic Singularities. So, it follows that hyper-K\"{a}hler singularities are necessarily Symplectic Singularities in the sense of \cite{beauville2000symplectic}. 

It is sometimes the case that there exists a \textit{family} of hyper-K\"{a}hler structures on a manifold. To discuss such a family, it is convenient to fix ourselves to one complex structure, say $I$. Now, the family of hyper-K\"{a}hler structures could correspond to varying the (cohomology class of) $\omega_I$ while leaving $\Omega_I$ fixed, varying (cohomology class of) the holomorphic symplectic form $\Omega_I$ while leaving $\omega_I$ fixed or varying both $\omega_I$ and $\Omega_I$.  In the present context, such families of hyper-K\"{a}hler structures will arise when one varies the boundary conditions in Nahm's equations. Such families of hyper-K\"{a}hler structures were first constructed by Kronheimer in \cite{kronheimer1990hyper}. It will turn out that the existence of mass deformations is related to the existence of such families of hyper-K\"{a}hler structures. One can, additionally, build a twistor space using the entire $\mathbb{CP}^1$ worth of complex structures that is available for a fixed hyper-K\"{a}hler structure. For the case of semi-simple adjoint orbits, such twistor spaces have been studied by  \cite{santacruz}.

When studying solutions to Nahm's equations, it is often convenient to fix a complex structure and then cast (\ref{nahmtwo}) as a pair of equations for $m = (a + ih)/2, n = (b + ic)/2$,
\begin{equation}
\begin{split}
 \frac{1}{2} \frac{dm}{ds} +  [m,n] &= 0 \\ 
 \frac{1}{2} \frac{d}{ds} (m + \overline{m}) + [m,\overline{m}] + [n,\overline{n}] &= 0
 \label{compelxnahm}
\end{split}
\end{equation}
\noindent
The boundary values reached by $m(s),n(s)$ as $s \rightarrow \infty$ are $m(\infty) = \alpha, n(\infty) = \beta + i \gamma $.

Now, we want to consider how the moduli space changes when the boundary values $\alpha,\beta,\gamma$ vary in a \textit{fixed} sheet of the complex Lie algebra. Sheets are relevant for us because these are precisely the deformations of the Hitchin system that will turn out to satisfy the conditions (\ref{omega} and \ref{dimensioncondition}) required for being a mass-like deformation. We will briefly introduce sheets below and delve into the theory in greater detail in \S\ref{sheets}.

\subsection{Sheets : A first look}

For present purposes, it is sufficient to recall some basic definitions about sheets in complex Lie algebras. Let $U_d$ be the union of adjoint orbits of complex dimension $d$ in a complex Lie algebra $\mathfrak{g}$. The irreducible components of $U_d$ are called \textbf{sheets}. As we will review in the next section, a sheet in a Lie algebra always contains exactly one nilpotent orbit. A sheet may additionally contain non-nilpotent orbits. When a sheet contains orbits that are semi-simple, then the corresponding sheet is called a Dixmier sheet. When studying sheets, it is often simpler to first consider the case of Dixmier sheets and then generalize to an arbitrary sheet. 

Let $\mathcal{O}_e$ denote the unique nilpotent orbit in a Dixmier sheet and let $e$ be a nilpotent element in this orbit. Let $s$ denote a semi-simple element in one of the infinite number of semi-simple orbits in this Dixmier sheet. Let $P$ be one of the parabolic subgroups associated to the Dixmier sheet. Such a choice of $P$ is usually termed a choice of a polarization for the nilpotent orbit $\mathcal{O}_e$.  Let us denote this moduli space by $\mathcal{M}^{loc}(\alpha, \beta + i\gamma)$ since this is the local moduli space for the tame Hitchin system.

Then, from the works of \cite{biquard1996equations,kovalev1996nahm} (see also \cite{Gukov:2006jk}), it follows that the moduli space of solutions to (\ref{compelxnahm}) has the following structure,
\begin{equation}
\begin{split}
\mathcal{M}^{loc}(0,s) &= G_\mathbb{C}/L \\
\mathcal{M}^{loc}(s,e) &=  T^*(G_\mathbb{C}/P) \\
\mathcal{M}^{loc}(0,e) &= \overline{\mathcal{O}_{e}} 
\end{split}
\end{equation}
where $L$ is the Levi subgroup corresponding to the centralizer of $\beta+ i \gamma$,  $\mathfrak{l} \equiv Z_{\mathfrak{g}}(\beta + i \gamma)$.  The parabolic subalgebra $\mathfrak{p}$ has a  decomposition $\mathfrak{p}=\mathfrak{l}+n$ where $\mathfrak{l}$ is a reductive subalgebra and $\mathfrak{n}$ is its radical. The radical $\mathfrak{n}$ is nilpotent and is thus called the nilradical of $\mathfrak{p}$. The algebra $\mathfrak{l}$ is called the Levi factor of the parabolic subalgebra. \footnote{That such a decomposition exists follows from Levi's Theorem. If there are two or more such decompositions, then the corresponding Levi factors are conjugate. So, it makes sense to speak of \textit{the} Levi factor of a parabolic subalgebra. For proofs of these statements, see Chap III of \cite{jacobson1979lie}.} The condition that $\mathfrak{p}$ be a polarization for the nilpotent orbit $\mathcal{O}_e$ further implies that the nilpotent orbit $\mathcal{O}_e$ occurs as a dense orbit in the nilradical $n$ of $\mathfrak{p}$. In the context of surface operators in $\mathcal{N}=4$ SYM, the parameters $(\alpha,\beta,\gamma)$ are usually called Gukov-Witten parameters and the compact group associated to the Lie algebra $\mathfrak{l}$ is called the Levi type of the surface operator \cite{Gukov:2006jk}.


Our choice of boundary conditions breaks $SU(2)$ symmetry acting on the space of complex structures to a $U(1)$. This $U(1)$ leaves fixed a particular complex structure and this is usually called the complex structure $I$. In this complex structure, the map $\mu = T^*(G_\mathbb{C}/P) \rightarrow \overline{n} $ is a resolution of the singularities of $\overline{n}$. The construction automatically produces a holomorphic symplectic form $\Omega_I = \omega_J + i \omega_K$ which is non-degenerate on the singular and the resolved spaces. Such resolutions are termed symplectic resolutions \cite{fu2003symplectic}. We have recovered a particular symplectic resolution of a nilpotent orbit $n$ using the local geometry of the Hitchin system and sheets.  

While the subject of solutions to Nahm's equations (with the above boundary conditions) has been well studied over the years, we believe that a transparent use of the language of sheets clarifies several subtle aspects. In any case, our starting point of wanting to describe the local geometry of the Hitchin system when the residue is deformed naturally leads us to use the theory of sheets.

The canonical symplectic form on $T^*(G_\mathbb{C}/P)$ can be the viewed as the real part of a holomorphic symplectic form on the semi-simple orbit of, say, $\alpha + i \beta$.  Note that for a given sheet, there can exist different choices of parabolics $P$. In the mathematical literature, this freedom is often denoted as a choice of a polarization \cite{hesselink1978polarizations}. A given Richardson orbit may have different Dixmier sheets associated to it. But, once we fix a Dixmier sheet, then the hyper-K\"{a}hler metric on $\mathcal{M}^{loc}(s,s,s)$ is fixed. But, on the other hand, the choice of a polarization is not fixed.  The choice of a polarization is encoded in the breaking of the $SU(2)_R$ symmetry to a $U(1)_R$ and the choice of the real slice in which the real mass $m_1$ is taken to live. From a holomorphic symplectic point of view, the choice of this Parabolic enters the geometry in a crucial way. But aspects of the geometry like the  hyper-K\"{a}hler structure on the associated semi-simple orbit $G_\mathbb{C}/L_{\mathbb{C}}$ depend only on the choice of a Levi and not on the choice of a particular Parabolic $P$ whose Levi decomposition contains the Levi $L$. 

While we will almost exclusively study deformations of symplectic singularities, we wish to point out that these results are  in agreement with a theorem of B. Fu that Richardson orbits are the only orbits that can have symplectic resolutions \cite{fu2003symplectic}. Fu's proof \cite{fu2003symplectic} proceeds in two steps. First, he shows that any symplectic resolution of the closure of a nilpotent orbits has to be a Springer map $\mu : T^\star(G_\mathbb{C}/P) \rightarrow \bar{\mathcal{O}}$ for some $P$. Then, he uses prior results to classify the orbit closures for which the Springer map is a resolution. The orbits which occur in the image of the Springer map are precisely the Richardson orbits. So, the ones that admit symplectic resolutions are a necessarily a subset of the Richardson orbits. For Richardson orbits that fail to have symplectic resolutions, the Springer map $\mu : T^\star(G_\mathbb{C}/P) \rightarrow \bar{\mathcal{O}}$ still exists but it fails to be one-to-one away from the singularities of $\mathcal{O}$.\footnote{Recall that a resolution $\mu:\tilde{M} \rightarrow M$ is a birational map such that $\tilde{M}$ is smooth and $\mu$ is one-to-one away from the singularities of $M$.} This failure is related the possibility of nontrivial component groups for centralizers of a Richardson orbit $\mathcal{O}$. The more striking failure is the complete absence of a symplectic resolution (or a symplectic deformation) for the other (non-Richardson) nilpotent orbits.

To incorporate these cases, we turn to instances where we have a sheet that does not contain any semi-simple elements. In other words, it is not a Dixmier sheet.  As we explain in \S\ref{sheets}, this can happen outside of Type A. If $(\alpha',\beta',\gamma')$ are non-nilpotent values in a non-Dixmier sheet, then we first apply the additive Jordan-Chevalley Decomposition to these elements
\begin{equation}
\begin{split}
\alpha' &= \alpha_{ss} + \alpha_{n}\\
\beta' &= \beta_{ss} + \beta_{n} \\
\gamma' &= \gamma_{ss} + \gamma_{n} 
 \end{split}
 \end{equation}
such that $[\alpha_{ss},\alpha_{n}]=0$ and so on. This ensures that $\dim(\mathcal{O}_{\alpha'})=\dim(\mathcal{O}_{\alpha_{ss}}) + \dim(\mathcal{O}_{\alpha_n})$.
In this case, $\mathcal{M}^{loc}(\alpha',e)$ is not smooth since $\alpha'$ is not semi-simple. Instead, it still has a singularity which can locally be described using a Hitchin system associated to a group $L^\vee$ where $L^\vee$ is a centralizer of the semi-simple part of $\alpha'$. The local geometry of this singularity is encoded in $M^{loc}_{L^\vee}(\alpha_n,\beta_n, \gamma_n)$. This can be argued  based on the fact that the boundary values can be thought of as inducing a boundary symmetry breaking. In the realization of the defect theories as boundary conditions in 4d $N=4$ SYM or as codimension two defects in $5d \mathcal{N}=2$ SYM, this breaks the gauge symmetry at the boundary to the stabilizer of the semi-simple part of the boundary values $(\alpha, \beta, \gamma)$. At the end of the RG flow in the boundary that is triggered by these semi-simple parts, what remains can be described by the Hitchin system associated to $M_{L^\vee}$.

We will denote $M^{loc}_{L^\vee}(\alpha_n,\beta_n, \gamma_n)$ as the \textbf{residual singularity}. By virtue of the fact that it is obtained as the moduli space of solutions to a Nahm system (now for $L^\vee$), it is automatically hyper-K\"{a}hler.

\subsection{The holomorphic symplectic structure}

Let us briefly focus on the moduli space $\mathcal{M}^{loc} (0,s=\beta + i \gamma) = G_\mathbb{C}/L$. The construction of this moduli space as the space of solutions to the complex Nahm's equations endows it with a holomorphic symplectic form, the Kirillov-Kostant-Souriau (KKS) symplectic form $\Omega_{KKS}$ \cite{biquard1996equations,kovalev1996nahm}. The symplectic form $\Omega_{KKS}$ exists canonically on any co-adjoint orbit. Using the Killing form, we identify adjoint and co-adjoint orbits and hence think of them interchangeably. On a co-adjoint orbit, $\Omega_{KKS}$ can be described explicitly in the following way. First, we recall that there is the following action of $G$ on the dual of the Lie algebra $\mathfrak{g} ^\star$. For any $R \in \mathfrak{g}^\star, X \in \mathfrak{g}$, the co-adjoint action $C(g)$ acts in the following way,
\begin{equation}
\langle C(g) R, X \rangle = \langle R , Ad(g^{-1} X) \rangle.
\end{equation}

Let $\mathcal{O}$ be an orbit of this co-adjoint action and let $r\in \mathcal{O}$ be a representative of this orbit. Then, the orbit can be viewed as the homogeneous space $G/Z(r)$, where $Z(r)$ is the centralizer (in $G$) of the element $r$. The tangent space to this orbit is $\mathfrak{g} / \mathfrak{z} (r)$, where $\mathfrak{z}(r)$ is the complex Lie algebra associated to $Z(r)$. Now, given two tangent vectors $x.r$ and $y.r$ on the orbit (for $x,y \in \mathfrak{g}$, we have the following two form,
\begin{equation}
\Omega_{KKS} (x.r, y.r) = r([x,y])
\label{kksform}
\end{equation}

This is in fact a closed, non-degenerate over the entire orbit $\mathcal{O}$. When the orbit is one of semi-simple elements, it is of the form $G_\mathbb{C}/L$ for some Levi subgroup $L$ and its tangent space is $\mathfrak{g}/\mathfrak{l}$, where $\mathfrak{l}$ is a Levi subalgebra. When we fix the centralizer but vary the semi-simple element such its eigenvalues change, then the resulting new orbit can be related to the original orbit in a straightforward way. These two orbits share an underlying real manifold $G_\mathbb{C}/L$ but the holomorphic symplectic form $\Omega_{KKS}$ is different. In particular, $\Omega_{KKS}$ varies \textit{linearly} with the eigenvalues. This follows from its definition (\ref{kksform}). In our present case, it means that $\mathcal{M}^{loc}(0,s)$ is endowed with a holomorphic symplectic form which varies linearly with the eigenvalues of $s=\beta + i \gamma$. This is precisely what we expect of a mass-like deformation  of a Seiberg-Witten integrable system. Since $\Omega_{KKS}$ is non-degenerate, we have

\begin{equation}
\begin{split}
\mathrm{rank}(\Omega_{KKS}) &= \dim(\mathcal{O}) \\ &= \dim(\mathfrak{g}) - \dim(\mathfrak{z}(f))
\end{split}
\end{equation}

\textbf{An aside about notation} : Let $\alpha,\beta,\gamma$ be semi-simple elements of the Lie algebra and let $\lambda_\alpha, \lambda_\beta, \lambda_\gamma (\in \mathfrak{h}/W)$ to be the eigenvalues of $\alpha,\beta,\gamma$. In this case, one could parameterize the moduli space varying just as a function of the eigenvalues.  But, when $\alpha,\beta,\gamma$ are non semi-simple elements of the Lie algebra, the eigenvalues would make sense only for the semi-simple parts of $\alpha,\beta,\gamma$. A particular extreme case is if $\beta + i \gamma$ is nilpotent. This is a case that will be important to us. In this case, the eigenvalues $\lambda_{\beta + i \gamma}$ would all be identically zero \textit{independent} of which nilpotent orbit $\beta + i \gamma$ is valued in. So, in this case, the eigenvalues of the boundary values cease to be a useful way to keep track of what happens at the boundary and hence of the moduli space. In the literature, one sometimes finds that the boundary values $\alpha,\beta,\gamma$ and the eigenvalues of their semi-simple parts are sometimes given the same labels. We find this confusing and will choose to always draw a clear distinction between the two.

While the geometry in complex structure $I$ looks different for different choices of $P$ that share the same Levi factor $L$, the hyper-K\"{a}hler structure on the semi-simple orbits in a fixed sheet \textit{does not} depend on the choice of the parabolic $P$. This observation is essentially contained in \cite{Gukov:2006jk} but we believe it is better elucidated in the language of sheets. The choice of the parabolic $P$ is encoded in a reality condition that one imposes on the complex eigenvalues of the semi-simple orbits. As we recalled above, we have the standard KKS holomorphic symplectic form $\Omega_{KKS}$ on any simple-simple orbit. Let $\Omega_{KKS} =\omega_I + i \omega_J $. Now, imposing a reality condition on the eigenvalues allows one to write obtain a real symplectic form on a semi-simple orbit. For example, if the eigenvalues are purely real, we obtain $\omega_I$ as a real symplectic form. Once such a real symplectic form is obtained, it is possible to exhibit a symplectomorphism between the semi-simple orbit with the real symplectic form and the cotangent bundle $T^\star(G_{\mathbb{C}}/P)$ with its canonical symplectic form $\omega_{can}$ \cite{azada2008symplectic,martinez2016semisimple}, where $P$ is a parabolic corresponding to the real semi-simple element. 

\section{Sheets and  mass-like deformations of the Hitchin system}
\label{sheets}
We saw in \S\ref{SymplecticSingularities} that the theory of sheets enters the study of the local geometry of the Hitchin System in a crucial way. In this section, we expand on this by first introducing sheets in more detail and then using sheets to classify all mass-like deformations of the tame Hitchin system. One of the striking features is that the theory of sheets is substantially simpler in Lie algebras of type A. One has to study sheets in more general Lie algebras to appreciate the various subtle aspects of the theory. In this section, we only present the usual theory of sheets. This will be sufficient for the purposes of classifying mass-like deformations of the Hitchin system. But, in the next section, where we study mass deformations of the $T^\rho[G]$ theories, we will need to make an important modification to the usual theory of sheets. We will leave the explanation of this modification to \S\ref{massdeformations}.

\subsection{Theory of sheets}\label{theory_of_sheets}

As before, let $U_d$ be the union of adjoint orbits of complex dimension $d$ in a complex Lie algebra $\mathfrak{g}$. The irreducible components of $U_d$ are called \textbf{sheets}. They were first introduced and studied in  \cite{borho1979bahnen,borho1981schichten}. Useful reviews of the theory can be found in \cite{elashvili2009induced,premet2016rigid}. In this section, we will keep the discussion completely general and consider sheets for any complex Lie algebra $\mathfrak{g}$. But, we will end up applying this theory to the Langlands dual algebra $\mathfrak{g}^\vee$ in \S\ref{massdeformations}.

To understand what kinds of sheets can occur in a Lie algebra, consider the Jordan-Chevalley decomposition \footnote{For $GL_n$, this is the usual Jordan Decomposition for matrices. In the setting of arbitrary algebraic groups, the decomposition can be described directly using root data and in this more general setting, it is called the Jordan-Chevalley decomposition\cite{humphreys2012linear}.} for an arbitrary element $\tilde{a}$ in the Lie algebra.
\begin{equation}
\tilde{a} = a_{ss} + a_n,
\end{equation}
where $a_{ss}$ is the semi-simple part and $a_n$ is the nilpotent part and the decomposition is \emph{defined} such that they obey the important condition $[a_{ss},a_n]=0$. It immediately follows that $\dim(\mathcal{O}_{\tilde{a}})= \dim(\mathcal{O}_{a_{ss}}) + \dim(\mathcal{O}_{a_n})$. An arbitrary element in the Lie algebra could have a JC decomposition that is either purely semi-simple, a mix of semi-simple and nilpotent parts or purely nilpotent. These three possibilities will be important for us. 

It is known that any sheet in a complex Lie algebra contains a unique nilpotent orbit \cite{borho1979bahnen}. So, to classify sheets, we take as the starting point the classification of nilpotent orbits in $\mathfrak{g}$ and then we then look for the sheets that contain each of these nilpotent orbits. Away from the nilpotent orbit, we have the following three possibilities,

\begin{enumerate}[label=(\alph*)]
\item \textit{Dixmier sheets}: These are sheets which contain semi-simple elements,
\item \textit{Mixed sheets}: These are sheets where the non-nilpotent elements have non-zero semi-simple and nilpotent parts,
\item \textit{Rigid sheets}: The sheet contains a unique orbit which is a nilpotent orbit.
\end{enumerate}
At the boundary of each of these types of sheets, we have the following type of nilpotent orbits,
\begin{enumerate}[label=(\alph*)]
\item \textit{Richardson orbits}: They occur at the boundary of a Dixmier sheet,
\item \textit{Induced but not Richardson orbits}: They occur at the boundary of a mixed sheet,
\item \textit{Rigid orbits}: The unique orbit in a rigid sheet. 
\end{enumerate}
Each of the above three possibilities will turn out to be important for the study of deformations of the Hitchin system. But we will first describe how to classify all sheets that occur in a complex algebra and then return to the Hitchin system in \S\ref{masslikeclassification}.

In Fig \ref{sheetschematic}, we give a schematic presentation of what it means to move along a sheet.

\begin{center}
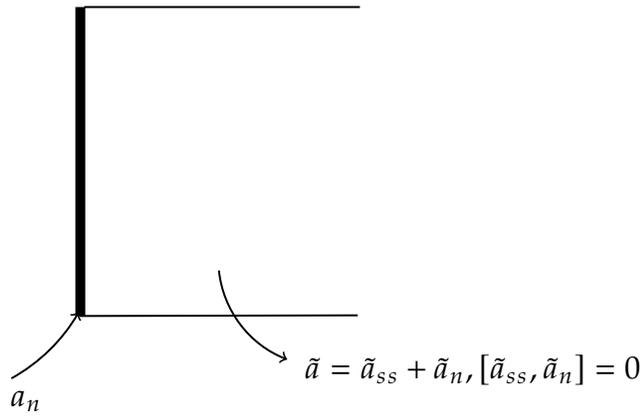
\begin{figure}
\hspace{2in}
\ifx\du\undefined
  \newlength{\du}
\fi
\setlength{\du}{15\unitlength}
\begin{tikzpicture}
\pgftransformxscale{1.000000}
\pgftransformyscale{-1.000000}
\definecolor{dialinecolor}{rgb}{0.000000, 0.000000, 0.000000}
\pgfsetstrokecolor{dialinecolor}
\definecolor{dialinecolor}{rgb}{1.000000, 1.000000, 1.000000}
\pgfsetfillcolor{dialinecolor}
\pgfsetlinewidth{0.250000\du}
\pgfsetdash{}{0pt}
\pgfsetdash{}{0pt}
\pgfsetbuttcap
{
\definecolor{dialinecolor}{rgb}{0.000000, 0.000000, 0.000000}
\pgfsetfillcolor{dialinecolor}
\definecolor{dialinecolor}{rgb}{0.000000, 0.000000, 0.000000}
\pgfsetstrokecolor{dialinecolor}
\draw (18.050000\du,8.050000\du)--(18.050000\du,15.850000\du);
}
\pgfsetlinewidth{0.050000\du}
\pgfsetdash{}{0pt}
\pgfsetdash{}{0pt}
\pgfsetbuttcap
{
\definecolor{dialinecolor}{rgb}{0.000000, 0.000000, 0.000000}
\pgfsetfillcolor{dialinecolor}
\definecolor{dialinecolor}{rgb}{0.000000, 0.000000, 0.000000}
\pgfsetstrokecolor{dialinecolor}
\draw (18.150000\du,8.050000\du)--(25.100000\du,8.050000\du);
}
\pgfsetlinewidth{0.050000\du}
\pgfsetdash{}{0pt}
\pgfsetdash{}{0pt}
\pgfsetbuttcap
{
\definecolor{dialinecolor}{rgb}{0.000000, 0.000000, 0.000000}
\pgfsetfillcolor{dialinecolor}
\definecolor{dialinecolor}{rgb}{0.000000, 0.000000, 0.000000}
\pgfsetstrokecolor{dialinecolor}
\draw (18.029289\du,15.847487\du)--(25.044297\du,15.830798\du);
}
\pgfsetlinewidth{0.050000\du}
\pgfsetdash{}{0pt}
\pgfsetdash{}{0pt}
\pgfsetbuttcap
{
\definecolor{dialinecolor}{rgb}{0.000000, 0.000000, 0.000000}
\pgfsetfillcolor{dialinecolor}
\pgfsetarrowsend{to}
\definecolor{dialinecolor}{rgb}{0.000000, 0.000000, 0.000000}
\pgfsetstrokecolor{dialinecolor}
\pgfpathmoveto{\pgfpoint{16.311190\du}{17.421981\du}}
\pgfpatharc{61}{30}{4.444977\du and 4.444977\du}
\pgfusepath{stroke}
}
\pgfsetlinewidth{0.050000\du}
\pgfsetdash{}{0pt}
\pgfsetdash{}{0pt}
\pgfsetbuttcap
{
\definecolor{dialinecolor}{rgb}{0.000000, 0.000000, 0.000000}
\pgfsetfillcolor{dialinecolor}
\pgfsetarrowsend{to}
\definecolor{dialinecolor}{rgb}{0.000000, 0.000000, 0.000000}
\pgfsetstrokecolor{dialinecolor}
\pgfpathmoveto{\pgfpoint{21.544102\du}{14.699281\du}}
\pgfpatharc{174}{111}{2.706101\du and 2.706101\du}
\pgfusepath{stroke}
}
\definecolor{dialinecolor}{rgb}{0.000000, 0.000000, 0.000000}
\pgfsetstrokecolor{dialinecolor}
\node[anchor=west] at (24.231125\du,21.770495\du){};
\definecolor{dialinecolor}{rgb}{0.000000, 0.000000, 0.000000}
\pgfsetstrokecolor{dialinecolor}
\node[anchor=west] at (20.094550\du,19.967373\du){};
\definecolor{dialinecolor}{rgb}{0.000000, 0.000000, 0.000000}
\pgfsetstrokecolor{dialinecolor}
\node[anchor=west] at (15.993331\du,17.987474\du){$a_n$};
\definecolor{dialinecolor}{rgb}{0.000000, 0.000000, 0.000000}
\pgfsetstrokecolor{dialinecolor}
\node[anchor=west] at (23.417952\du,17.209656\du){$\tilde{a}=\tilde{a}_{ss}+\tilde{a}_n, [\tilde{a}_{ss},\tilde{a}_n]=0$};
\end{tikzpicture}
\caption{A schematic view of a sheet in a complex Lie algebra. The boundary of a sheet is a nilpotent orbit containing the element $a_n$. Points away from the boundary are non-nilpotent orbits that obey the condition $\dim(\mathcal{O}_{a_n}) = \dim(\mathcal{O}_{\tilde{a}})$. Along a sheet, the eigenvalues $\lambda_{a_{ss}}^i$ of $a_{ss}$ increase in magnitude.  If it is a Dixmier sheet, then $\tilde{a}_n=0$ and the sheet contains semi-simple elements. }
\label{sheetschematic}
\end{figure}
\end{center}

\subsection{Classification of sheets and Orbit Induction}

To classify all sheets, we will need to describe the ways in which one can deform a given nilpotent orbit to a non-nilpotent orbit such that the dimension of the orbit is preserved. A systematic way to study such possibilities is through the study of \textit{orbit induction} \cite{lusztig1979induced}. A nilpotent orbit $\mathcal{O}_e$ in $\mathfrak{g}$ is said to be induced from a nilpotent orbit $\mathcal{O}_{\mathfrak{l}}$ in a proper Levi subalgebra $\mathfrak{l}$ iff there exists a sheet containing the orbit $\mathcal{O}_e$ that additionally has non-nilpotent elements with a Jordan decomposition of the form $\tilde{a}_{ss} + \tilde{a}_n$ where is the Levi $\mathfrak{l}$ is the centralizer in $\mathfrak{g}$ of the adjoint orbit $\mathcal{O}_{\tilde{a}_{ss}}$ and $\tilde{a}_n \in \mathcal{O}_l$, a nilpotent orbit in the Levi subalgebra $\mathfrak{l}$. We express the existence of such an orbit induction by,
\begin{equation}
\mathcal{O}_e = Ind_{\mathfrak{l}}^{\mathfrak{g}} (\mathcal{O}_{\mathfrak{l}})
\end{equation}

For every sheet, we thus obtain a pair $(\mathfrak{l},\mathcal{O})$, where $\mathfrak{l}$ is a Levi subalgebra and $\mathcal{O}$ is a nilpotent orbit in $\mathfrak{l}$. The pair $(\mathfrak{l},\mathcal{O})$ is called the \textit{sheet label}. To classify all sheets, one proceeds in the following way. First, we make use of the fact that nilpotent orbits in any complex Lie algebra are classified \cite{collingwood1993nilpotent}. For the classical Lie algebras, the classification is in the form of partition labels. More generally, one can classify nilpotent orbits using the Bala-Carter method \cite{bala1976classes1,bala1976classes2}. Using this classification and the explicit knowledge of how orbit induction works, one can deduce the sheets attached to each nilpotent orbit. This was first done for classical Lie algebras by Kempken-Spaltenstein \cite{kempken1983induced,spaltenstein2006classes} and for exceptional Lie algebras by Elashvili (\cite{spaltenstein2006classes}, \cite{elashvili2009induced}). In classical Lie algebras, it is possible to provide explicit combinatorial description of orbit induction. We recall this in \S \ref{inductionappendix}. In the exceptional cases, no such combinatorial algorithm exists and one has to work case by case using the tables in \cite{elashvili2009induced}.

We will now highlight some important lessons that one learns from carrying out the explicit classification of sheets in a complex Lie algebra. We encourage the reader to deduce these for herself using \S \ref{inductionappendix} and the tables in \cite{elashvili2009induced}.

\begin{enumerate}[label=(\alph*)]
\item \textit{Dixmier sheets}: There are a large class of sheets which contain semi-simple orbits. These are called Dixmier sheets. From the definition, it follows that these have sheet labels of the form $(\mathfrak{l},0)$ where $0$ denotes the trivial (zero) orbit in $\mathfrak{l}$. The nilpotent orbits occurring at their boundaries are precisely the Richardson nilpotent orbits. It turns out that every sheet in Lie algebras of type A is a Dixmier sheet.  When the residue of the Higgs field is valued in a Dixmier sheet, the local moduli space is completely smoothed. This follows from discussion in \S\ref{SymplecticSingularities}.
\item \textit{Mixed sheets and the residual singularity:} More generally, mixed sheets have sheet labels $(\mathfrak{l},\mathcal{O})$ for some non-trivial nilpotent orbit $\mathcal{O}$. When the residue of the Higgs field is valued in a mixed sheet, the local moduli space is partially smoothed. A \textit{residual singularity} remains and local form of this residual singularity is the orbit closure $\overline{\mathcal{O}}$ in the Lie algebra $\mathfrak{l}$. This follows from the discussion in \S\ref{SymplecticSingularities} where we introduced the term \textit{residual singularity}.
\item \textit{Rigid sheets:} Rigid sheets are those which contain a single nilpotent orbit as its only constituent. They exist whenever there are nilpotent orbits which do not have any non-nilpotent orbits of the same dimension. They turn out to exist in every Lie algebra other than $\mathfrak{sl}_N$. Constituents of rigid sheets are called rigid nilpotent orbits. For example, the minimal nilpotent orbit in Cartan types $B,C,D,E,F,G$ is always rigid. 
\item \textit{Sheets can meet}: One of the basic properties of a sheet in a complex Lie algebra is that it has a unique nilpotent orbit occurring at its boundary. The converse, however, need not be true. It is possible to have the same nilpotent orbit at the boundary of two or more sheets. In other words, it is possible to have distinct sheets meeting at their boundaries. Such examples do not occur in Lie algebras of type A but they are fairly common in other types. This makes the study of mass-like deformations of Hitchin systems dramatically more interesting outside of Cartan type A.
\item \textit{Richardson orbits and non-Dixmier sheets}: While it is true that every Dixmier sheet has a Richardson nilpotent orbit at its boundary, not every sheet attached to a Richardson orbit is Dixmier. Since non-Dixmier sheets do not occur in type A, this too is a phenomenon that occurs only in Lie algebras outside of type A.  
\item \textit{Special Sheets vs Non-Special sheets:}  We call a sheet attached to an orbit $\mathcal{O}_H$ a \textit{special sheet} if the residual singularity under such a deformation is a special orbit in a Levi subalgebra in the sense of Lusztig (see \cite{collingwood1993nilpotent}). The zero orbit is always a special orbit. So, it follows that all Dixmier sheets are special sheets.  Similarly, we denote a sheet attached to an orbit $\mathcal{O}_H$ to be a \textit{non-special sheet} if the residual singularity is a non-special nilpotent orbit in a Levi subalgebra. It is possible to show that non-special nilpotent orbits are attached only to non-special sheets attached and that every special nilpotent orbit is attached to at least one special sheet.\footnote{To show this, one can use a compatibility property between orbits induction and the Springer correspondence due to \cite{lusztig1979induced} and then use properties of special representations of the Weyl group \cite{lusztig1982class}. This is recalled in greater detail in \cite{Balasubramanian:2014jca}. }

But one finds that certain special nilpotent orbits do occur at the boundary of some non-special sheets. This observation will play a crucial role for us when we identify those mass-like deformation of the Hitchin that do turn out to be related to mass deformations of $T^\rho[G]$ theories. Since this is an important point, we will return to it in \S\ref{specialvsnonspecial}.
 
\item \textit{Stratification by sheets is not Whitney: } Unlike the stratification of the nilpotent cone by nilpotent orbits, the stratification of the Lie algebra $\mathfrak{g}$ by sheets does \textit{not} obey the so called \textit{frontier condition} \cite{mather2012notes}. The frontier condition requires that the closure of stratum is a union of smaller strata. But, the closure of a sheet need not be a union of smaller sheets. This is because the closure of a sheet can meet other sheets but not fully contain them. Since the frontier condition is one of the requirements for a stratification to be a Whitney stratification, it follows that the stratification of $\mathfrak{g}$ by sheets is not Whitney. 

Such a breakdown of the frontier condition occurs already in type A (see M. Bulois's appendix in \cite{carnovale2013lusztig} and \S\ref{typeAexample}). The presence of rigid nilpotent orbits leads to more such examples outside of type A.  For example, the closure of a rigid nilpotent orbit could contain several other non-rigid nilpotent orbits. But, the closure does not contain the entire sheet(s) attached to these non-rigid nilpotent orbits. 

\end{enumerate}

\subsection{Classifying mass-like deformations of the tame Hitchin system}
\label{masslikeclassification}

Having reviewed the classification of sheets, we can return to the original problem in \S\ref{localhitchin} that motivated the study of sheets in the first place. We would like to parameterize all deformation of the tame Hitchin system (for the lie algebra $\mathfrak{g}^\vee$) that obey the conditions \ref{omega} and \ref{dimensioncondition} and can thus be called a mass-like deformation of the complex integrable system. We argued in \S\ref{localhitchin} that deforming the residue of the Higgs field of the Hitchin system in such a way that it varies along a fixed sheet of the Lie algebra leads precisely to a mass-like deformation. In this section, we have learned how to classify all sheets in any given Lie algebra. Such sheets are parameterized by sheet labels $(\mathfrak{l}^\vee,\mathcal{O})$ for certain Levi subalgebras $\mathfrak{l}^\vee$ and nilpotent orbits $\mathcal{O}$ in the Levi subalgebras. Given a tame Hitchin system of type $\mathfrak{g}^\vee$ on a punctured disc where the Higgs field has a simple pole with the residue living in a nilpotent orbit $\mathcal{O}_H$, then the set of all mass-like deformations are classified by the set of all sheets whose boundary is the nilpotent orbit $\mathcal{O}_H$. If $(\mathfrak{l}^\vee,\mathcal{O})$ is the sheet label for one of those sheets, then the following should hold,
\begin{equation}
\mathcal{O}_H = Ind_{\mathfrak{l}^\vee}^{\mathfrak{g}^\vee} (\mathcal{O}) 
\label{inductionsondition}
\end{equation}
In practice, one uses (\ref{inductionsondition}) to enumerate all the possible mass-like deformations. We will take this as the starting point for \S\ref{massdeformations} where we will further identify the subset of the mass-like deformations that admit an interpretation as a mass deformation of the $T^\rho[G]$ theories.

\subsection{A finite group action on mass-like parameters}
\label{finitegroupQ}

But, before turning to a study of mass deformations of $T^\rho[G]$ theories, we take a short detour to make note of an important feature enjoyed by all mass-like deformations of the tame Hitchin integrable system. From the description of the mass-like deformations, it follows that there is a natural finite group that acts on the space of mass-like deformations and we will explicitly identify this finite group. This finite group $Q$ is identified in the following way. First, note that the normalizer of $L^\vee$ in $G^\vee$, denoted by $N_{G^\vee}(L^\vee)$, acts naturally on every nilpotent orbit in $\mathfrak{l}^\vee$. Denote by $\mathcal{Q}$, the following finite group
\begin{equation}
\mathcal{Q} = N_{G^\vee} (L^\vee, \mathcal{O}) / L^\vee
\end{equation}
where $N_{G^\vee} (L^\vee, \mathcal{O})$ denotes the subgroup in $N_{G^\vee}(L^\vee)$ that stabilizes the nilpotent orbit $\mathcal{O}$. For a sheet parameterized by $(\mathfrak{l}^\vee,\mathcal{O})$, this finite group acts on the eigenvalues of the semi-simple parts of the non-nilpotent elements in the sheet. In other words, $\mathcal{Q}$ acts on $Z(\mathfrak{l}^\vee)$ \cite{losev2016deformations}.  

In special cases, this is straightforward to see. For example, when $\mathfrak{l}^\vee$ is the trivial Levi subalgebra, then $\mathcal{Q} = N_{G^\vee}(T)/T$, for a Cartan torus $T$. But, $N_{G^\vee}(T)/T$ is nothing but $W(G^\vee)$, the Weyl group. So, we recover the statement that, for the principal/regular sheet in a Lie algebra $\mathfrak{g}^\vee$, the eigenvalues of the regular semi-simple elements carry a natural $W(G^\vee)$ action. The other extreme case is  when $L^\vee = G^\vee$, corresponding to the zero orbit (or the trivial sheet). Here, we have $\mathcal{Q}=\{ Id \}$, the trivial subgroup of $W(G^\vee)$. For all the other sheets that contain non-nilpotent elements, the finite group $\mathcal{Q}$ will be a non-trivial proper subgroup of $W(G^\vee)$. 

From the point of view of the complex integrable system, any mass-like deformation that is parameterized by an element $m_i \in Z(\mathfrak{l}^\vee)$ is equivalent to a deformation with a parameter $\lambda m_i$, where $\lambda \in \mathcal{Q}$. 

\section{Mass Deformations of $T^\rho[G]$ theories}
\label{massdeformations}

In the previous section, we studied \textit{mass-like} deformations of the tame Hitchin system on a punctured disc and obtained an explicit classification of such deformation using sheets. In this section, we investigate which among these mass-like deformations admit an interpretation as a mass deformation of one of the $T^\rho[G]$ theories. As noted in the Introduction,  a relation between the theory of sheets and mass deformed $T^\rho[G]$ theories was already proposed in \cite{Chacaltana:2012zy}. This proposal fully describes mass deformations in type A. What follows can be taken to be a precise realization of this proposal outside of type A. 

\subsection{Recollections about 3d $\mathcal{N}=4$ SCFTs}
\label{recollections}

Here, we recall some important aspects of 3d $\mathcal{N}=4$ theories SCFTs. The 3d $\mathcal{N}=4$ superconformal algebra is $\mathfrak{osp}(4 \mid 4)$. The $R$-symmetry group is $Spin(4) \simeq SU(2)_H \times SU(2)_C$. Typically, a 3d $\mathcal{N}=4$ SCFT has an exact moduli space of vacua of the form $\mathcal{H}_i \times \mathcal{C}_i$, where $\mathcal{H}_0$ (called the Higgs branch) is the component of the moduli space on which $SU(2)_C$ acts trivially while $\mathcal{C}_0$ (called the Coulomb branch) is the component on which $SU(2)_H$ acts trivially. The other components of the moduli space of vacua that are of the form $\mathcal{H}_i \times \mathcal{C}_i, i \neq 0$ are usually termed mixed branches. They carry non-trivial actions of both $SU(2)_H$ and $SU(2)_C$. Due to the existence of eight real supercharges, both the Higgs and Coulomb branches, when non-trivial, are guaranteed to be hyper-K\"{a}hler spaces. 

In theories with UV Lagrangians, the classical Higgs branch is parameterized by vacuum expectation values of the hypermultiplet scalars. The geometry on the Higgs branch, including the Riemannian metric on it, is not corrected in the quantum theory. The case of the Coulomb branch is quite different. When we have a UV gauge theory flowing in the IR to a 3d $\mathcal{N}=4$ SCFT, the classical Coulomb branch is parameterized by vacuum expectation values of the vector multiplet scalars \textit{and} the vacuum expectation values of the dual photons \cite{Intriligator:1996ex}. In the quantum theory, the VEVs of dual photons are replaced by VEVs of monopole operators \cite{Borokhov:2002cg}. At a generic point of the Coulomb branch, the gauge group is Higgsed to $U(1)^r$, where $r$ is the rank of the gauge group. Both the Higgs and Coulomb branches are singular spaces. A singularity in the vacuum moduli space in the IR effective field theory signals a break down of the EFT and the appearance of additional massless states.

Several examples of 3d $\mathcal{N}=4$ theories have been studied and their Higgs and Coulomb branches have been identified using various techniques. See \cite{Hanany:1996ie,deBoer:1996ck,Hori:1997zj} for some of the early work in this direction. Subsequent works are too numerous to admit a full recollection here. But we note that a common theme in the literature has been the identification of the Higgs and Coulomb branches using geometric realizations of the 3d $\mathcal{N}=4$ theories using String/M-theory. The verification of dualities like 3d Mirror symmetry, which, among other things, exchanges the Higgs and Coulomb branches, has been another important motivation for the study of 3d $\mathcal{N}=4$ SCFTs. In such investigations, it is useful to be able to calculate certain observables on both sides of the duality to check that they match. Among the natural observables to calculate on the quantum Coulomb branch are its ring of holomorphic functions, called the \textit{chiral ring} and the exact hyper-K\"{a}hler metric. For a large class of examples that includes the $T^\rho[G]$ theories, this chiral ring has been studied in detail using the Hilbert Series technique in \cite{Cremonesi:2013lqa,Cremonesi:2014vla,Cremonesi:2014uva,Carta:2016fjb,Hanany:2016gbz,Hanany:2017ooe}. More recently, there has also been progress in providing a mathematically precise definition of the Coulomb branch chiral ring for many of these theories \cite{Nakajima:2015txa,2015arXiv151003908N,braverman2016towards}. 

 Some of these known $\mathcal{N}=4$ SCFTs have UV Lagrangian descriptions while there many that are not known to have a UV Lagrangian. In the instances where UV Lagrangians are not known, one can still study the IR theory if there is an independent way to identify its Higgs and Coulomb branches. This is the approach we take in this paper to study the $T^\rho[G]$ theories. However, when one has a UV Lagrangian, it is desirable to have a general prescription for deriving the geometry of the quantum Coulomb branch together with its hyper-K\"{a}hler metric. Such a prescription has been provided recently in \cite{Bullimore:2015lsa}. Their prescription involves a map between the VEVs in the UV non-abelian gauge theory to the VEVs of scalars in the IR theory, an abelian gauge theory. This map, called an \textit{abelianization map}, is then used in conjunction with localization techniques to define the hyper-K\"{a}hler metric on the IR moduli space for gauge theories with unitary gauge groups \cite{Bullimore:2015lsa}. One expects to have a generalization of this for the case with SO/USp gauge groups as well. The theories that we consider in this paper are known to have UV Lagrangians in some special cases. For instance, the $T^\rho[SU(N)]$ theories can be described as certain Quiver gauge theories with fundamental and bi-fundamental matter. For $G=SO/USp$,  $T^\rho[G]$ admit Lagrangian descriptions for certain $\rho$. We provide a short summary of these results in \S \ref{uvlagrangians}. But, we note here that for general $G$ and $\rho$, there is no known Lagrangian description for the $T^\rho[G]$ theories. In arriving at the results in this paper, we have not assumed the existence of a UV Lagrangian. In the instances where the $T^\rho[G]$ do have Lagrangians,\footnote{ In purely mathematical terms, this amounts to a question of when certain non-compact hyper-K\"{a}hler spaces can be obtained using finite dimensional hyper-K\"{a}hler quotient constructions.}  it would be interesting to obtain these results using these Lagrangians descriptions.

\subsection{Relevant Deformations for 3d $\mathcal{N}=4$ SCFTs}
\label{relevantrecollections}
In this paper, we want to study not just the superconformal theory $T^\rho[G]$ but the theory $T^\rho[G]$  together with a possibly non-zero mass deformation. From \cite{Cordova:2016xhm}, we note that the possible $\mathcal{N}=4$ preserving relevant deformations of any $\mathcal{N}=4$ SCFTs fall into the following three categories (see also \cite{Bhattacharya:2008zy,Dolan:2008vc}). 

\begin{enumerate}[label=(\alph*)]
\item \textit{Mass deformations} : These are the relevant deformations in the $B_1[(2,0),1]$ multiplet of $\mathfrak{osp}(4 \mid 4)$. They occur in the same multiplet as the flavour current for Higgs branch. They are charged under $SU(2)_C$ and neutral under $SU(2)_H$.
\item \textit{Twisted Mass deformations} : These are the deformations in the $B_1[(0,2),1]$ multiplet of $\mathfrak{osp}(4 \mid 4)$.  They occur in the same multiplet as the flavour current for Coulomb branch. They are charged under $SU(2)_H$ and neutral under $SU(2)_C$.
\item \textit{Universal Mass Deformation} : These are the deformations in the multiplet $A_2[(0,0),1]$ of $\mathfrak{osp}(4 \mid 4)$. They occur in the same multiplet as the stress tensor.
\end{enumerate}

In the present paper, we are focused only on mass deformations of the $T^\rho[G]$ theories and their effect on the geometry of the vacuum moduli spaces. Twisted mass deformation and the universal mass deformation of $T^\rho[G]$ do not play any role in our discussion.

\subsection{Deformations of hyper-K\"{a}hler Singularities and 3d $\mathcal{N}=4$ SCFTs}

As we noted in \S\ref{recollections}, the vacuum moduli spaces $\mathcal{H}_0$ and $\mathcal{C}_0$ are singular hyper-K\"{a}hler spaces. The most singular stratum (usually termed \textit{the SCFT point}) is common to the two moduli spaces. Each of the moduli spaces is an example of a symplectic singularity \cite{beauville2000symplectic}. These are singular space that admit a non-degenerate holomorphic symplectic form $\Omega$. These singularities and their possible resolutions and deformations is an actively studied mathematical subject. We refer to \cite{fu2006survey,kaledin2009geometry,2017arXiv170100713O} for some surveys of the mathematical literature. 

When a mass deformation is added to UV Lagrangian, the flow to the IR is modified and this affects the geometry of the IR vacuum moduli space. The resulting Higgs branch is smaller and the Coulomb branch is less singular than the one for the case without mass deformation. When the resulting Coulomb Branch is smooth, it implies that the mass deformation followed by a flow to the IR leaves us with a family of free theories parameterized by the smoothed Coulomb Branch. The Higgs branch is fully lifted in this case. But, as we will see in several examples, this possibility of fully lifting the Higgs branch by a mass deformation does not always exist for a 3d $\mathcal{N}=4$ SCFT.

The flavour symmetry $F$ acts as continuous hyper-K\"{a}hler isometries of the Higgs branch. The mass deformations is valued in the Cartan subalgebra $\mathfrak{h}(F)$. More accurately the mass parameters of a 3d $\mathcal{N}=4$ SCFT form a $SU(2)_C$ triplet $(m_1,m_2,m_3)$ where each of the $m_i$ are valued in the Cartan subalgebra $\mathfrak{h}(F)$. If we fix ourselves to one of the complex structures of $\mathcal{C}_0$, this breaks the $SU(2)_C$ R-symmetry to a $U(1)_C$. Then the masses are naturally split into a real mass $m_1$ which is not charged under the $U(1)_C$ and a complex mass $m_{\mathbb{C}} = m_2+ i m_3$ which is charged under the $U(1)_C$. We confine ourselves to the study of the effect of a non-zero complex mass deformation $m_{\mathbb{C}}$. The case of more general mass deformation can always to related to a case with a pure complex mass deformation by a suitable $SU(2)_C$ rotation on the space of mass deformation parameters. This is equivalent to a choice of a distinguished complex structure on the deformed geometry. We will call this distinguished complex structure to be the complex structure $J$.

In mathematical terms, a complex mass deformation can be interpreted as a deformation of a holomorphic symplectic structure in the following way. Recall that any hyper-K\"{a}hler manifold is equipped with a family of holomorphic symplectic forms $\Omega^\zeta$, where $\zeta \in \mathbb{CP}^1$ is a co-ordinate on the associated twistor space. Let $\Omega_J$ correspond to the holomorphic symplectic form in the complex structure $J$. As the complex mass parameter is varied, we obtain a family of Coulomb branches $\mathcal{C}_0 [m_{\mathbb{C}}]$ that are fibered over the space of mass parameters $\mathcal{S}$. Computing the de-Rham cohomology class $[\Omega_{m_\mathbb{C}}]$ over the space of mass parameters gives a map $\mathcal{P} : \mathcal{S} \rightarrow  H^2(\mathcal{C}_0)$. This map $\mathcal{P}$ is often called a \textit{period map} \cite{kaledin2002period} and is used to study the moduli space of complex structure deformations of $(\mathcal{C}_0,\Omega)$.

Based on the study of Coulomb branches in Lagrangian theories, we expect the $[\Omega_{m_\mathbb{C}}]$ to vary in the following way,
\begin{equation}
\begin{split}
[\Omega_{m_\mathbb{C}}] &= \sum_{i} m^a [C_a] \\ 
\lim_{m_{\mathbb{C}} \rightarrow 0}[\Omega_{m_\mathbb{C}}] &= [\Omega_0] 
\label{masslinear}
\end{split}
\end{equation}
where $m^a$ are natural co-ordinates on $\mathcal{S}$, $a=1,2,\ldots, \mathrm{rank}(F)$ and $[C_i]$ are some standard representatives for $H^2(\mathcal{C}_0)$ and $\Omega_0$ is the holomorphic symplectic structure on the singular, undeformed Coulomb branch. The decomposition (\ref{masslinear}) is not unique and should be understood up to the action of $W(F)$, the Weyl group of $F$, on $\mathcal{S}$ and the action of any non-trivial dualities under which $[\Omega]$ is expected to be invariant \cite{Seiberg:1994rs,Seiberg:1996nz}. Now, (\ref{masslinear}) gives an embedding of the space of mass parameters $\mathcal{S}$ into the space of complex structure deformations of $(\mathcal{C}_0,\Omega_0)$.

\subsubsection{Deformation Quantization}

Although we do not consider this in this paper, we note in passing an interesting feature of these vacuum moduli spaces. Since the moduli spaces in question are holomorphic symplectic, it is possible to consider the question of whether the Poisson structure on these manifolds can be further deformed in a way that mimics passage from classical mechanics to quantum mechanics. Such a deformation is usually termed a deformation quantization of the holomorphic symplectic space in question.  At least three seemingly distinct such quantizations have been studied from a physical standpoint. The first is a 3d analog of the $\Omega_{\epsilon}$ background \cite{Yagi:2014toa,Bullimore:2015lsa}, the second  is a deformation that becomes available  at the superconformal fixed point \cite{Beem:2016cbd}\footnote{We thank Wolfger Peelaers for a related discussion.} and the third is the quantization arising from studying sphere partition functions \cite{Dedushenko:2017avn}. It would be of interest to study each of these deformation quantizations for Coulomb branches of an arbitrary $T^\rho[G]$ theory. The Coulomb branches of the Rigid SCFTs that we introduce in the \S\ref{taxonomy} below would be particularly interesting examples to study in the above frameworks. Quantizations also play an important role in the original formulation of \textit{Symplectic Duality} in \cite{braden2014quantizations}. We discuss the relation between our setup and that in \cite{braden2014quantizations} in \S\ref{SymplecticDuality}.

\subsection{$T^\rho[G]$ theories under mass deformations}
\label{taxonomy}

We finally turn to discussing the behaviour of $T^\rho[G]$ theories under mass deformations. We take as a starting point the classification in \S\ref{masslikeclassification} of mass-like deformations of the tame Hitchin integrable system by sheets in the Lie algebra $\mathfrak{g}^\vee$. However, in order for such a deformation to be an actual mass deformation of a $T^\rho[G]$ theory, it has to additionally obey the Flavour Condition \ref{flavour}. This constraint follows from the fact that mass deformations occur in the same superconformal multiplet as the flavour current (we reviewed this in \S\ref{relevantrecollections} ). This is the final step in identifying the mass deformed geometry of the Coulomb branch. For every defect $(\mathcal{O}_N,{\mathcal{O}_H})$, the geometry of the mass deformed SCFT is captured by a triplet $(\mathcal{O}_N,\mathcal{O}_H, (\mathfrak{l}^\vee,\mathcal{O}_H^{l^\vee}))$, where $(\mathcal{O}_N,\mathcal{O}_H)$ are the Nahm and Hitchin labels as before and $(\mathfrak{l}^\vee,\mathcal{O}_H^{l^\vee})$ is the label for a special sheet/sub-sheet attached to the Hitchin orbit $\mathcal{O}_H$. We summarize here all the steps involved in arriving at this triplet. 

\begin{enumerate}
\item We first identify the Bala-Carter Levi $\mathfrak{l}_{BC}$ associated to the Nahm orbit $\mathcal{O}_N$ which is a nilpotent orbit in $\mathfrak{g}$.
\item If $\mathcal{O}_N$ corresponds to a nilpotent orbit of principal Levi type (see \S\ref{pltype} for the definition), then the mass deformed Coulomb branch is described by the sheet in the dual Lie algebra $\mathfrak{g}^\vee$ with label $(\mathfrak{l}^\vee_{BC},0)$. In this case, one can verify that imposing the Flavour condition does not lead to any further restrictions of the space of allowed mass parameters. The massless limit of the Coulomb branch is given by the Richardson nilpotent orbit at the boundary of the sheet $(\mathfrak{l}^\vee_{BC},0)$. One can also check that this obeys $\mathcal{O}_H = d_{BV}(\mathcal{O}_N)$ as conjectured in \cite{Chacaltana:2012zy}.

\item If $\mathcal{O}_N$ is not of principal Levi type, we look at all sheets attached to $\mathcal{O}_H = d_{BV}(\mathcal{O}_N)$ and identify the sheet whose associated Levi is $\mathfrak{l}^\vee_{BC}$. If such a sheet exists, we observe that it is unique. This sheet then describes the mass deformed Coulomb branch of $T^\rho[G]$.
\item If such a sheet does not exist in step 3 above, then we find that it is always possible to find a unique inclusion $\mathfrak{l}^\vee_{sheet} \subset \mathfrak{l}^\vee_{BC}$ such that $Z(\mathfrak{l}^\vee_{BC}) \subset Z(\mathfrak{l}^\vee_{sheet})$. In these instances, the deformed Coulomb branch is a subspace of an ordinary sheet and not the entire sheet. We call such subspaces \textit{sub-sheets}. The appearance of sub-sheets is the basic reason why the story of mass deformations leads to a slight refinement of the usual theory of sheets. 
\item Corresponding to the sheet or sub-sheet from step 2, 3 or 4, we find that
\begin{equation}
\mathcal{O}_H = Ind_{\mathfrak{l}^\vee}^{\mathfrak{g}^\vee} (\mathcal{O}_H^{\mathfrak{l}^\vee}),
\end{equation}
for an unique special nilpotent orbit $\mathcal{O}_H^{\mathfrak{l}^\vee}$ in the Lie algebra $\mathfrak{l}^\vee_{BC}$. So, it follows that the sheet/sub-sheet that we identified at the end of step 4 is actually a \textit{special} sheet/sub-sheet (in accordance with what is means for a sheet to be a special sheet as defined in \S\ref{sheets}). This is how we arrive at the third entry  $(\mathfrak{l}^\vee,\mathcal{O}_H^{l^\vee})$ in the triplet $(\mathcal{O}_N,\mathcal{O}_H, (\mathfrak{l}^\vee,\mathcal{O}_H^{l^\vee}))$.
\end{enumerate}

By carrying out the above steps for all simple $G$, we find that $T^\rho[G]$ fall into the following three \textit{deformations classes}. 

\begin{enumerate}[label=(\alph*)]
\item Theories where the flavour symmetry $F$ is non-trivial and the sheet label for the mass deformation is of the form $(\mathfrak{l}^\vee, 0)$. The Coulomb branches of these theories are completely smoothed by turning on all the available mass parameters. We denote such theory to be a \textbf{Smoothable SCFT}. 
\item Theories for which $F$ is trivial. We denote such a theory to be a \textbf{Rigid SCFT}. The Higgs branch of a Rigid SCFT (among the $T^\rho[G]$ theories) is always a Slodowy Slice to a distinguished nilpotent orbit in $\mathfrak{g}$. 
\item Theories where the flavour symmetry $F$ is non-trivial and the sheet label for the mass deformation is of the form $(\mathfrak{l}^\vee, \mathcal{O})$ for some non-zero nilpotent orbit $\mathcal{O}$ in the Levi $\mathfrak{l}^\vee$.  The Coulomb branches of such theories have a residual singularity even after all the available mass deformations and turned on (and we flow to the IR). We denote such theory to be a \textbf{Malleable SCFT}. In these cases, RG flow after a full mass deformation leaves behind (in the IR) a non-trivial smaller SCFT that is Rigid. 
\end{enumerate}

We illustrate the existence of the three deformation classes in a schematic way in Fig \ref{deformationclassesfig}. For rigid and malleable $T^\rho[G]$ theories, our identification of the mass deformed Coulomb branch depends on the conjecture about the massless Coulomb branch in \cite{Chacaltana:2012zy}. However, for smoothable cases, our results can be viewed a proof of this conjecture.

 While our focus in this paper has been about $T^\rho[G]$ theories, much of the geometry carries over to more general 3d $\mathcal{N}=4$ theories. In particular, the Coulomb branches of \textit{any} 3d $\mathcal{N}=4$ theory is a symplectic singularity and depending on the behaviour of this symplectic singularity under deformations, the theory can be slotted into one of the above deformation classes. 

\begin{changemargin}{1.5cm}{1.5cm}
\begin{center}
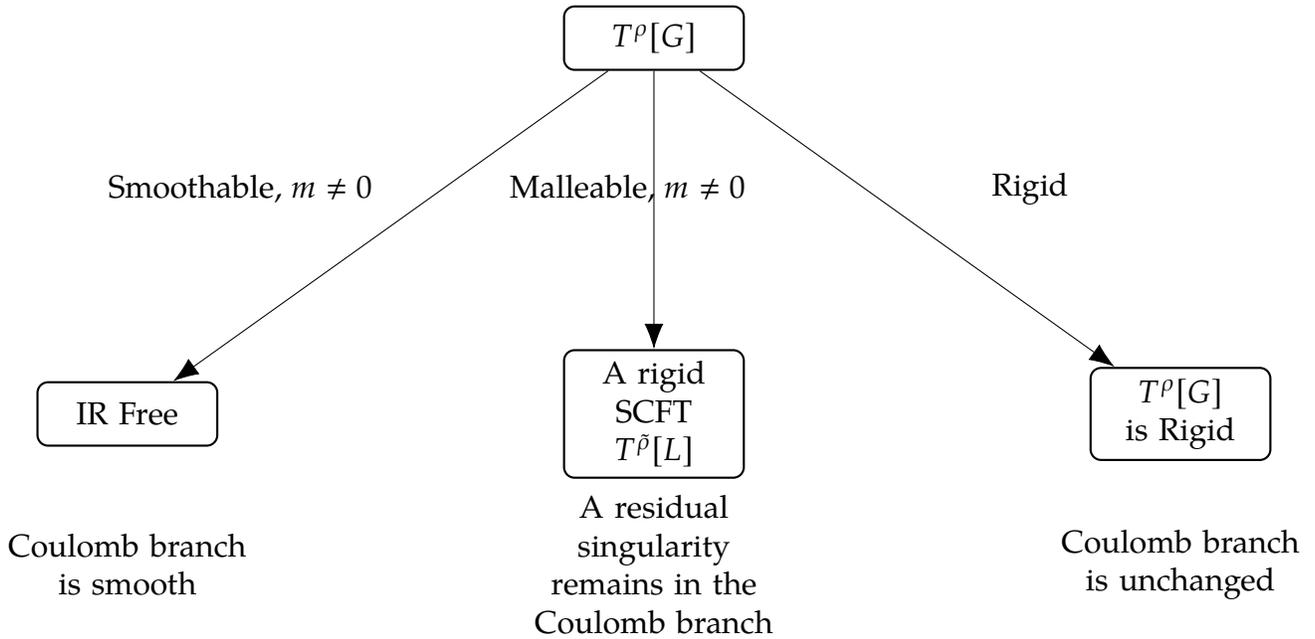
\begin{figure}
\begin{tikzpicture}[%
    auto,
    block/.style={
      rectangle,
      draw=black,
      thick,
      text width=5em,
      align=center,
      rounded corners,
      minimum height=2em
    },
     block1/.style={
      rectangle,
      draw=white,
      thick,
      text width=8em,
      align=center,
      rounded corners,
      minimum height=2em
    },
    line/.style={
      draw,thick,
      -latex',
      shorten >=2pt
    },
    cloud/.style={
      draw=red,
      thick,
      ellipse,
      fill=red!20,
      minimum height=1em
    }
  ]
    \draw (7,0) node[block] (A) {$T^\rho[G]$};
    \draw (0,-5) node[block] (S) {IR Free};
     \draw (0,-7) node[block1] (S') {Coulomb branch is smooth};
    \draw (7,-5) node[block] (M) {A rigid SCFT $T^{\tilde{\rho}}[L]$};
    \draw (7,-7) node[block1] (M') {A residual singularity remains in the Coulomb branch};
    \draw (14,-5) node[block] (R) {$T^\rho[G]$ is Rigid};
     \draw (14,-7) node[block1] (R') {Coulomb branch is unchanged};
  \draw[-{Latex[black,scale=2]}] (A) -- (S); 
  \draw[-{Latex[black,scale=2]}] (A) -- (M);
  \draw[-{Latex[black,scale=2]}] (A) -- (R);  
 \draw (1.5,-2) node (SS) {Smoothable, $m \neq 0$} ;  
 \draw (6.65,-2) node (MM) {Malleable, $m \neq 0$} ;
 \draw (12,-2) node (RR) {Rigid } ;
 
\end{tikzpicture}
\caption{The three deformation classes into which $T^\rho[G]$ theories fall and the corresponding IR end points under a full mass deformation $m \in \mathfrak{h}(F)$ (when $F$ is non-trivial).}
\label{deformationclassesfig}
\end{figure}
\end{center}
\end{changemargin}

The separation of the defects into the three deformation classes is best understood by way of working out explicit examples. We do this in \S\ref{tables} and \S \ref{MoreTaxnonomy} in the form of tables. To compile the tables, we begin with the pair of nilpotent orbits $(\mathcal{O}_N, \mathcal{O}_H)$ describing the massless limit and then impose the  conditions (\ref{omega}),(\ref{dimensioncondition}) and (\ref{flavour}). We recall these conditions here for convenience. We will denote the residue of the Higgs field before mass deformation to be $a_0$ and the residue after mass deformation to be $a_M$. We will denote the corresponding adjoint orbits in $\mathfrak{g}^\vee$ by $\mathcal{O}_{a_0}$ and $\mathcal{O}_{a_M}$.  

Now, condition (\ref{omega}) requires that
\begin{equation}
\boxed{[\Omega]_{m \neq 0} \propto m_i}
\end{equation}
\noindent
and the dimension condition (\ref{dimensioncondition}) requires that
\begin{equation}
\boxed{\dim(\mathcal{O}_{a_M}) = \dim(\mathcal{O}_{a_0})}
\end{equation}
\noindent
and finally, the Flavour Condition (\ref{sheetbclevisame}) (which is equivalent to \ref{flavour}) requires that
\begin{equation}
\boxed{\mathfrak{l}^\vee = \mathfrak{l}_{BC}^{\vee}}
\end{equation}

We find that the above conditions have a \textit{unique} solution and it is this solution that is depicted in the tables in \S\ref{tables} and Appendix \ref{MoreTaxnonomy}. From this solution, one also observes, on a case-by-case basis, that $\mathcal{O}_H^{l^\vee}= d_{BV}(\mathcal{O}_N^l)$ where $\mathcal{O}_N^l$ is the distinguished nilpotent orbit in $\mathfrak{l}$ that is part of the Bala-Carter label for the Nahm orbit $\mathcal{O}_N$. We did \textit{not} impose this in the three conditions above. So, the fact that this comes out automatically is a  highly non-trivial observation. What we have just discovered for ourselves is the Spaltenstein-Barbasch-Vogan theorem about how orbit induction interacts with the order-reversing duality and Bala-Carter theory (see Theorem 8.3.1 in \cite{collingwood1993nilpotent}). What we have seen here is a physical interpretation of this theorem.

\subsection{A physically motivated refinement of the theory of sheets}

 In the usual mathematical theory, the parameterization of sheets is done by pairs $(\mathfrak{l}^\vee,\mathcal{O})$, where $\mathcal{O}$ is a rigid nilpotent orbit in the Levi subalgebra $\mathfrak{l}^\vee$. However, in the context of the present paper, it is Rigid SCFTs that play a role analogous to the one played by rigid orbits in the usual theory. Since Rigid SCFTs can have Coulomb branches that are non-rigid orbits, we are led to a slight refinement of the usual theory of sheets.  We will retrieve an ordinary sheet label $(\mathfrak{l},\mathcal{O})$ as the residual singularity only if the residual singularity happens to involve a rigid Hitchin orbit (in the usual structure theory sense). In the more general situation, mass deformations will turn out to be parameterized by a sub-sheet and not an entire sheet. So, by a refined theory of sheets, we really mean a theory of sheets attached directly to the pair $(O_N,O_H)$, where $\mathcal{O}_N$ is any nilpotent orbit, $\mathcal{O}_H$ is a special nilpotent orbit such that $\mathcal{O}_H = d_{BV}(\mathcal{O}_N)$ and $\mathfrak{l}^\vee_{\mathrm{sheet}/\text{sub-sheet}} = \mathfrak{l}^\vee_{BC}$, where $\mathfrak{l}_{BC}$ is the Bala-Carter Levi associated to $\mathcal{O}_N$.
 
 If one compares the stratification of $\mathfrak{g}^\vee$ by spaces of mass deformations to the stratification of $\mathfrak{g}^\vee$ by sheets, we observe that the stratification by spaces of mass deformations is a finer stratification. This is why we have termed this a refinement of the usual theory of sheets. In our setup, we needed to only consider this refinement for special sheets attached to special nilpotent orbits. It would be interesting to know if a natural refinement also exists for non-special sheets attached to special/non-special nilpotent orbits. 
 
 In the usual theory, $\mathrm{rank}(\mathrm{sheet})$ plays an important role since it counts the number of distinct eigenvalues of the semi-simple part of any non-nilpotent orbit in a sheet.  In our setup, the role of $\mathrm{rank}(\mathrm{sheet})$ in the usual theory of sheets is replaced by $\mathrm{rank}(F)$, where $F$ is the flavour symmetry associated to the defect. In instances where a non-trivial sub-sheet parameterizes the mass deformed Coulomb branch, the following holds
 \begin{equation}
 \mathrm{rank}(F) = \mathrm{rank}(\text{sub-sheet}) < \mathrm{rank}(\mathrm{sheet}).
 \end{equation}
 
 From the examples in \S\ref{tables}, one observes that it is entirely possible for a full mass deformation to leave behind a residual SCFT with a Coulomb Branch that is not a rigid nilpotent orbit (see \S\ref{e6examples} for a specific case in $E_6$). But, it will always be true that we have $\mathrm{rank}(F) =0$ for the residual SCFT. It is this condition that defines a Rigid SCFT. 
\subsection{A finite group action on mass parameters}
\label{finitegrouponmasses}

In \S\ref{finitegroupQ}, we noted that the space of mass-like deformations of the tame Hitchin system admits a natural action of a finite group $\mathcal{Q}$. This finite group can be obtained as $\mathcal{Q} = N_{G^\vee} (L^\vee, \mathcal{O}) / L^\vee$, where $(\mathfrak{l}^\vee,\mathcal{O})$ is the sheet label associated to the mass-like deformation and $N_{G^\vee} (L^\vee, \mathcal{O})$ is the subgroup of the normalizer $N_{G^\vee} (L^\vee)$ that leaves fixed the nilpotent orbit $\mathcal{O}$.  For describing actual mass deformations, we have seen that it might be necessary to restrict to a sub-sheet of a full sheet. So, in general, the sheet label corresponding to a mass deformation could be a pair $(\mathfrak{l}^\vee,\mathcal{O})$, where $\mathcal{O}$ is not a rigid orbit. One can define a finite group $\mathcal{Q}$ in an identical fashion in these instances and $\mathcal{Q}$ will act in a similar on the space of actual mass parameters $m_i \in Z(\mathfrak{l}^\vee)$. In the case of an actual mass deformation, this finite group action is nothing but the action of $W(F)$, the Weyl group of the Flavour symmetry group $F$ on the space of mass parameters $m_i$. One can \textit{derive} this fact by nothing that the definition of $\mathcal{Q}$ together with the Flavour Condition \S\ref{sheetbclevisame} implies that that $\mathcal{Q}$ is equivalent, as a Coxeter group, to $W(F)$. Such finite group actions (and their categorifications) play an important role in the relation between 3d $\mathcal{N}=4$ theories and \textit{Symplectic Duality}. We discuss this further in \S\ref{SymplecticDuality}.

\subsection{Mass deformations and the local AGT correspondence}

In the relation between sphere partition functions of 4d Class $\mathcal{S}[\mathfrak{j}]$ SCFTs and 2d Toda correlations functions, known as the Alday-Gaiotto-Tachikawa (AGT) correspondence \cite{Alday:2009aq}, one of the important elements in the dictionary is a map from each codimension two defect of the six dimensional theory and certain semi-degenerate primaries of the 2d Toda CFT \cite{Kanno:2009ga}. This is the \textit{local} content of the AGT correspondence. It was first noted in \cite{Balasubramanian:2014jca} that these semi-degenerate primaries can be described by null vectors at level one precisely when the Nahm label associated to the defect is principal Levi type. Using this observation, the primary map was described for such defects in \cite{Balasubramanian:2014jca}. See also \cite{Haouzi:2016ohr} in this regard. But, we have just observed that such defects are precisely the ones which correspond smoothable SCFTs. In other words, these are the SCFTs that admit a mass deformation such that the resulting IR theory is free. So, the smoothable SCFTs admit the most straightforward primary map. This further raises a natural question of what the Toda primaries associated to Malleable and Rigid SCFTs are. A closely related phenomenon is that fact that, to every such semi-degenerate primary labeled by the Nahm orbit $\mathcal{O}_N$, one can associated a \textit{pair} of W-algebras, $W(\mathfrak{g},\mathcal{O}_N)$ and $W(\mathfrak{g}^\vee,\mathcal{O}_H)$, where $\mathcal{O}_H$ is the corresponding Hitchin orbit, using quantum Hamiltonian reduction for the corresponding Jacobson-Morozov $\mathfrak{sl}_2$ embeddings $\rho_N,\rho_H$. It would be intriguing to study such pairs of W-algebras corresponding to the Rigid SCFTs among $T^\rho[G]$. It would also be interesting to compare this with other associations between 3d $\mathcal{N}=4$ SCFTs and Vertex Operator Algebras \cite{Gaiotto:2017euk,Creutzig:2017uxh}.

\subsection{More about Deformation Classes of $T^\rho[G]$ theories}

In this section, we outline in a bit more detail the relationship of the deformation class of a $T^\rho[G]$ theory and notions that appear in the structure theory of nilpotent orbits in complex semi-simple Lie algebras.\footnote{For this section, we benefited from conversations with Pramod Achar.}  A physically minded reader can skip this section and proceed to the subsequent section where examples are treated.

In what follows, we use several properties of nilpotent orbits. The corresponding definitions are recalled in  \S \ref{definitionsappendix}. We will additionally use `pL' to denote nilpotent orbits that are of \textit{principal Levi type} (see \ref{pltype} for the definition) and `npL' to denote nilpotent orbits that not of principal Levi type. 

\begin{itemize}%
\item \textbf{Type I:} These are the Smoothable SCFTs. Mass deformations exist for these SCFTs and a full mass deformation leads to a trivial SCFT in the IR. 

Within this class, we find the following sub-classes

\begin{itemize}%
\item \textbf{Type Ia:} $(\mathcal{O}_N, \mathcal{O}_H)$ = (pL +Special, Richardson, a Dixmier sheet of $\mathcal{O}_H$).
\item \textbf{Type Ib:} $(\mathcal{O}_N, \mathcal{O}_H)$ = (pL+Non-Special, Richardson, a Dixmier sheet of $\mathcal{O}_H$).
\end{itemize}
For defects of type I, sub-sheets do not occur in the description of the mass deformed Coulomb branch.

\item \textbf{Type II}:  These are the Malleable SCFTs. Mass deformations exist these SCFTs but a full mass deformation leads to a non-trivial Rigid SCFT in the IR.

Within this class, we find the following sub-classes

\begin{itemize}%
\item \textbf{Type IIa:} $(O_N,O_H)$ = (npL +Special, Special Induced, special mixed sheet/sub-sheet)
\item \textbf{Type IIb:} $(O_N,O_H )$ = (npL + Non-Special, Special Induced, special mixed sheet/sub-sheet)
\item \textbf{Type IIc:} $(O_N,O_H)$ = (npL +Special, Richardson, special mixed sheet/sub-sheet)
\item \textbf{Type IId:} $(O_N,O_H )$ = (npL + Non-Special, Richardson, special mixed sheet/sub-sheet)

\end{itemize}

\item \textbf{Type III:} These are the Rigid SCFTs. By definition, no mass deformations exist for these theories. 

Within this class, we find the following sub-classes
\begin{itemize}
\item \textbf{Type IIIa:} $(O_N,O_H)$ = (Distinguished, Rigid)
\item \textbf{Type IIIb:} $(O_N,O_H)$ = (Distinguished, special Induced)
\item \textbf{Type IIIc:} $(O_N,O_H)$ = (Distinguished, Richardson)
\end{itemize}
\end{itemize}

Using the Spaltenstein-Barbasch-Vogan Theorem (Theorem 8.3.1 in \cite{collingwood1993nilpotent}), it is possible to prove that the above list exhausts all possible pairs $(O_N,O_H)$ which obey $\mathcal{O}_H = d_{BV}(\mathcal{O}_N)$. A nice corollary of this result is a simple characterization of Smoothable SCFTs using Nahm data. Looking at the list above, we note that the Smoothable SCFTs are precisely the ones which have a principal Levi type orbit as its Nahm orbit! Under the duality, these get paired with Dixmier sheets attached to Richardson nilpotent orbits. The other sub-classes appearing in the list above do not appear to imply the existence of additional physical information. For example, from a physical standpoint, the Rigid SCFTs of type IIIa are no different from Rigid SCFTs of type IIIb. The existence of these types is just another reflection of the fact that the natural notion of Rigidity arising in the physics of $T^\rho[G]$ theories is different from the usual mathematical notion of a rigid nilpotent orbit.

\section{Examples}
\label{tables}

We finally turn to providing several explicit examples.  We collect the examples in the form of several tables in this section and in \S \ref{MoreTaxnonomy}. From the tables collected in this section and \S \ref{MoreTaxnonomy}, one can read off the mass deformed Hitchin data in the following way.

\begin{enumerate}%
\item From the refined Sheet label $(\mathfrak{l}^\vee,\mathcal{O}_H^{l^\vee})$, one can read off a Levi subalgebra $\mathfrak{l}^\vee$ of $g^\vee$ and $\mathcal{O}_H^{l^\vee}$ is a special nilpotent orbit in this Levi subalgebra.
\item The Hitchin residue $a_M$ for the mass deformed case (if it exists) has the following Jordon decomposition
\end{enumerate}
\begin{equation}
a_M = a_{ss} + a_n, [a_{ss},a_n]=0,
\end{equation}
where $a_{ss}$ is a semi-simple element in $g^\vee$ that actually sits in $Z(\mathfrak{l}^\vee)$, the center of the Levi subalgebra $\mathfrak{l}^\vee$ and $a_n$ is a representative of the nilpotent orbit $\mathcal{O}_H^{l^\vee}$.

To simplify notation, we use $a_{ss}$ to denote both a generic element in $Z(\mathfrak{l}^\vee)$ and the corresponding element in $\mathfrak{g}^\vee$ which is obtained by the natural inclusion $Z(\mathfrak{l}^\vee) \hookrightarrow \mathfrak{g}^\vee$ (Similarly for $a_n$ : It denotes both an element in $\mathfrak{l}^\vee$ and its natural inclusion in $\mathfrak{g}^\vee$).

One can check that the dimension condition is obeyed using the following formulas.

\begin{subequations}
\begin{align}
\dim(\mathcal{O}_{a_M}) &= \dim(\mathcal{O}_{a_{ss}}^{g^\vee}) + 
 \dim(\mathcal{O}_H^{l^\vee})\\
\dim(\mathcal{O}_{a_ss}^{g^\vee}) &= \dim(\mathfrak{g}^\vee) - 
 (\dim(\mathfrak{l}^\vee_{ss}) + \dim(Z(\mathfrak{l}^\vee)))
 \label{ssdimension}
\end{align}
\end{subequations}

\subsection{Type $A$}
\label{typeAexample}

We start with the tame defects in type $A$. Mass deformations for these are well understood \cite{Gaiotto:2009we,Gaiotto:2009hg}. We nevertheless review what is known in the language of the present paper. Nilpotent orbits in the Lie algebra $A_{n-1}$ are labeled by partitions of $n$. Let $[p_i]$ be such a partition label for a nilpotent orbit of $A_{n-1}$. Any element of the orbit $\mathcal{O}_{[p_i]}$ can be cast in its Jordan canonical form. The parts $p_i$ correspond to the sizes of the Jordan blocks. Let $[r_i]=[p_i]^t$ denote the transpose partition. The complex dimension of a nilpotent orbit corresponding to the partition $[p_i]$ is given by \cite{collingwood1993nilpotent},
\begin{equation}
\dim_{\mathbb{C}}(\mathcal{O}_{[p_i]}) = \dim(\mathfrak{sl}_n) - (\sum_{i}r_i^2 -1).
\label{typeAdimension}
\end{equation}

It is also helpful to recall the Flavour symmetry acting on the Slodowy Slice to an orbit $\mathcal{O}_{[p_i]}$. It is given by \cite{Chacaltana:2012zy},

\begin{equation}
F = S(\prod_i U(r_i)).
\end{equation}

In any Lie algebra $A_{n-1}$, every sheet is a Dixmier sheet. Dixmier sheets carry labels of the form $(\mathfrak{l},0)$ for some Levi subalgebra $\mathfrak{l}$. In a Dixmier sheet, the boundary is a Richardson nilpotent orbit. Every (non-zero) nilpotent orbit in $A_n$ occurs at the boundary of a Dixmier sheet. In other words, all nilpotent orbits in $A_n$ are Richardson. Furthermore, every nilpotent orbit in type A belongs to a \textit{unique} sheet. This simplifies the study of mass deformations remarkably. In type A, every defect is Smoothable in sense of the terminology introduced in \S\ref{taxonomy}. 

The dimension of the semi-simple orbits in a Dixmier sheet can be calculated using the formula (\ref{ssdimension}).

Now, we can describe the mass deformed in the language of sheets. Let the Nahm label for the defect be $p_N = [p_i]$. The Hitchin label for the defect is $p_H=[p_i]^T$. 

The Hitchin nilpotent orbit $[p_i]^T$ is at the boundary of the Dixmier sheet $(\mathfrak{l}_{[p_i]},0)$, where $\mathfrak{l}_{[p_i]}$ is the Levi factor of the standard parabolic corresponding to the partition $[p_i]$.  This is a consequence of the fact that $Ind_{\mathfrak{l}_{[p_i]}}^{\mathfrak{g}}(0) = [p_i]^T$ (see \ref{inductionappendix} for the summary of orbit induction ). It is straightforward to check that Flavour condition is automatically obeyed for every sheet. So, it does not amount to an additional restriction. Hence, there is no refinement of sheets in type A. Once the sheet corresponding to the mass deformation is know, the mass deformed Hitchin system is straightforward to write down. One has to merely require that the residue of the Higgs field $Res(\phi) \in Z(\mathfrak{l}_{[p_i]}) $, where $Z(\mathfrak{l}_{[p_i]})$ denotes the center of the Levi subalgebra associated to $[p_i]$. 

To illustrate all of this in greater detail, we include below the data for mass deformed defects in the Lie algebra $A_6$. In Fig \ref{hassesl7}, we depict the nilpotent orbits of $A_6$ together with the sheets they are part of.

\begin{center}
\vspace*{\fill}
\begin{minipage}{.95\textwidth}
\subsubsection{$\mathfrak{g}=A_6$} 
\begin{tabular}{c|c|c|c|c|c|c}
\label{a6defects}
$O_N$&$O_H$&$\dim(\mathcal{O}_H)$&$\dim(\mathcal{O}_{a_{ss}}^{g^\vee})$&$F$& $(\mathfrak{l}^\vee,\mathcal{O}_H^{l^\vee})$&$\dim(Z(\mathfrak{l}^\vee))$\\
\toprule[0.75mm] 
$[1^7]$&$[7]$& 42 & 42 &$SU(7)$&$(0,0)$&6\\
$[2,1^5]$&$[6,1]$&40&40&$SU(5)\times U(1)$&$(A_1,0)$& 5\\
$[2^2,1^3]$&$[5,2]$&38&38& $SU(4) \times U(1)$&$(A_1+A_1,0)$& 4\\
$[3,1^4]$&$[5,1^2]$&36&36&$SU(4) \times U(1)$&$(A_2,0)$& 4\\
$[2^3,1]$&$[4,3]$&36&36&$SU(3)\times U(1)$&$(3A_1,0)$& 3\\
$[3,2,1^2]$&$[4,2,1]$&34&34&$SU(3)\times U(1)$&$(A_2+A_1,0)$& 3\\
$[4,1^3]$&$[4,1^3]$&30&30&$SU(3)\times U(1)$&$(A_3,0)$& 3\\
$[3,2^2]$&$[3^2,1]$&30&30&$SU(2)\times U(1)$&$(A_2+2A_1,0)$& 2\\
$[3^2,1]$&$[3,2^2]$& 32 & 32& $SU(2)\times SU(2)$&$(2A_2,0)$& 2\\
$[4,2,1]$&$[3,2,1^2]$& 28 & 28&$SU(2)\times U(1)$&$(A_3+A_1,0)$& 2\\
$[5,1^2]$&$[3,1^4]$& 22 & 22 &$SU(2)\times U(1)$&$(A_4,0)$& 2\\
$[4,3]$&$[2^3,1]$& 24 & 24 & $U(1)$&$(A_3+A_2,0)$& 1\\
$[5,2]$&$[2^2,1^3]$& 20 & 20 &$U(1)$&$(A_4+A_1,0)$& 1\\
$[6,1]$&$[2,1^5]$& 12 & 12 & $U(1)$& $(A_5,0)$ & 1 \\
$[7]$&$[1^7]$& 0 & - &-& $(A_6,0)$ &  0 \\
\toprule[0.75mm]
\end{tabular}
\end{minipage}
\vfill
\clearpage
\end{center}

Let us look a couple of examples from table \ref{a6defects} in some more detail. As a first example, consider the defect with Nahm label $[6,1]$. The 3d SCFT $T^{[6,1]}[\mathfrak{sl}_7]$ corresponding to this defect has a Higgs branch which is the Slodowy slice to the subregular orbit $[6,1]$. It carries a $U(1)$ Flavour symmetry. So, we expect a mass deformation of rank one. The dual Hitchin orbit is the minimal nilpotent orbit $[2,1^5]$. The dimension of the minimal nilpotent orbit is 12 (calculated using \ref{typeAdimension}). Now, there is a Dixmier sheet $(A_5,0)$ whose boundary is precisely the minimal nilpotent orbit $[2,1^5]$. One can check this using the fact that

\begin{equation}
Ind_{A_5}^{A_6} (0) = [2,1^5].
\end{equation}

To help check the above statement about orbit induction, we have summarized orbit induction in \S \ref{inductionappendix}. 

The dimension of the semi-simple orbits in this sheet can be calculated using (\ref{ssdimension}),
\begin{equation}
\dim(\mathcal{O}_{a_{ss}}^{g^\vee}) = (48) - (35 + 1) = 12,
\end{equation} 
where we have used the fact that $\dim(A_6)=48,dim(A_5)=35$ and $\dim(Z(\mathfrak{l}_{A_5}))= \mathrm{rank}(A_6) - \mathrm{rank}(A_5)=1$. We see that $\dim(\mathcal{O}_{a_{ss}}^{g^\vee}) $ is exactly equal to the dimension of the nilpotent orbit $[2,1^5]$.\footnote{The fact that there is a family of semi-simple  orbits with the same dimension as the minimal nilpotent orbit is a special feature of Lie algebras of type $A$.} This shows that the sheet $(A_5,0)$ does correspond to a mass-like deformation. 

To further confirm that this sheet parameterizes the mass deformed Coulomb branch of the theory $T^{[6,1]}[\mathfrak{sl}_7]$, we also note that $A_5$ is indeed the Bala-Carter Levi of the Nahm orbit $[6,1]$.\footnote{We have summarized the procedure to deduce the Bala-Carter Levi in \S \ref{balacarterappendix}.} Finally, for clarity, we also note that
\begin{equation}
\mathrm{\mathrm{rank}}(F) = \dim(Z(\mathfrak{l}_{A_5}) ) = 1.
\label{flavoura1}
\end{equation}
This, however, is not an independent check. The condition (\ref{flavoura1}) was guaranteed to be satisfied once we ensured that $A_5$, the Levi appearing in the sheet label, is the BC Levi of the Nahm orbit $[6,1]$. So, we conclude that we have correctly identified the mass deformed family of Coulomb branches.

As a second example, consider the defect corresponding to the 3d SCFT $T^{[3,2^2]}[\mathfrak{sl}_7]$. Its Higgs branch is the Slodowy Slice to $[3,2^2]$ intersected with the nilpotent cone and its Coulomb branch is the nilpotent orbit $[3^2,1]$. The Flavour symmetry acting on the Higgs branch is $SU(2) \times U(1)$. So, we expect a rank two mass deformation.

 Using (\ref{typeAdimension}), we see that $\dim([3^2,1])=32$. Using known facts about orbit induction (see \S \ref{inductionappendix}), one can see that
\begin{equation}
Ind_{A_2+2A_1}^{A_6} (0) = [3^2,1].
\end{equation}
This shows that the nilpotent orbit $[3^2,1]$ is Richardson and occurs at the boundary of the Dixmier sheet $(A_2+2A_1,0)$. The dimension of the semi-simple orbits in this sheet can be calculated using (\ref{ssdimension}),
\begin{equation}
\dim(\mathcal{O}_{a_{ss}}^{g^\vee}) = (48) - (14 + 2) = 32,
\end{equation}
where we have used the fact that $\dim(A_2+2A_1)=14$ and $\dim(Z(\mathfrak{l}_{A_2+2A_1}))=2$. We again notice that $\dim(\mathcal{O}_{a_{ss}}^{g^\vee}) $ is exactly equal to the dimension of the nilpotent orbit $[3^2,1]$. As a final check, we note that $A_2 + 2A_1$ is the BC Levi of the Nahm orbit $[3,2^2]$. This guarantees that $\mathrm{rank}(F) = \dim(Z(\mathfrak{l}_{A_2+2A_1}) ) = 2$, as required. 

 To get a better overall perspective on all the mass deformed defects at once, we depict the sheets attached each of the nilpotent orbits in $A_6$ in Fig (\ref{hassesl7}) together with the closure ordering on the nilpotent orbits. 

\begin{center}
\begin{figure} [!h]
\begin{tikzpicture}
  \node (max) at (0,17) {$[7]$};
   \node (max+) at (4,17) {$(0,0)$};
    \node (max-name) at (6.5,17) {``Principal sheet''};
    
  \node (a) at (0,16) {$[6,1]$};
   \node (a+) at (4,16) {$(A_1,0)$};
   
  \node (b) at (0,15) {$[5,2]$};
  \node (b+) at (4,15) {$(2A_1,0)$};
  
   \node (bb) at (-2,14) {$[5,1^2]$};
  \node (bb+) at (-6,14) {$(A_2,0)$};
  
   \node (bbb) at (2,14) {$[4,3]$};
  \node (bbb+) at (6,14) {$(3A_1,0)$};
  
    \node (cc) at (0,13) {$[4,2,1]$};
  \node (cc+) at (4,13) {$(A_2+A_1,0)$};
  
  \node (c) at (-2,10.75) {$[4,1^3]$};
   \node (c+) at (-6,10.75) {$(A_3,0)$};
   
     \node (dd) at (0,11.5) {$[3^2,1]$};
  \node (dd+) at (4,11.5) {$(2A_2,0)$};
     
  \node (d) at (0,10) {$[3,2^2]$};
  \node (d+) at (4,10) {$(A_2+2A_1,0)$};
  
  \node (e) at (0,8) {$[3,2,1^2]$};
  \node (e+) at (4,8) {$(A_3+A_1,0)$};
  
  \node (f) at (-2,7) {$[3,1^4]$};
  \node (f+) at (-6,7) {$(A_4,0)$};
  
  \node (g) at (2,7) {$[2^3,1]$};
   \node (g+) at (6,7) {$(A_3+A_2,0	)$};
   
  \node (h) at (0,6) {$[2^2,1^3]$};
  \node (h+) at (4,6) {$(A_4+A_1,0)$};
  
  \node (i) at (0,5) {$[2,1^5]$};
  \node (i+) at (4,5) {$(A_5,0)$};
  \node (j) at (0,4){$\mathbf{[1^7]}$};
   \draw[preaction={draw=white, -,line width=10pt}] (max) -- (a)--(b)--(bb)--(cc)--(dd)--(d)--(e);
    \draw[preaction={draw=white, -,line width=10pt}] (h)--(i)--(j);
  \draw[preaction={draw=white, -,line width=10pt}] (d)--(e);
  \draw[preaction={draw=white, -,line width=10pt}] (e) -- (g);
  \draw[preaction={draw=white, -,line width=10pt}] (e) -- (f) --(h) --(g);
  \draw[preaction={draw=white, -,line width=10pt}] (b)--(bbb) -- (cc) --(c) --(e);
  \draw[|-{Latex[black,scale=2]}, dashed] (max) -- (max+);
   \draw[|-{Latex[black,scale=2]}, dashed](a) -- (a+);
     \draw[|-{Latex[black,scale=2]}, dashed] (b) -- (b+);
      \draw[|-{Latex[black,scale=2]}, dashed] (bb) -- (bb+);
       \draw[|-{Latex[black,scale=2]}, dashed] (bbb) -- (bbb+);
   \draw[|-{Latex[black,scale=2]}, dashed](c) -- (c+);
      \draw[|-{Latex[black,scale=2]}, dashed](cc) -- (cc+);
      \draw[|-{Latex[black,scale=2]}, dashed](d) -- (d+);
      \draw[|-{Latex[black,scale=2]}, dashed](dd) -- (dd+);
      \draw[|-{Latex[black,scale=2]}, dashed](e) -- (e+);
      \draw[|-{Latex[black,scale=2]}, dashed] (f) -- (f+);
      \draw[|-{Latex[black,scale=2]}, dashed] (g) -- (g+);
      \draw[|-{Latex[black,scale=2]}, dashed] (h) -- (h+);
       \draw[|-{Latex[black,scale=2]}, dashed] (i) -- (i+);
 \end{tikzpicture}

\caption{This diagram shows the sheets for the Lie algebra $\mathfrak{sl}_7$ and the nilpotent orbits that occur at their boundary. In this case, every sheet is a special sheet. Each dashed line is a sheet. The solid lines encode the closure ordering on the nilpotent orbits. They do \textit{not} imply closure ordering for the entire sheet(s) attached to the nilpotent orbits. }
\label{hassesl7}
\end{figure}
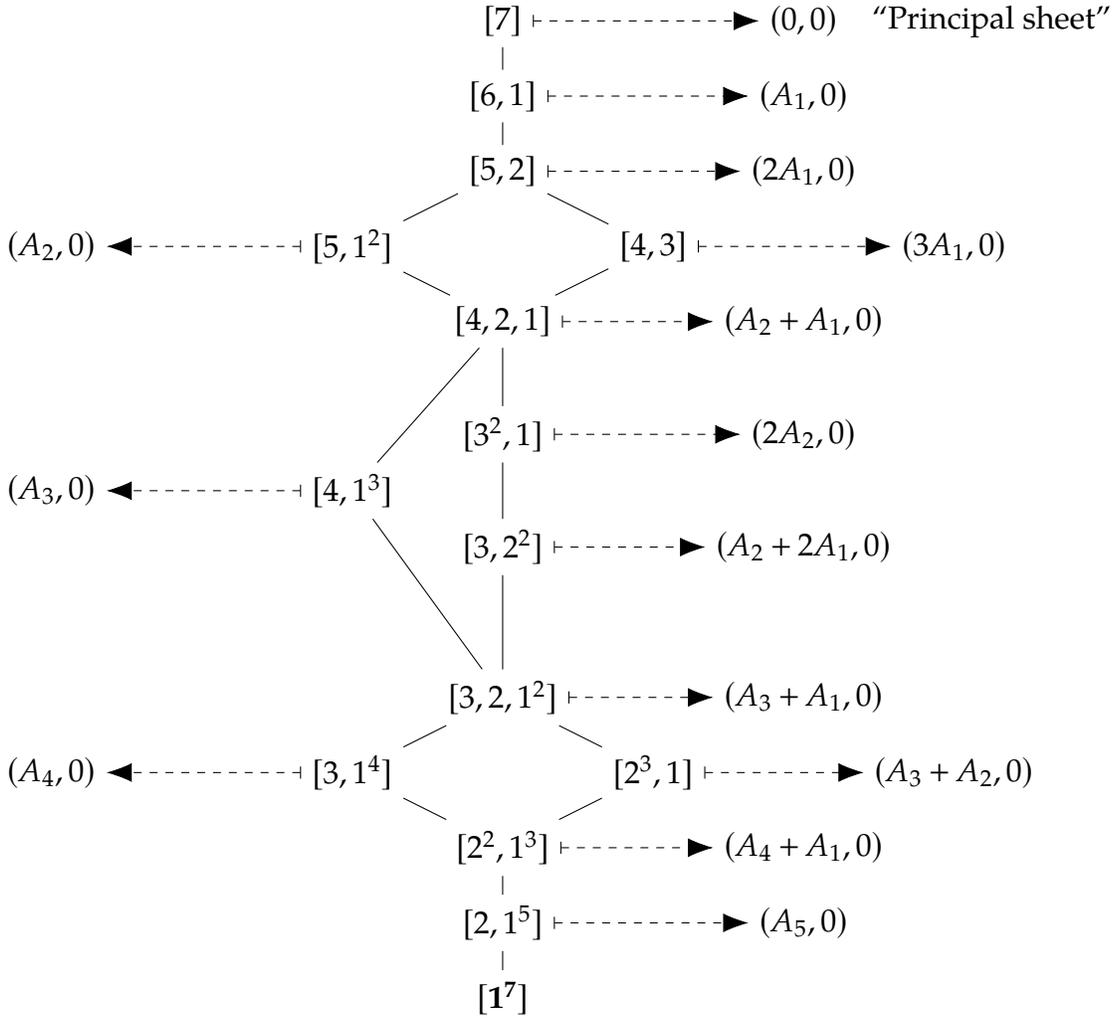

\end{center}

\subsubsection{Sheet closures vs Nilpotent Orbit closures}

Before moving to examples in other Cartan types, we wish to discuss a feature that reveals that the stratification of $\mathfrak{g}$ by sheets is quite subtle. Already in type $A$, we encounter the feature that containment relations among sheets is \textit{not} the same as the closure ordering among the nilpotent orbits occurring at their boundaries. For example, consider two Dixmier sheets $S = (\mathfrak{l},0)$ and $S'=(\mathfrak{l}',0)$ and let $\mathcal{O}$ and $\mathcal{O}'$ be the Richardson nilpotent orbits occurring at their respective boundaries. Now, $S \subset \overline{S'}$ iff the opposite inclusion $\mathfrak{l}' \subset \mathfrak{l}$ holds between the Levi subalgebras (see, for example, the Appendix of \cite{carnovale2013lusztig}). This is a stronger condition than $\mathcal{O} \subset \overline{\mathcal{O}'}$. There are several examples in which $\mathcal{O} \subset \overline{\mathcal{O}'}$ but $\mathfrak{l}' \not\subset \mathfrak{l}$. As an example from Figure \ref{hassesl7}, note that while $\mathcal{O}_{[2^2,1^3]} \subset \overline{\mathcal{O}_{[2^3,1]}}$,  $A_3 + A_2$ is not a subalgebra of $A_4+A_1$. So, we see the closure of a sheet is not necessarily a union of smaller sheets. In other words, the stratification of $\mathfrak{g}$ by sheets does not obey the \textit{frontier condition} \cite{mather2012notes}. It would be interesting to explore if this failure of the frontier condition has any physical consequences.

\subsection{Type $D$}

For defects of type $D$, we provide the general solution at first and then use examples from $D_4$ and $D_5$ to explain some new features that were not seen for defects in type $A$. 

\subsubsection{$\mathfrak{g}=D_n$}

Let $[p_i]$ be the partition label for the Nahm orbit associated to the defect. Let $[q_i]$ be the partition label for the Hitchin orbit associated to the defect. As we reviewed earlier, $[q_i]$ is the Barbasch-Vogan dual \footnote{For type $D$, the Barbasch-Vogan dual is the same as the Spaltenstein dual.} to $[p_i]$,
\begin{equation}
\mathcal{O}_{[q_i]} = d_{BV} (\mathcal{O}_{[p_i]}).
\end{equation}

Let $[r_i]$ be $[p_i]^T$, the transpose of $[p_i]$ and let $[s_i]$ be $[q_i]^T$, the transpose of $[q_i]$. The dimension of the Coulomb branch of $T^{[p_i]}[D_n]$ is given by the dimension of the nilpotent orbit $\mathcal{O}_{[q_i]}$. This can be calculated using
\begin{equation}
\dim(\mathcal{O}_{[q_i]} = \dim(\mathfrak{so}_{2n}) - \frac{1}{2}(\sum_i s_i^2 - \sum_{i \in odd} m_i),
\label{dimensionD}
\end{equation}
where $m_i$ is the multiplicity of the number $i$ in the partition $[q_i]$.

The Flavour symmetry associated to the defect can be deduced from the Nahm orbit $[p_i]$ using (\cite{Chacaltana:2012zy})
\begin{equation}
F = \prod_{i \in odd} SO(t_i) \times \prod_{i \in even} Sp(t_i/2),
\label{flavourD}
\end{equation}
where $t_i$ is the multiplicity of the number $i$ in the partition $[p_i]$.

One of the features that is new for defects in type D is the possibility of \textit{Special Pieces}. These refer to collections of defects which have different 3d Higgs branches but have identical 3d Coulomb branches \cite{Chacaltana:2012zy}. The Higgs branches of such defects will be Slodowy Slices to certain orbits $\mathcal{O}_N^i$, for $i \in 1,2,\ldots, k$ all of which obey $d_{BV}(\mathcal{O}_N^i)=\mathcal{O}_H$ for a fixed special nilpotent orbit $\mathcal{O}_H$. We will denote such a collection of defects by the notation $\{ \mathcal{O}_N^i \}, i \in 1,2,\ldots, k$. Here, the number $k$ denotes the size of the special piece. As we explain below, the mass deformations for defects that belong to a special piece $\{ \mathcal{O}_N^i \}$ will end up being related to special sheets attached to the same nilpotent orbit $\mathcal{O}_H$. 

To determine the special sheet associated to the mass deformation of $T^{[p_i]}[D_n]$, we first calculate the BC Levi $\mathfrak{l}_{BC}$ of the Nahm orbit. This can be done using the procedure summarized in \S \ref{balacarterappendix}. Then, from among the special sheets attached to the orbit $[q_i]$, we pick the one whose sheet label contains the $\mathfrak{l}_{BC}$ as the Levi. In some instances, it may be required to place an additional restriction on the special sheet. Such an additional restriction happens whenever there is no special sheet with $\mathfrak{l}_{BC}$ as its sheet Levi but there is a special sheet with a restriction such that the Levi associated to such a restriction is the Bala-Carter Levi $\mathfrak{l}_{BC}$. This restricted special sheet is what we denoted a refined sheet in \S\ref{massdeformations} . By inspection, we observe any non-special sheet attached to the special nilpotent orbit $[q_i]$ does not ever have the BC Levi of one Nahm orbits in $\mathcal{O}_{N}^i$ as part of its sheet label (or a refined sheet label). From this, we conclude that non-special sheets do not obey the Flavour condition. 

Defects in type D can fall into any one of the three possible deformation classes of \S\ref{taxonomy}. We now use examples from $D_4$ and $D_5$ that bring out some of these features.

\subsubsection{$\mathfrak{g}=D_4$}
\label{d4defects}

Below, we described the mass deformed SCFTs $T^\rho[\mathfrak{so}_8]$ in a table. When compared with the theories $T^\rho[A_n]$, we see some new features appearing here. Let us first consider the 3d SCFT $T^{[5,3]}[\mathfrak{so}_8]$. The Higgs branch of this theory is the Slodowy slice to the orbit $[5,3]$ intersected with nilpotent cone. This Higgs branch has a trivial Flavour symmetry since orbit $[5,3]$ is a distinguished nilpotent orbit. \footnote{See \S\ref{definitionsappendix} for a definition of a distinguished nilpotent orbit.} If the partition label for a nilpotent orbit does not have repeated parts, then the label corresponds to a distinguished orbit. Since the Higgs branch has a trivial Flavour symmetry, this is an example of a \textit{Rigid SCFT}. The corresponding Coulomb branch is the minimal nilpotent orbit $[2^2,1^4]$. This is a nilpotent orbit of complex dimension 10. If there were to be a sheet attached to $[2^2,1^4]$, then it should be possible to find non-nilpotent orbit whose dimension is 10. A straightforward calculation shows that this is impossible. It is possible to check that the smallest semi-simple orbits in $D_4$ have dimension 12. Any mixed orbit would have to have an even larger dimension. So, this rules out the possibility that $[2^2,1^4]$ has any nearby orbits of the same dimension. In other words, $[2^2,1^4]$ is a rigid orbit. This, in turn, implies that there are no mass-like deformations of the tame Hitchin system on a punctured disc with a nilpotent singularity of type $[2^2,1^4]$. This is in complete agreement with the fact that the Flavour symmetry $F$ is trivial for this defect. 

As a second example, consider the pair of defects with Nahm label $\{[3^2,1^2],[3,2^2,1] \}$. They constitute a special piece since the dual Hitchin label is $[3^2,1^2]$ for both defects. The flavour symmetry associated to $T^{[3^2,1^2]}[\mathfrak{so}_8]$ is $U(1)^2$ and flavour symmetry associated to $T^{[3,2^2,1]}[\mathfrak{so}_8]$ is $SU(2)$. The Bala-Carter Levis of $[3^2,1^2], [3,2^2,1]$ are $A_2, A_1+D_2$ (respectively). 

There are two sheets attached to the nilpotent orbit $[3^2,1^2]$. Both of these are Dixmier sheets (and hence special) and their sheet labels are $(\mathfrak{l}_{(A_2)},0)$ and $(\mathfrak{l}_{(A_1+D_2)},0)$. The sheet labels imply that the centralizer of the semi-simple elements in these sheets are, respectively, the Levi subalgebras $\mathfrak{l}_{(A_2)}$,$\mathfrak{l}_{(A_1+D_2)}$. The Levis occurring in the sheet labels are precisely the BC Levis of the two Nahm orbits. So, imposing the Flavour Condition picks out the $(\mathfrak{l}_{(A_2)},0)$ sheet as the mass deformed Coulomb branch of $T^{[3^2,1^2]}[\mathfrak{so}_8]$ and the $(\mathfrak{l}_{(A_1+D_2)},0)$ sheet as the mass deformed Coulomb branch of $T^{[3,2^2,1]}[\mathfrak{so}_8]$. 

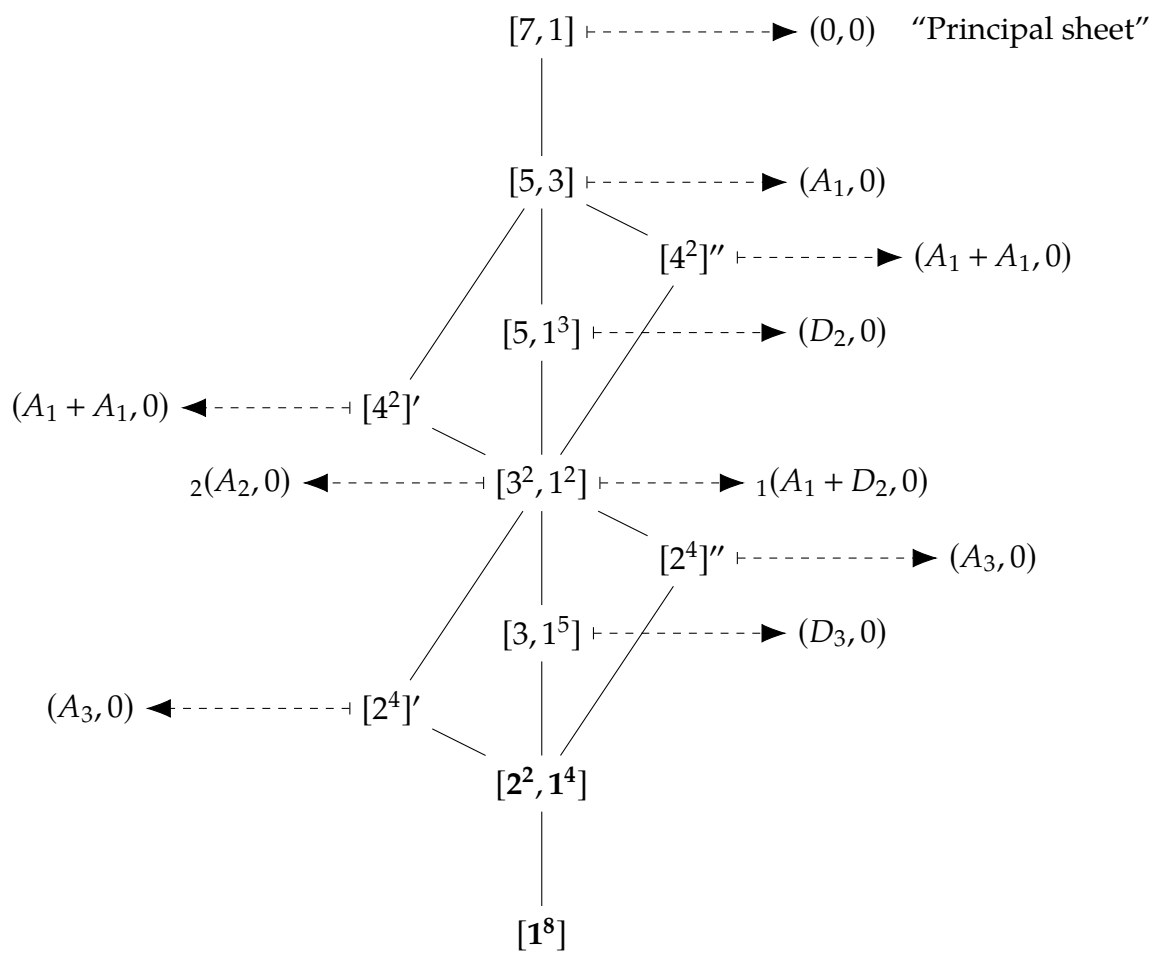
\begin{figure} [!h]
\begin{center}
\begin{tikzpicture}
  \node (max) at (0,12) {$[7,1]$};
   \node (max+) at (4,12) {$(0,0)$};
    \node (max-name) at (6.5,12) {``Principal sheet''};
  \node (a) at (0,10) {$[5,3]$};
   \node (a+) at (4,10) {$(A_1,0)$};
  \node (b) at (0,8) {$[5,1^3]$};
  \node (b+) at (4,8) {$(D_2,0)$};
  \node (c) at (-2,7) {$[4^2]'$};
   \node (c+) at (-6,7) {$(A_1+A_1,0)$};
  \node (d) at (2,9) {$[4^2]''$};
  \node (d+) at (6,9) {$(A_1+A_1,0)$};
  \node (e) at (0,6) {$[3^2,1^2]$};
  \node (e+) at (4,6) {$_1(A_1+D_2,0)$};
  \node (e++) at (-4,6) {$_2(A_2,0)$};
  \node (f) at (-2,3) {$[2^4]'$};
  \node (f+) at (-6,3) {$(A_3,0)$};
  \node (g) at (2,5) {$[2^4]''$};
   \node (g+) at (6,5) {$(A_3,0	)$};
  \node (h) at (0,4) {$[3,1^5]$};
  \node (h+) at (4,4) {$(D_3,0)$};
  \node (i) at (0,2) {$\mathbf{[2^2,1^4]}$};
  \node (j) at (0,0){$\mathbf{[1^8]}$};
  \draw[preaction={draw=white, -,line width=10pt}] (max) -- (a)--(b)--(e)--(h)--(i)--(j);
  \draw[preaction={draw=white, -,line width=10pt}] (d)--(e);
  \draw[preaction={draw=white, -,line width=10pt}] (e) -- (g);
  \draw[preaction={draw=white, -,line width=10pt}] (e) --(c) --(a) --(d);
  \draw[preaction={draw=white, -,line width=10pt}] (e) -- (f) --(i) --(g);
  \draw[|-{Latex[black,scale=2]}, dashed] (max) -- (max+);
   \draw[|-{Latex[black,scale=2]}, dashed](a) -- (a+);
     \draw[|-{Latex[black,scale=2]}, dashed] (b) -- (b+);
   \draw[|-{Latex[black,scale=2]}, dashed](c) -- (c+);
      \draw[|-{Latex[black,scale=2]}, dashed](d) -- (d+);
      \draw[|-{Latex[black,scale=2]}, dashed](e) -- (e+);
   \draw[|-{Latex[black,scale=2]}, dashed](e) -- (e++);
      \draw[|-{Latex[black,scale=2]}, dashed] (f) -- (f+);
      \draw[|-{Latex[black,scale=2]}, dashed] (g) -- (g+);
      \draw[|-{Latex[black,scale=2]}, dashed] (h) -- (h+);
 \end{tikzpicture}
\end{center}
\caption{This diagram shows the twelve special sheets for the Lie algebra $\mathfrak{so}_8$ and the nilpotent orbits that occur at their boundary. Each dashed line is a sheet.  The rigid nilpotent orbits are shown in bold. They form sheets by themselves. As before, the solid lines encode just the closure ordering on the nilpotent orbits.}
\label{hasseso8}
\end{figure}

In Fig (\ref{hasseso8}), we depict all the special sheets in the Lie algebra $\mathfrak{so}_8$ together with the nilpotent orbits occurring at their boundaries and the closure ordering on those nilpotent orbits. In $\mathfrak{so}_8$, there are no non-special sheets with special nilpotent orbits at their boundaries. When there is more than one special sheet attached to the same Hitchin orbit, we additionally add a numerical label to both the sheet and corresponding dual Nahm orbit.

\begin{center}
\vspace*{\fill}
\begin{minipage}{.95\textwidth}
\begin{tabular}{c|c|c|c|c|c|c}
$O_N$&$O_H$&$\dim(\mathcal{O}_H)$&$\dim(\mathcal{O}_{a_{ss}}^{g^\vee})$&$F$& $(\mathfrak{l}^\vee,\mathcal{O}_H^{l^\vee})$&$\dim(Z(\mathfrak{l}^\vee))$\\
\toprule[0.75mm]
$[1^8]$&$[7,1]$&24&24&Spin(8)&$(0,0)$&4\\
$[2^2,1^4]$&$[5,3]$&22&22&$SU(2)^3$&$(A_1,0)$&3\\
$[3,1^5]$&$[5,1^3]$&20&20&$Sp(2)$&$(D_2,0)$&2\\
$[2^4]^I$&$[4^2]I$&20&20&$Sp(2)$&$(A_1+A_1,0)$&2\\
$[2^4]^{II}$&$[4^2]^{II}$&20&20&$Sp(2)$&$(A_1+A_1,0)$&2\\ \hline
$_1 [3,2^2,1]$&$[3^2,1^2]$&18&18&$SU(2)$&$(A_1+D_2,0)$&1\\
$_2 [3^2,1^2]$&$[3^2,1^2]$&18&18&$U(1)^2$&$(A_2,0)$&2\\ \hline
$[5,1^3]$&$[3,1^5]$&12&12&$SU(2)$&$(D_3,0)$&1\\
$[4^2]^I$&$[2^4]^I$&12&12&$SU(2)$&$(A_3,0)$&1\\
$[4^2]^{II}$&$[2^4]^{II}$&12&12&$SU(2)$&$(A_3,0)$&1\\
$[5,3]$&$[2^2,1^4]$&10&0&-&$(D_4,[2^2,1^4])$&0\\
$[7,1]$&$[1^8]$&0&0&-&$(D_4,[1^8])$&0\\ \toprule[0.75mm]
\end{tabular}
\end{minipage}
\vfill
\clearpage
\end{center}

\vfill
\clearpage

\newpage

\subsubsection{$\mathfrak{g}=D_5$}

Some of the new features that appeared in $D_4$ continue to appear in the case of $D_5$. For example, the theory $T^{[7,3]}[\mathfrak{so}_{10}]$ corresponds to a \textit{Rigid SCFT}. Its Higgs branch is the Slodowy slice to the subregular orbit $[7,3]$ and its Coulomb branch is the minimal nilpotent orbit $[2^2,1^6]$, a rigid nilpotent orbit. We also have the special pieces with Nahm labels $\{[3^2,1^4],[3,2^2,1^3]\}$ and $\{ [5,3,1^2],[5,2^2,1]\}$. The details for the special piece $\{[3^2,1^4],[3,2^2,1^3]\}$ closely parallel the special piece from $D_4$. 

The special piece $\{ [5,3,1^2],[5,2^2,1]\}$, on the other hand, reveals a new feature. The dual Hitchin orbit for this special piece is $[3^2,1^4]$. There are two sheets that have the orbit $[3^2,1^4]$ at its boundary. These sheets can be identified using the following facts about orbit induction,
\begin{equation}
\begin{split}
Ind_{D_4}^{D_5} ([2^2,1^4]) &= [3^2,1^4] ,\\
Ind_{A_1+D_3}^{D_5}[0] &= [3^2,1^4].
\end{split}
\end{equation}
It follows that $[3^2,1^4]$ occurs at the boundary of two sheets $(\mathfrak{l}_{A_1+D_3},0)$ and $(\mathfrak{l}_{D_4},[2^2,1^4])$. The former is a Dixmier sheet while the latter is not. Both sheets are, however, special sheets.

The Bala-Carter Levis associated to the Nahm orbits $\{ [5,3,1^2],[5,2^2,1]\}$ are, respectively, $D_4,A_1+D_3$. So, imposing the Flavour condition identifies the sheet $(\mathfrak{l}_{A_1+D_3},0)$ as the one corresponding to the mass deformation of $T^{[5,2^2,1]}[\mathfrak{so}_{10}]$ and  $(\mathfrak{l}_{D_4},[2^2,1^4])$ as the sheet corresponding to the mass deformation of $T^{[5,3,1^2]}[\mathfrak{so}_{10}]$. When a non-Dixmier sheet parameterizes the mass deformed Coulomb branches, there is a residual singularity in the IR theory (see discussion in \S\ref{taxonomy}). In the case of $T^{[5,2^2,1]}[\mathfrak{so}_{10}]$, the residual singularity is the nilpotent orbit $[2^2,1^4]$ in $\mathfrak{so}_8$. This is an indication that upon mass deforming and flowing to the IR, we hit a non-trivial SCFT. In this case, it is the SCFT $T^{[5,3]}[\mathfrak{so}_8]$. 3d $\mathcal{N}=4$ that flow (in the IR) to a smaller but non-trivial SCFT upon mass deformation are the ones that we denoted as \textit{Malleable SCFTs}. The discussion above shows that $T^{[5,3,1^2]}[\mathfrak{so}_{10}]$ is a malleable SCFT.

We collect the defects in $D_5$ along with sheets corresponding to each of them in the table below. In Fig (\ref{hasseso10}), we depict all the special sheets in $\mathfrak{so}_8$. 

\begin{center}
\vspace*{\fill}
\begin{minipage}{.95\textwidth}
\begin{tabular}{c|c|c|c|c|c|c}
$O_N$&$O_H$&$\dim(\mathcal{O}_H)$&$\dim(\mathcal{O}_{a_{ss}}^{g^\vee})$&$F$&$(\mathfrak{l}^\vee,\mathcal{O}_H^{l^\vee})$&$\dim(Z(\mathfrak{l}^\vee))$\\
\toprule[0.75mm] 
$[1^{10}]$&$[9,1]$&40&40&$Spin(10)$&$(0,0)$&5\\
$[2^2,1^6]$&$[7,3]$&38&38&$SU(4)xSU(2)$&$(A_1,0)$&4\\
$[3,1^7]$&$[7,1^3]$&36&36&$Spin(7)$&$(D_2,0)$&3\\
$[2^4,1^2]$&$[5^2]$&36&36&$Sp(2) \times U(1)$&$(A_1+A_1,0)$&3\\ \hline 
$_1[3^2,1^4]$&$[5,3,1^2]$&34&34&$SU(2)^2\times U(1)$&$(A_2,0)$&3\\
$_2[3,2^2,1^3]$&$[5,3,1^2]$&34&34&$SU(2)\times SU(2)$&$(A_1+D_2,0)$&2\\ \hline 
$[3^2,2^2]$&$[4^2,1^2]$&32&32&$SU(2) \times U(1)$&$(A_2+A_1,0)$&2\\
$[3^3,1]$&$[3^3,1]$&30&30&$SU(2)$&$(A_2+D_2,0)$&1\\
$[5,1^5]$&$[5,1^5]$&28&28&$Sp(2)$&$(D_3,0)$&2\\
$[4^2,1^2]$&$[3^2,2^2]$&28&28&$SU(2)\times U(1)$&$(A_3,0)$&2\\ \hline 
$_3[5,3,1^2]$&$[3^2,1^4]$&26&16&$U(1)$&$(D_4,[2^2,1^4])$&1\\
$_4[5,2^2,1]$&$[3^2,1^4]$&26&26&$SU(2)$&$(A_1+D_3,0)$&1\\ \hline 
$[5^2]$&$[2^4,1^2]$&20&20&$U(1)$&$(A_4,0)$&1\\
$[7,1^3]$&$[3,1^7]$&16&16&$SU(2)$&$(D_4,0)$&1\\
$[7,3]$&$[2^2,1^6]$&14&-&-&$(D_5,[2^2,1^6])$&0\\
$[9,1]$&$[1^{10}]$&0&0&-&$(D_5,[1^{10}])$&0\\ \toprule[0.75mm]
\end{tabular}
\end{minipage}
\vfill
\clearpage
\end{center}

\begin{center}

\begin{figure} [!h]

\begin{tikzpicture}
 \node (a) at (0,10) {$[9,1]$};
 \node (a+) at (4,10) {$(0,0)$};
 
 \node (b)  at (0,9) {$[7,3]$};
 \node (b+)  at (4,9) {$(A_1,0)$};
 
 \node (c) at (-2,8) {$[7,1^3]$};
 \node (c+)  at (-6,8) {$(D_2,0)$};
 
 \node (cc) at (2,8) {$[5^2]$};
  \node (cc+) at (6,8) {$(2A_1,0)$};
 
 \node (d) at (0,7) {$[5,3,1^2]$} ; 
  \node (d+) at (4,7) {$_1(A_2,0)$} ; 
   \node (d++) at (-4,7) {$_2(A_1+A_2,0)$} ; 
 
 \node (e) at (-2,6) {$[4^2,1^2]$} ;
  \node (e+) at (-6,6) {$(A_2+A_1,0)$} ;
 
 \node (ee) at (2,5) {$[5,1^5]$} ;
  \node (ee+) at (6,5) {$(D_3,0)$} ;
  
 \node (f) at (-2,5) {$[3^3,1]$} ;
  \node (f+) at (-6,5) {$(A_2+D_2,0)$} ;
 
 \node (g) at (-2,4) {$[3^2,2^2]$} ;
  \node (g+) at (-6,4) {$(3^2,2^2)$} ;
  
 \node (h) at (0,3) {$[3^2,1^4]$} ;
  \node (h+) at (4,3) {$_3(D_4,[2^2,1^4])$} ;
    \node (h++) at (-4,3) {$_4(A_1+D_3,0)$} ;
 
 \node (i) at (-2,2) {$[2^4,1^2]$} ;
 \node (i+) at (-6,2) {$(A_4,0)$} ;
  
 \node (ii) at (2,2) {$[3,1^7]$} ;
 \node (ii+) at (6,2) {$(D_4,0)$} ;
 
 \node (j) at (0,1) {$\mathbf{[2^2,1^6]}$} ;
  
 \node (k) at (0,0) {$\mathbf{[1^{10}]}$} ;

  \draw[preaction={draw=white, -,line width=10pt}] (a)--(b)--(c)--(d)--(e)--(f)--(g)--(h)--(i)--(j)--(k);
   \draw[preaction={draw=white, -,line width=10pt}] (b)--(cc)--(d)--(ee)--(h)--(ii)--(j);
  
   \draw[|-{Latex[black,scale=2]}, dashed](a) -- (a+);
     \draw[|-{Latex[black,scale=2]}, dashed] (b) -- (b+);
   \draw[|-{Latex[black,scale=2]}, dashed](c) -- (c+);
      \draw[|-{Latex[black,scale=2]}, dashed](cc) -- (cc+);
      \draw[|-{Latex[black,scale=2]}, dashed](d) -- (d+);
        \draw[|-{Latex[black,scale=2]}, dashed](d) -- (d++);
      \draw[|-{Latex[black,scale=2]}, dashed](e) -- (e+);
        \draw[|-{Latex[black,scale=2]}, dashed](ee) -- (ee+);
      \draw[|-{Latex[black,scale=2]}, dashed] (f) -- (f+);
      \draw[|-{Latex[black,scale=2]}, dashed] (g) -- (g+);
      \draw[|-{Latex[black,scale=2]}, dashed] (h) -- (h+);
       \draw[|-{Latex[black,scale=2]}, dashed] (h) -- (h++);
        \draw[|-{Latex[black,scale=2]}, dashed] (i) -- (i+);
  \draw[|-{Latex[black,scale=2]}, dashed] (ii) -- (ii+);       
     
 \end{tikzpicture}

\caption{This diagram shows the special sheets for the Lie algebra $\mathfrak{so}_{10}$.}
\label{hasseso10}
\end{figure}
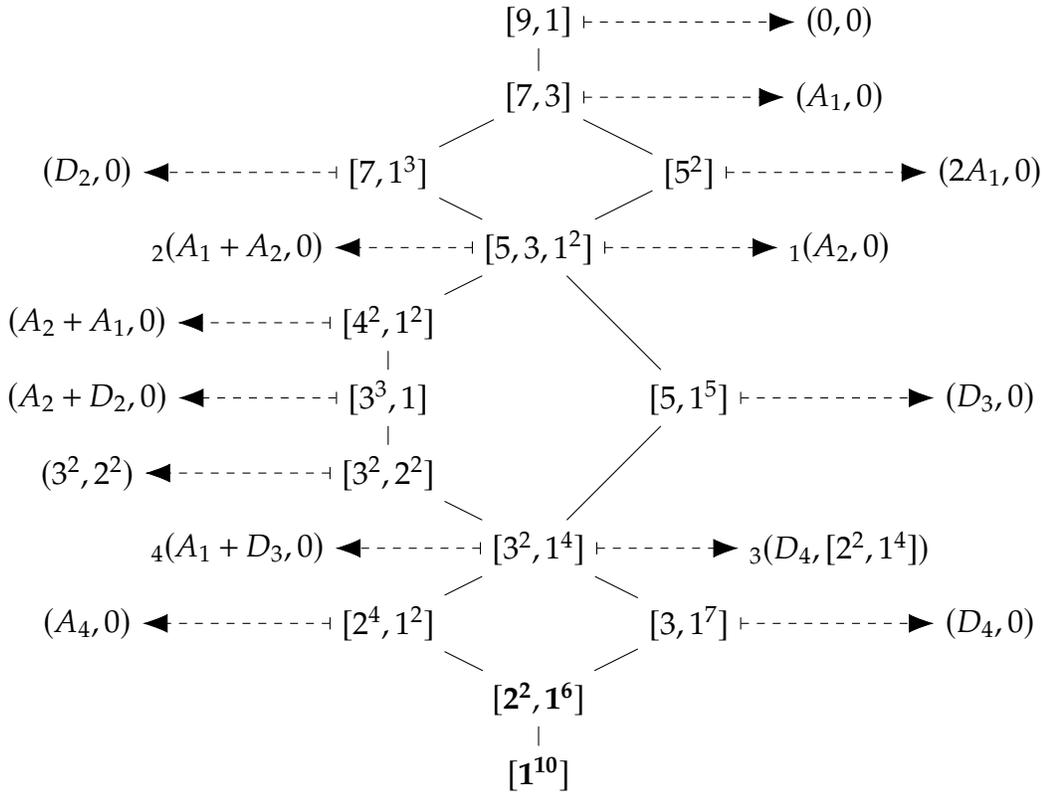
\end{center}

\subsection{Type $E$}

In this section, we describe the mass deformed defects in $E_6,E_7,E_8$. In order to be able to describe the mass deformed defects, it is important to first identify the sheets attached to each of the special nilpotent orbits. Unlike the case of nilpotent orbits in classical Lie algebras, there is no general algorithm. One has to proceed case-by-case. 

One can deduce this information using the known results about orbit induction for nilpotent orbit in exceptional Lie algebras. These were first obtained by Elashvili and were included in \cite{spaltenstein2006classes}. They have also been summarized in \cite{elashvili2009induced}.\footnote{We warn the reader that the notation used for nilpotent orbits in \cite{spaltenstein2006classes} is, sometimes, different from the one used in \cite{elashvili2009induced}. Whenever there is a conflict, we have adopted the notation used in \cite{elashvili2009induced}. } Results about dimension of nilpotent orbits in the exceptional Lie algebras is available in \cite{collingwood1993nilpotent} and the their associated Flavour symmetries (when these orbits occur as the Nahm label) can be deduced from the tables in \cite{carterfinite} and has been summarized in \cite{Chacaltana:2012zy}.

\subsubsection{$\mathfrak{g}=E_6$}
\label{e6examples}

\begin{figure} [!h]
\begin{center}
\begin{tikzpicture}
 \node (a) at (0,14) {$E_6$};
  \node (a+) at (4,14) {$(0,0)$};
 
 \node (b) at (0,13) {$E_6(a_1)$};
  \node (b+) at (4,13) {$(A_1,0)$};
 
 \node (c) at (0,12) {$D_5$};
   \node (c+) at (4,12) {$(2A_1,0)$};
   
 \node (d) at (0,11) {$E_6(a_3)$};
 \node (d+) at (4,11) {$_1(3A_1,0)$};
  \node (d++) at (-4,11) {$_2(A_2,0)$};
 
 \node (e) at (0,10) {$D_5(a_1)$};
 \node (e+) at (4,10) {$(A_2+A_1,0)$};  
  
 \node (f) at (-2,9) {$A_4+A_1$};
  \node (f+) at (-6,9) {$(A_2+2A_1,0)$};
 
 \node (ff) at (2,9) {$D_4$};
  \node (ff+) at (6,9) {$(2A_2,0)$};
  
 \node (g) at (-2,8) {$A_4$};
  \node (g+) at (-6,8) {$(A_3,0)$};
 
 \node (h) at (0,6) {$D_4(a_1)$};
  \node (h+) at (4,7) {$_1(2A_2+A_1,0)$};
  \node (h++) at (-4,7) {$_2(A_3+A_1,0)$}; 
  \node (h+++) at (-4,5) {$_3(D_4,[2^2,1^4])$};

 \node (i) at (-2,4) {$A_3$};
  \node (i+) at (-6,4) {$(A_4,0)$};
 
 \node (ii) at (2,2) {$2A_2$};
  \node (ii+) at (6,2) {$(D_4,0)$};
  
 \node (j) at (-2,2) {$A_2+2A_1$};
  \node (j+) at (-6,2) {$(A_4+A_1,0)$};
 
 \node (k) at (0,1) {$A_2+A_1$};
  \node (k+) at (4,1) {$(D_5,[2^2,1^6])$};
  
 \node (l) at (0,0) {$A_2$};
  \node (l+) at (4,0) {$_1(A_5,0)$};
    \node (l++) at (-4,0) {$_2\mathbf{(E_6,A_2)}$};
 
 \node (m) at (0,-1) {$2A_1$};
  \node (m+) at (4,-1) {$(D_5,0)$};
  
 \node (n) at (0,-2) {$\mathbf{A_1}$};
 
 \node (o) at (0,-3) {$\mathbf{0}$};
  
\draw[preaction={draw=white, -,line width=10pt}] (a)--(b)--(c)--(d)--(e)--(f)--(g)--(h)--(i)--(j)--(k)--(l)--(m)--(n)--(o);
\draw[preaction={draw=white, -,line width=10pt}] (e)--(ff)--(h)--(ii)--(k);
  
   \draw[|-{Latex[black,scale=2]}, dashed](a) -- (a+);
     \draw[|-{Latex[black,scale=2]}, dashed] (b) -- (b+);
   \draw[|-{Latex[black,scale=2]}, dashed](c) -- (c+);
      \draw[|-{Latex[black,scale=2]}, dashed](d) -- (d+);
        \draw[|-{Latex[black,scale=2]}, dashed](d) -- (d++);
      \draw[|-{Latex[black,scale=2]}, dashed](e) -- (e+);
        \draw[|-{Latex[black,scale=2]}, dashed](ff) -- (ff+);
      \draw[|-{Latex[black,scale=2]}, dashed] (f) -- (f+);
    
      \draw[|-{Latex[black,scale=2]}, dashed] (g) -- (g+);
      \draw[|-{Latex[black,scale=2]}, dashed] (h) -- (h+);
       \draw[|-{Latex[black,scale=2]}, dashed] (h) -- (h++);
         \draw[|-{Latex[black,scale=2]}, dashed] (h) -- (h+++);
        \draw[|-{Latex[black,scale=2]}, dashed] (i) -- (i+);
  \draw[|-{Latex[black,scale=2]}, dashed] (ii) -- (ii+);  
  \draw[|-{Latex[black,scale=2]}, dashed] (j) -- (j+);     
  \draw[|-{Latex[black,scale=2]}, dashed] (k) -- (k+);
  \draw[|-{Latex[black,scale=2]}, dashed] (l) -- (l+);
  \draw[|-{Latex[black,scale=2]}, dashed] (l) -- (l++);
  \draw[|-{Latex[black,scale=2]}, dashed] (m) -- (m+);
     
 \end{tikzpicture}
\end{center}
\caption{This diagram shows the special sheets for the Lie algebra $E_6$.}
\label{hasseE6}
\end{figure}
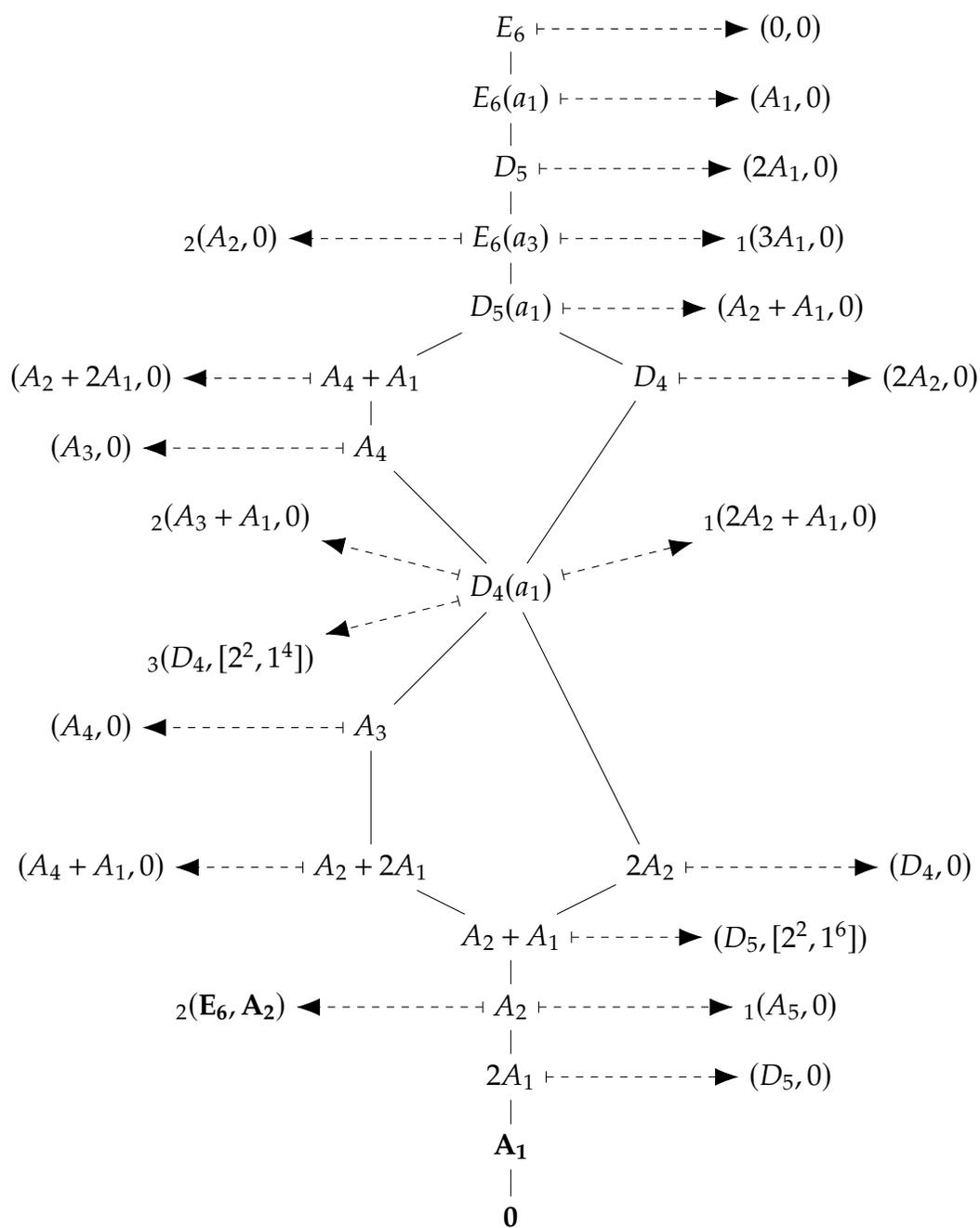

In $E_6$, we have a total of 20 defects out of which 16 are smoothable, 2 are malleable and 2 are rigid. The sheets corresponding to the mass deformation of the corresponding $T^\rho[E_6]$ theories are summarized below. The special sheets in $E_6$ together with the nilpotent orbits at their boundaries are depicted in Fig (\ref{hasseE6}). 

In $E_6$, many of the features noted in examples in type $D$ exist. We have two rigid SCFTs $T^{E_6(a_1)}[E_6]$ and $T^{E_6(a_3)}[E_6]$. And we have three non-trivial special pieces,
\begin{equation}
\begin{split}
d_{BV} (\{ 3A_1,A_2 \}) &= E_6(a_3) \\
d_{BV} (\{ A_5,E_6(a_3) \}) &= A_2 \\
d_{BV} (\{ 2A_2+A_1,A_3+A_1,D_4(a_1) \} ) &= D_4(a_1).
\end{split}
\end{equation}

But, we a further feature present in $E_6$ that was not seen in $D_4,D_5$. This is the possibility of a Rigid SCFT whose Coulomb branch is not a rigid orbit. The theory $T^{[E_6(a_3)]}[E_6]$ is an example of such a rigid SCFT. The Higgs branch of the theory is the Slodowy slice to $E_6(a_3)$, a distinguished orbit. Hence, it has a trivial Flavour symmetry. The dual Hitchin orbit, on the other hand, is $A_2$. This orbit is not a rigid orbit. It does occur at the boundary of a Dixmier sheet $(A_5,0)$. But, this mass-like deformation does \textit{not} obey the Flavour condition for the defect $T^{[E_6(a_3)]}[E_6]$.\footnote{Instead, this Dixmier sheet corresponds to the mass deformation of the defect $T^{A_5}[E_6]$ whose Coulomb branch is also the orbit $A_2$. As noted, the Nahm orbits $\{E_6(a_3),A_5 \}$ are part of the same special piece.} So, this in an instance where although the Coulomb branch admits mass-like deformations, they do not correspond to an honest mass deformation since the Flavour condition obstructs it. This is why the notion of a \textit{Rigid SCFT} is different from the notion of a rigid orbit. 

In this particular case, the Flavour condition was that $F$ is trivial. But, even in cases where $F$ is non-trivial, certain mass-like deformations may not satisfy the Flavour condition (only a subset might satisfy). It is for this reason that we denoted the stratification of $\mathfrak{g}$ arising from the study of mass deformation to be a refinement of the usual theory of sheets where every stratum is necessarily labeled by $(\mathfrak{l},\mathcal{O})$ for a rigid orbit $\mathcal{O}$ in $\mathfrak{l}$.

\begin{center}
\vspace*{\fill}
\begin{minipage}{.95\textwidth}
\begin{tabular}{c|c|c|c|c|c|c}
$O_N$&$O_H$&$\dim(\mathcal{O}_H)$&$\dim(\mathcal{O}_{a_{ss}}^{g^\vee})$&$F$&$(\mathfrak{l}^\vee,\mathcal{O}_H^{l^\vee})$&$\dim(Z(\mathfrak{l}^\vee))$\\
\toprule[0.75mm]
$0$&$E_6$&72&72&$E_6$&$(0,0)$&6\\
$A_1$&$E_6(a_1)$&70&70&$A_5$&$(A_1,0)$&5\\
$2A_1$&$D_5$&68&68&$B_3+T_1$&$(2A_1)$&4\\ \hline 
$_13A_1$&$E_6(a_3)$&66&66&$A_2+A_1$&$(3A_1,0)$&3\\
$_2A_2$&$E_6(a_3)$&66&66&$2A_2$&$(A_2,0)$&4\\ \hline 
$A_2 + A_1$&$D_5(a_1)$&64&64&$A_2+T_1$&$(A_2+A_1,0)$&3\\
$A_2 + 2A_1$&$A_4+A_1$&62&62&$A_1+T_1$&$(A_2+2A_1,0)$&2\\
$2A_2$&$D_4$&60&60&$G_2$&$(2A_2)$&2\\
$A_3$&$A_4$&60&60&$B_2+T_1$&$A_3$&3\\ \hline 
$_12A_2+A_1$&$D_4(a_1)$&58&58&$A_1$&$(2A_2+A_1,0)$&1\\
$_2A_3+A_1$&$D_4(a_1)$&58&58&$A_1+T_1$&$(A_3+A_1,0)$&2\\
$_3D_4(a_1)$&$D_4(a_1)$&58&48&$T_2$&$(D_4,[2^2,1^4])$&2\\ \hline 
$A_4$&$A_3$&52&52&$A_1+T_1$&$(A_4,0)$&2\\
$D_4$&$2A_2$&48&48&$A_2$&$(D_4,0)$&2\\
$A_4+A_1$&$A_2+2A_1$&50&50&$T_1$&$(A_4+A_1,0)$&1\\
$D_5$&$2A_1$&32&32&$T_1$&$(D_5,0)$&1\\
$D_5(a_1)$&$A_2+A_1$&46&32&$T_1$&$(D_5,[2^2,1^6])$&1\\ \hline 
$_1A_5$&$A_2$&42&42&$A_1$&$(A_5,A_2)$&1\\
$_2E_6(a_3)$&$A_2$&42&-&-&$(E_6,A_2)$&0\\ \hline 
$E_6(a_1)$&$A_1$&22&-&&$(E_6,A_1)$&0\\
$E_6$&$0$&0&0&-&$(E_6,0)$&0\\ \toprule[0.75mm]
\end{tabular}
\end{minipage}
\vfill
\clearpage
\end{center}

\vfill
\clearpage
\newpage

\subsubsection{$\mathfrak{g}=E_7$}

In $E_7$, we have a total of 44 defects out of which 31 are smoothable, 8 are malleable and 5 are rigid. The sheets corresponding to the mass deformation of the corresponding $T^\rho[E_7]$ theories are summarized below. The features noted in $D_4,D_5,E_6$ continue to exist in this case and it should be possible to find the examples from the data collected in the table below. Additionally, in $E_7$, there exist certain non-special sheets attached to special nilpotent orbits. But, we discuss these case separately in \S\ref{specialvsnonspecial}.

\begin{center}
\begin{longtable}{c|c|c|c|c|c|c}
$O_N$&$O_H$&$\dim(\mathcal{O}_H)$&$\dim(\mathcal{O}_{a_{ss}}^{g^\vee})$&$F$&$(\mathfrak{l}^\vee,\mathcal{O}_H^{l^\vee})$&$\dim(Z(\mathfrak{l}^\vee))$\\
\toprule[0.75mm]
\endhead
$0$&$E_7$&126&126&$E_7$&$(0,0)$&7\\
$A_1$&$E_7(a_1)$&124&124&$D_6$&$(A_1,0)$&6\\
$2A_1$&$E_7(a_2)$&122&122&$B_4+A_1$&$(2A_1,0)$&5\\ \hline 
$3A_1'$&$E_7(a_3)$&120&120&$C_3+A_1$&$(3A_1',0)$&4\\
$A_2$&$E_7(a_3)$&120&120&$A_5$&$(A_2,0)$&5\\ \hline 
$(3A_1)''$&$E_6$&120&120&$F_4$&$(3A_1'',0)$&4\\ \hline 
$4A_1$&$E_6(a_1)$&118&118&$C_3$&$(4A_1,0)$&3\\
$A_2+A_1$&$E_6(a_1)$&118&118&$A_3+T_1$&$(A_2+A_1,0)$&4\\ \hline 
$A_2 + 2A_1$&$E_7(a_4)$&116&116&$3A_1$&$(A_2+2A_1,0)$&3\\
$A_2+3A_1$&$A_6$&114&114&$G_2$&$(A_2+3A_1,0)$&2\\
$A_3$&$D_6(a_1)$&114&114&$B_3+A_1$&$(A_3,0)$&4\\
$2A_2$&$D_5+A_1$&114&114&$G_2+A_1$&$(2A_2,0)$&3\\ \hline 
$2A_2+A_1$&$E_7(a_5)$&112&112&$2A_1$&$(2A_2+A_1,0)$&2\\
$(A_3+A_1)'$&$E_7(a_5)$&112&112&$3A_1$&$((A_3+A_1)',0)$&3\\
$D_4(a_1)$&$E_7(a_5)$&112&102&$3A_1$&$(D_4,[2^2,1^4])$&3\\ \hline 
$(A_3+A_1)''$&$D_5$&112&112&$3A_1$&$((A_3+A_1)'',0)$&3\\ \hline 
$A_3+2A_1$&$E_6(a_3)$&110&110&$2A_1$&$(A_3+2A_1,0)$&2\\
$D_4(a_1) +A_1$&$E_6(a_3)$&110&100&$2 A_1$&$(D_4+A_1,[2^2,1^4])$&2\\ \hline 
$A_3+A_2$&$D_5(a_1) + A_1$&108&108&$A_1 + T_1$&$(A_3+A_2,0)$&2\\
$D_4$&$A_5''$&102&102&$C_3$&$(D_4,0)$&3\\
$A_3+A_2+A_1$&$A_4+A_2$&106&106&$A_1$&$(A_3+A_2+A_1,0)$&1\\
$A_4$&$D_5(a_1)$&106&106&$A_2+T_1$&$(A_4,0)$&3\\
$A_4 +A_1$&$A_4+A_1$&104&104&$T_2$&$(A_4+A_1,0)$&2\\ \hline 
$D_4+A_1$&$A_4$&100&100&$B_2$&$(D_4+A_1,0)$&2\\
$D_5(a_1)$&$A_4$&100&86&$A_1+T_1$&$(D_5,[2^2,1^6])$&2\\ \hline 
$A_4 + A_2$&$A_3+A_2+A_1$&100&100&$A_1$&$(A_4+A_2,0)$&1\\
$A_5''$&$D_4$&96&96&$2A_1$&$(A_5'',0)$&2\\
$D_5(a_1) +A_1$&$A_3+A_2$&98&84&$A_1$&$(D_5,[2^2,1^6])$&1\\ \hline 
$A_5'$&$D_4(a_1) + A_1$&96&96&$2A_1$&$(A_5',0)$&2\\
$E_6(a_3)$&$D_4(a_1) + A_1$&96&54&$A_1$&$(E_6,2A_2+A_1)$&1\\ \hline 
$D_5$&$(A_3+A_1)''$&86&86&$2A_1$&$(D_5,0)$&2\\
$A_5+A_1$&$D_4(a_1)$&94&94&$A_1$&$(A_5+A_1,0)$&1\\
$D_6(a_2)$&$D_4(a_1)$&94&-&$A_1$&$(D_6,[2^4,1^4])$&1\\
$E_7(a_5)$&$D_4(a_1)$&94&-&-&$(E_7,D_4(a_1))$&0\\ \hline 
$D_5 + A_1$&$2A_2$&84&84&$A_1$&$(D_5+A_1,0)$&1\\
$D_6(a_1)$&$A_3$&84&66&$A_1$&$(D_6,[2^2,1^8])$&1\\
$A_6$&$A_2+3A_1$&84&84&$A_1$&$(A_6,0)$&1\\
$E_7(a_4)$&$A_2+2A_1$&82&-&-&$(E_7,A_2+2A_1)$&0\\
$E_6(a_1)$&$A_2+A_1$&76&54&$T_1$&$(E_6,A_1)$&1\\
$E_6$&$(3A_1)''$&54&54&$A_1$&$(E_6,0)$&1\\ \hline 
$D_6$&$A_2$&66&66&$A_1$&$(D_6,0))$&1\\
$E_7(a_3)$&$A_2$&66&-&-&$(E_7,A_2))$&0\\ \hline 
$E_7(a_2)$&$2A_1$&52&-&-&$(E_7,2A_1))$&0\\
$E_7(a_1)$&$A_1$&34&-&-&$(E_7,A_1)$&0\\
$E_7$&$0$&0&-&-&$(E_7,0)$&0\\ \toprule[0.75mm]
\end{longtable}
\end{center}

\subsubsection{$\mathfrak{g}=E_8$}

In $E_8$, we have a total of 69 defects out of which 40 are smoothable, 19 are malleable and 10 are rigid. The sheets corresponding to the mass deformation of the corresponding $T^\rho[E_8]$ theories are summarized below. As in the case of $E_7$, there exist certain non-special sheets attached to special orbits in $E_8$. We discuss them in \S\ref{specialvsnonspecial}.

\begin{center}
\begin{longtable}{c|c|c|c|c|c|c}
$O_N$&$O_H$&$\dim(\mathcal{O}_H)$&$\dim(\mathcal{O}_{a_{ss}}^{g^\vee})$&$F$&$(\mathfrak{l}^\vee,\mathcal{O}_H^{l^\vee})$&$\dim(Z(\mathfrak{l}^\vee))$\\
\toprule[0.75mm]
\endhead
$0$&$E_8$&240&240&$E_8$&$(0,0)$&8\\
$A_1$&$E_8(a_1)$&238&238&$E_7$&$(A_1,0)$&7\\
$2A_1$&$E_8(a_2)$&236&236&$B_6$&$(2A_1,0)$&6\\
$3A_1$&$E_8(a_3)$&234&234&$F_4+A_1$&$(3A_1,0)$&5\\
$A_2$&$E_8(a_3)$&234&234&$E_6$&$(A_2,0)$&6\\ \hline 
$4A_1$&$E_8(a_4)$&232&232&$C_4$&$(4A_1,0)$&4\\
$A_2+A_1$&$E_8(a_4)$&232&232&$A_5$&$(A_2+A_1,0)$&5\\ \hline
$A_2+2A_1$&$E_8(b_4)$&230&230&$B_3+A_1$&$(A_2+2A_1,0)$&4\\
$A_3$&$E_7(a_1)$&228&228&$B_5$&$(A_3,0)$&5\\ \hline
$A_2+3A_1$&$E_8(a_5)$&228&228&$G_2+A_1$&$(A_2+3A_1,0)$&3\\
$2A_2$&$E_8(a_5)$&228&228&$2G_2$&$(2A_2,0)$&4\\ \hline
$2A_2+A_1$&$E_8(b_5)$&226&226&$G_2+A_1$&$(2A_2+A_1,0)$&3\\
$A_3+A_1$&$E_8(b_5)$&226&226&$B_3+A_1$&$(A_3 + A_1,0)$&4\\
$D_4(a_1)$&$E_8(b_5)$&226&216&$D_4$&$(D_4,[2^2,1^4])$&4\\  \hline
$2A_2+2A_1$&$E_8(a_6)$&224&224&$B_2$&$(2A_2+2A_1,0)$&2\\
$A_3+2A_1$&$E_8(a_6)$&224&224&$B_2+A_1$&$(A_3+2A_1,0)$&3\\ 
$D_4(a_1)+A_1$&$E_8(a_6)$&224&214&$3A_1$&$(D_4+A_1,[2^2,1^4])$&3\\ \hline
$A_3+A_2$&$D_7(a_1)$&222&222&$B_2+T_1$&$(A_3+A_2,0)$&3\\ \hline
$A_3+A_2+A_1$&$E_8(b_6)$&220&220&$2A_1$&$(A_3+A_2+A_1,0)$&2\\
$D_4(a_1)+A_2$&$E_8(b_6)$&220&210&$A_2$&$(D_4+A_2,[2^2,1^4])$&2\\ \hline
$A_4$&$E_7(a_3)$&220&220&$A_4$&$(A_4,0)$&4\\
$D_4$&$E_6$&216&216&$F_4$&$(D_4,0)$&4\\
$A_4+A_1$&$E_6(a_1)+A_1$&218&218&$A_2+T_1$&$(A_4+A_1,0)$&3\\ \hline
$2A_3$&$D_7(a_2)$&216&216&$B_2$&$(2A_3,0)$&2\\
$A_4+2A_1$&$D_7(a_2)$&216&216&$A_1+T_1$&$(A_4+2A_1,0)$&2\\ \hline
$A_4+A_2$&$D_5+A_2$&214&214&$2A_1$&$(A_4+A_2,0)$&2\\ \hline
$D_4+A_1$&$E_6(a_1)$&214&214&$C_3$&$(D_4+A_1,0)$&2\\
$D_5(a_1)$&$E_6(a_1)$&214&200&$A_3$&$(D_5,[2^2,1^6])$&3\\ \hline
$A_4+A_2+A_1$&$A_6+A_1$&212&212&$A_1$&$(A_4+A_2+A_1,0)$&1\\
$D_4+A_2$&$A_6$&210&210&$A_2$&$(D_4+A_2,0)$&2\\
$D_5(a_1)+A_1$&$E_7(a_4)$&212&198&$2A_1$&$(D_5+A_1,[2^2,1^6])$&2\\ \hline
$A_5$&$D_6(a_1)$&210&210&$G_2+A_1$&$(A_5,0)$&2\\
$E_6(a_3)$&$D_6(a_1)$&210&168&$G_2$&$(E_6,A_2)$&2\\ \hline
$D_5$&$D_5$&200&200&$B_3$&$(D_5,0)$&3\\ \hline
$A_4+A_3$&$E_8(a_7)$&208&208&$A_1$&$(A_4+A_3,0)$&1\\
$D_5(a_1)+A_2$&$E_8(a_7)$&208&-&$A_1$&$(D_5+A_2,[2^2,1^6])$&1\\
$A_5+A_1$&$E_8(a_7)$&208&208&$2A_1$&$(A_5+A_1,0)$&2\\
$E_6(a_3)+A_1$&$E_8(a_7)$&208&166&$A_1$&$(E_6+A_1,A_2)$&1\\
$D_6(a_2)$&$E_8(a_7)$&208&-&$2A_1$&$(D_6,[2^4,1^4])$&1\\
$E_7(a_5)$&$E_8(a_7)$&208&114&$A_1$&$(E_7,D_4(a_1))$&1\\
$E_8(a_7)$&$E_8(a_7)$&208&-&$-$&$(E_8,E_8(a_7))$&-\\ \hline
$D_5+A_1$&$E_6(a_3)$&198&198&$2A_1$&$(D_5+A_1,0)$&2\\
$D_6(a_1)$&$E_6(a_3)$&198&180&$2A_1$&$(D_6,[2^2,1^8])$&2\\ \hline
$E_7(a_4)$&$D_5(a_1)+A_1$&196&114&$A_1$&$(E_7,A_2+2A_1)$&1\\
$A_6$&$D_4+A_2$&198&198&$2A_1$&$(A_6,0)$&2\\
$A_6+A_1$&$A_4+A_2+A_1$&196&196&$A_1$&$(A_6+A_1,0)$&1\\
$E_6(a_1)$&$D_5(a_1)$&190&168&$A_2$&$(E_6,A_1)$&2\\
$D_5+A_2$&$A_4+A_2$&194&194&$T_1$&$(D_5+A_1,0)$&1\\
$D_7(a_2)$&$A_4+2A_1$&192&156&$T_1$&$(D_7,[2^4,1^6])$&1\\
$E_6(a_1)+A_1$&$A_4+A_1$&188&166&$T_1$&$(E_6+A_1,A_1+0)$&1\\
$E_6$&$D_4$&168&168&$G_2$&$(E_6,0)$&2\\ \hline
$D_6$&$A_4$&180&180&$B_2$&$(D_6,0)$&1\\
$E_7(a_3)$&$A_4$&180&114&$A_1$&$(E_7,A_2)$&1\\ \hline
$A_7$&$D_4(a_1)+A_2$&184&184&$A_1$&$(A_7,0)$&1\\
$E_8(b_6)$&$D_4(a_1)+A_2$&184&-&$-$&$(E_8,D_4(a_1)+A_2)$&-\\ \hline
$D_7(a_1)$&$A_3+A_2$&178&156&$T_1$&$(D_7,[2^2,1^10])$&1\\ \hline
$E_6+A_1$&$D_4(a_1)$&166&166&$A_1$&$(E_6+A_1,0)$&1\\
$E_7(a_2)$&$D_4(a_1)$&166&-&$A_1$&$(E_7,2A_1)$&1\\
$E_8(b_5)$&$D_4(a_1)$&166&-&$-$&$(E_8,D_4(a_1))$&-\\ \hline
$D_7$&$2A_2$&156&156&$A_1$&$(D_7,0)$&1\\
$E_8(a_5)$&$2A_2$&156&-&$-$&$(E_8,2A_2)$&-\\ \hline
$E_7(a_1)$&$A_3$&148&114&$A_1$&$(E_7,A_1)$&1\\
$E_8(b_4)$&$A_2+2A_1$&146&-&$-$&$(E_8,A_2+2A_1)$&-\\
$E_8(a_4)$&$A_2+A_1$&136&-&$-$&$(E_8,A_2+A_1)$&-\\ \hline
$E_7$&$A_2$&114&114&$A_1$&$(E_7,0)$&1\\
$E_8(a_3)$&$A_2$&114&-&$-$&$(E_8,A_2)$&-\\ \hline
$E_8(a_2)$&$2A_1$&92&-&$-$&$(0,E_8,2A_1)$&-\\
$E_8(a_1)$&$A_1$&58&-&$-$&$(E_8,A_1)$&-\\
$E_8$&$0$&0&-&$-$&$(E_8,0)$&-\\ \toprule[0.75mm]
\end{longtable}
\end{center}

For twisted defects, which are classified by nilpotent orbits in non-simply laced Lie algebras, we have included the tables detailing their mass deformations in the Appendix \S\ref{MoreTaxnonomy}. In the next section, we take up some further examples to discuss \textit{non-special sheets} and \textit{special pieces}.

\section{Further Discussion}
\label{futher}

\subsection{Special Pieces and Sheets}

In the examples treated in \S\ref{tables}, we have already encountered collections of defects called \textit{special pieces}. Let us briefly recall what these are and refer to \cite{Chacaltana:2012zy} for further details. A special piece is a collection of tame defects with Nahm labels $\{\mathcal{O}_N^i \},i=1,2,\ldots ,k$ such that the Coulomb branch of the each of the 3d SCFTs $T^{\rho_i}[G]$ is the same special nilpotent orbit $\mathcal{O}_H$ that is related to $\{ \mathcal{O}_N^i  \}$ by
\begin{equation}
d_{BV} (\mathcal{O}_N^i) = \mathcal{O}_H, i=1,2,\ldots,k.
\end{equation}

Among the collection of defects $\{\mathcal{O}_N^i \}$, there is a unique defect whose Nahm orbit is a special nilpotent orbit. This special orbit is the smallest special orbit whose closure contains every other non-special orbit belonging to the special piece \cite{spaltenstein2006classes,lusztig1997notes}. We will refer to this special orbit as the \textit{anchor} of the special piece $\{\mathcal{O}_N^i \}$.  The singularities within a special piece have a rich structure. They can be studied using the results of  \cite{kraft1980minimal,kraft1982geometry} for the classical Lie algebras and in the exceptional cases, this has been described more recently in the work of \cite{fu2017generic}. We also refer to  \cite{fu2017generic} for a more detailed review of the mathematical literature on special pieces. 

From our perspective, special pieces are of interest because mass deformations of the collection $\{T^{\rho_i}[G]\}$ offers an extremely interesting test of our proposal that mass deformations are always related to special sheets attached to the Hitchin orbit $\mathcal{O}_H$. In \textit{every} instance, we find that there is a unique special sheet (or) a restriction of a special sheet that satisfies the Flavour Condition and we identify this as these as the mass deformed Coulomb branch(es). We arrived at this result using the classification of nilpotent orbits, known results about nilpotent orbits occurring at boundaries of sheets and the additional physical input from the Flavour Condition. From a conceptual standpoint, it might be interesting to look for way to arrive at this result that does not utilize the classification as an intermediate step. In the rest of this section, we include several examples to illustrate the non-trivial nature of the interaction between our proposal for mass deformations and special pieces. 

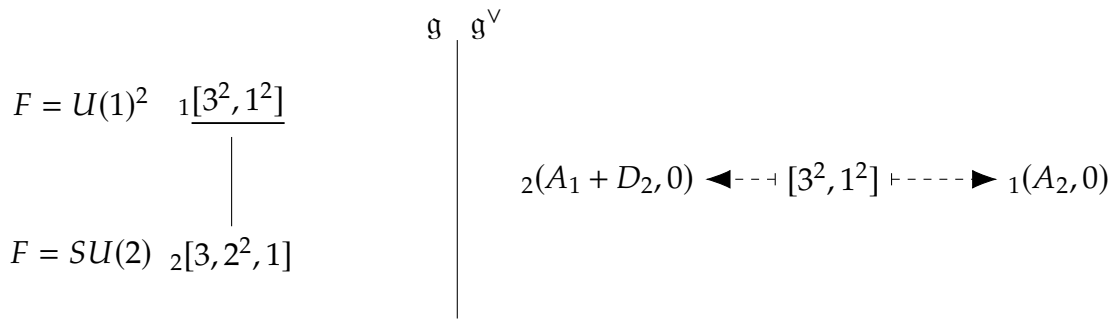
\begin{figure} [!h]
\begin{center}
\begin{tikzpicture}
 \node (a) at (0,0) {$_1\underline{[3^2,1^2]}$};
 \node (a+) at (-2,0) {$F=U(1)^2$};
 \node (b) at (0,-2) {$_2[3,2^2,1]$};
  \node (b+) at (-2,-2) {$F=SU(2)$};
  \draw[preaction={draw=white, -,line width=10pt}] (a)--(b);
 

\node (m) at (8,-1)  {$[3^2,1^2]$};
\node (m+) at (11,-1) {$_1(A_2,0)$};
 \node (m++) at (5,-1) {$_2(A_1+D_2,0)$} ;
   \draw[|-{Latex[black,scale=2]}, dashed](m) -- (m+);
    \draw[|-{Latex[black,scale=2]}, dashed] (m) -- (m++); 
\node (g) at (2.7,1) {$\mathfrak{g}$}  ;
\node (gd) at (3.4,1.065) {$\mathfrak{g}^\vee$}  ;
\node (gg) at (3.0,+1) {}  ;
\node (gg+) at (3.0,-3) {}  ;
     \draw[preaction={draw=white, -,line width=10pt}] (gg)--(gg+);
 \end{tikzpicture}
\end{center}
\caption{This diagram shows the duality in $\mathfrak{g}=D_4$ between Slodowy Slice to orbits in the special piece $\{ \underline{[3^2,1^2]},[3,2^2,1] \}$ and the special sheets attached to $[3^2,1^2]$.}
\label{specialpieced4}
\end{figure}

We begin with a special piece whose mass deformation was already discussed in \cite{Chacaltana:2012zy} and was also discussed in \S\ref{d4defects}. We depict this special piece in Fig \ref{specialpieced4}.

\begin{figure} [!h]
\begin{center}
\begin{tikzpicture}
 \node (a) at (0,0) {$_1\underline{E_8(a_7)}$};
 \node (a+) at (-2,0) {$F=0$};
 
  \node (aa) at (-1,-1) {$_2E_7(a_5)$};
 \node (aa+) at (-3,-1) {$F=A_1$};
 
 \node (b) at (0,-2) {$_4D_6(a_1)$};
  \node (b+) at (2,-2) {$F=2A_1$};
  
   \node (bb) at (-2,-2) {$_3E_6(a_3)+A_1$};
  \node (bb+) at (-4,-2) {$F=A_1$};

   \node (c) at (0,-4) {$_5D_5(a_1)+A_2$};
  \node (c+) at (2,-4) {$F=A_1$};
  
   \node (cc) at (-2.5,-5) {$_6A_5+A_1$};
  \node (cc+) at (-4.5,-5) {$F=2A_1$};

    \node (d) at (0,-6) {$_7A_4+A_3$};
  \node (d+) at (2,-6) {$F=A_1$};

  \draw[preaction={draw=white, -,line width=10pt}] (a)--(b)--(c)--(d);
    \draw[preaction={draw=white, -,line width=10pt}] (a)--(aa)--(bb)--(cc)--(d);
      \draw[preaction={draw=white, -,line width=10pt}] (aa)--(b)--(cc);
        \draw[preaction={draw=white, -,line width=10pt}] (bb)--(c);
 

\node (m) at (10,-1)  {$E_8(a_7)$};
\node (m1) at (6,-1) {$_7(A_4+A_3,0)$};
\node (m2) at (7,-2) {$_6(A_5+A_1,0)$}; 
\node (m3) at (8,-3) {$_5(D_5+A_2,[2^2,1^6])$};
\node (m4) at (10,-4) {$_4(D_6,[2^4,1^4])$};
\node (m5) at (12,-3) {$_3(E_6+A_1,A_2)$};
\node (m6) at (13,-2) {$_2(E_7,D_4(a_1))$};
 \node (m7) at (13,-1) {$_1(\mathbf{E_8,E_8(a_7))}$} ;
   \draw[|-{Latex[black,scale=2]}, dashed](m) -- (m1);
    \draw[|-{Latex[black,scale=2]}, dashed](m) -- (m2);
     \draw[|-{Latex[black,scale=2]}, dashed](m) -- (m3);
      \draw[|-{Latex[black,scale=2]}, dashed](m) -- (m4);
       \draw[|-{Latex[black,scale=2]}, dashed](m) -- (m5);
        \draw[|-{Latex[black,scale=2]}, dashed](m) -- (m6);
    \draw[|-{Latex[black,scale=2]}, dashed] (m) -- (m7); 
\node (g) at (3.7,0) {$\mathfrak{g}$}  ;
\node (gd) at (4.4,0.065) {$\mathfrak{g}^\vee$}  ;
\node (gg) at (4.0,+1) {}  ;
\node (gg+) at (4.0,-7) {}  ;
     \draw[preaction={draw=white, -,line width=10pt}] (gg)--(gg+);
 \end{tikzpicture}
\end{center}
\caption{This diagram shows the duality in $\mathfrak{g}=E_8$.}
\label{specialpiecee8-2}
\end{figure}
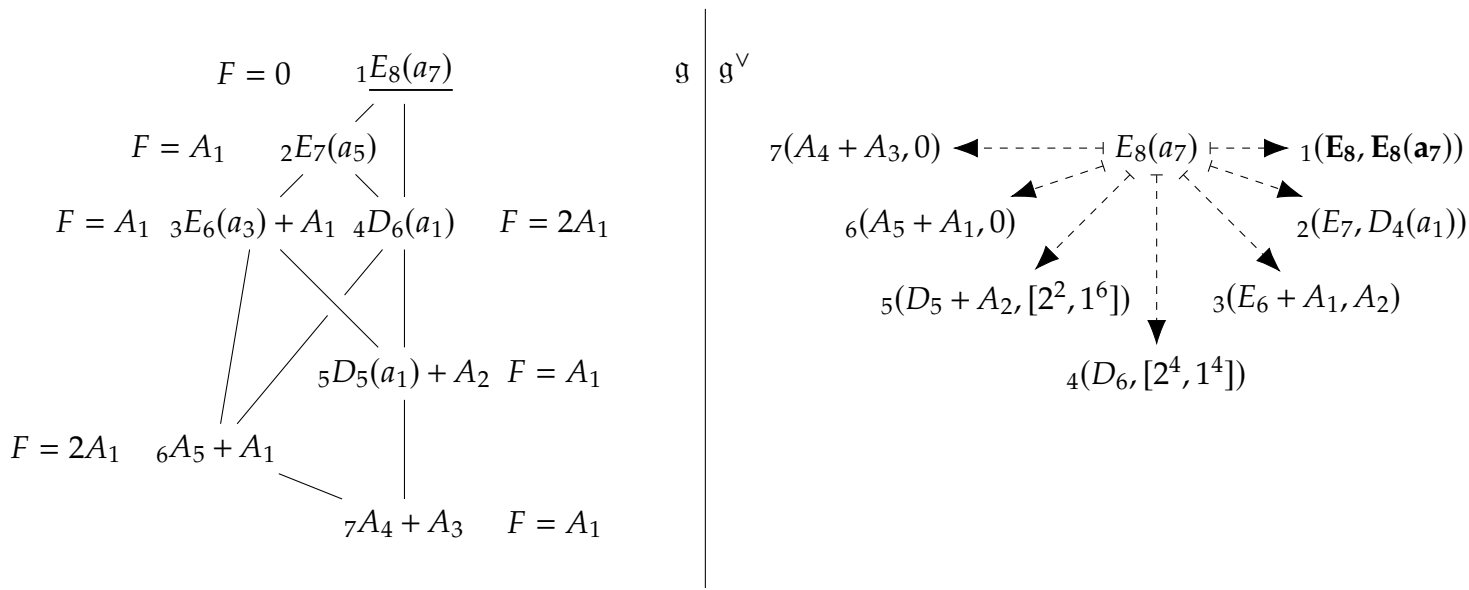

In Fig \ref{specialpiecee6}, we depict the special piece in $E_6$ that contains three defects. Finally, in Figs \ref{specialpiecee8-1} and \ref{specialpiecee8-2}, we depict two special pieces arising in $E_8$.
\begin{figure} [!h]
\begin{center}
\begin{tikzpicture}
 \node (a) at (1,0) {$_1\underline{D_4(a_3)}$};
 \node (a+) at (-2,0) {$F=U(1)^2$};
 \node (b) at (1,-2) {$_2A_3+A_1$};
  \node (b+) at (-2,-2) {$F=SU(2)\times U(1)$};
  \node (c) at (1,-4) {$_32A_2+A_1$};
  \node (c+) at (-2,-4) {$F=SU(2)$};
  \draw[preaction={draw=white, -,line width=10pt}] (a)--(b)--(c);
 

\node (m) at (8,-0)  {$D_4(a_1)$};
\node (m+) at (11,-3) {$_3(2A_2+A_1,0)$};
 \node (m++) at (5,-3) {$_2(A_3+A_1,0)$} ;
  \node (m+++) at (8,-3) {$_1(D_4,[2^2,1^4])$} ;
   \draw[|-{Latex[black,scale=2]}, dashed](m) -- (m+);
    \draw[|-{Latex[black,scale=2]}, dashed] (m) -- (m++); 
    \draw[|-{Latex[black,scale=2]}, dashed] (m) -- (m+++);
\node (g) at (2.7,1) {$\mathfrak{g}$}  ;
\node (gd) at (3.4,1.065) {$\mathfrak{g}^\vee$}  ;
\node (gg) at (3.0,+1) {}  ;
\node (gg+) at (3.0,-5) {}  ;
     \draw[preaction={draw=white, -,line width=10pt}] (gg)--(gg+);
 \end{tikzpicture}
\end{center}
\caption{This diagram shows the duality in $\mathfrak{g}=E_6$ between Slodowy slices to orbits in the special piece $\{\underline{D_4(a_1)},A_3+A_1,2A_2+A_1 \}$ and the special sheets attached to $D_4(a_1)$.}
\label{specialpiecee6}
\end{figure}
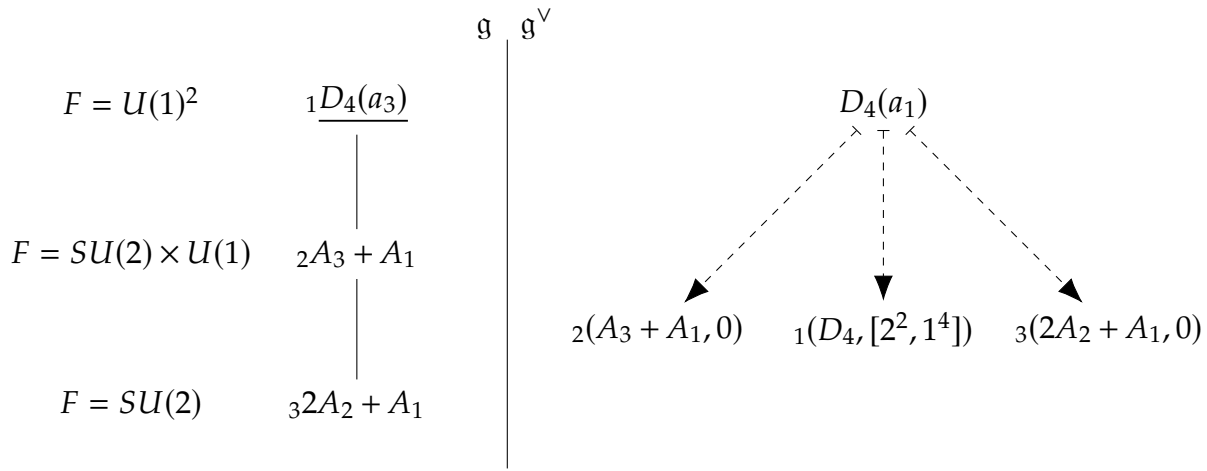

\begin{figure} [!h]
\begin{center}
\begin{tikzpicture}
 \node (a) at (0,0) {$_1\underline{E_7(a_3)}$};
 \node (a+) at (-2,0) {$F=SU(2)$};
 \node (b) at (0,-2) {$_2D_6$};
  \node (b+) at (-2,-2) {$F=SO(5)$};
  \draw[preaction={draw=white, -,line width=10pt}] (a)--(b);
 

\node (m) at (8,-1)  {$A_4$};
\node (m+) at (11,-3) {$_1(E_7,A_2)$};
 \node (m++) at (5,-3) {$_2(D_6,0)$} ;
   \draw[|-{Latex[black,scale=2]}, dashed](m) -- (m+);
    \draw[|-{Latex[black,scale=2]}, dashed] (m) -- (m++); 
\node (g) at (2.7,0) {$\mathfrak{g}$}  ;
\node (gd) at (3.4,0.065) {$\mathfrak{g}^\vee$}  ;
\node (gg) at (3.0,+1) {}  ;
\node (gg+) at (3.0,-3) {}  ;
     \draw[preaction={draw=white, -,line width=10pt}] (gg)--(gg+);
 \end{tikzpicture}
\end{center}
\caption{This diagram shows the duality in $\mathfrak{g}=E_8$ between Slodowy slices to orbits in the special piece $\{E_7(a_3), D_6 \}$ and the special sheets attached to $A_4$. As ordinary sheets, $(E_7,A_2)$ is actually a sub-sheet of $(D_6,0)$ but we treat them as two different refined sheets.}.
\label{specialpiecee8-1}
\end{figure}
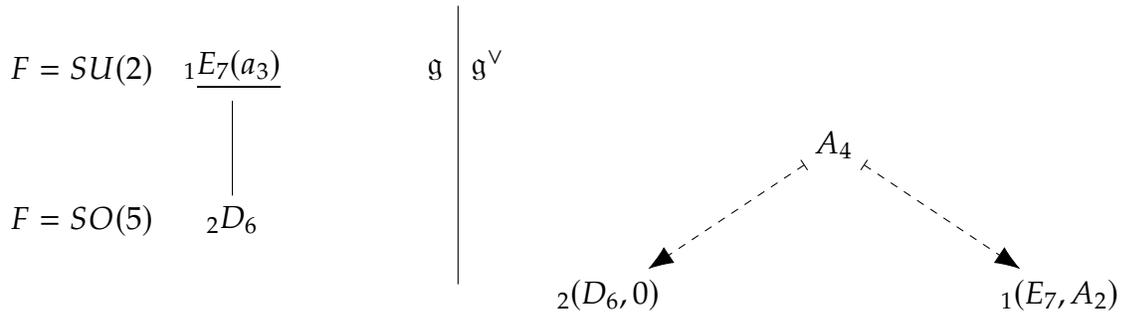

\clearpage
\subsection{Special Sheets vs Non-Special Sheets}
\label{specialvsnonspecial}

\textbf{Examples of non-special sheets attached to special orbits}

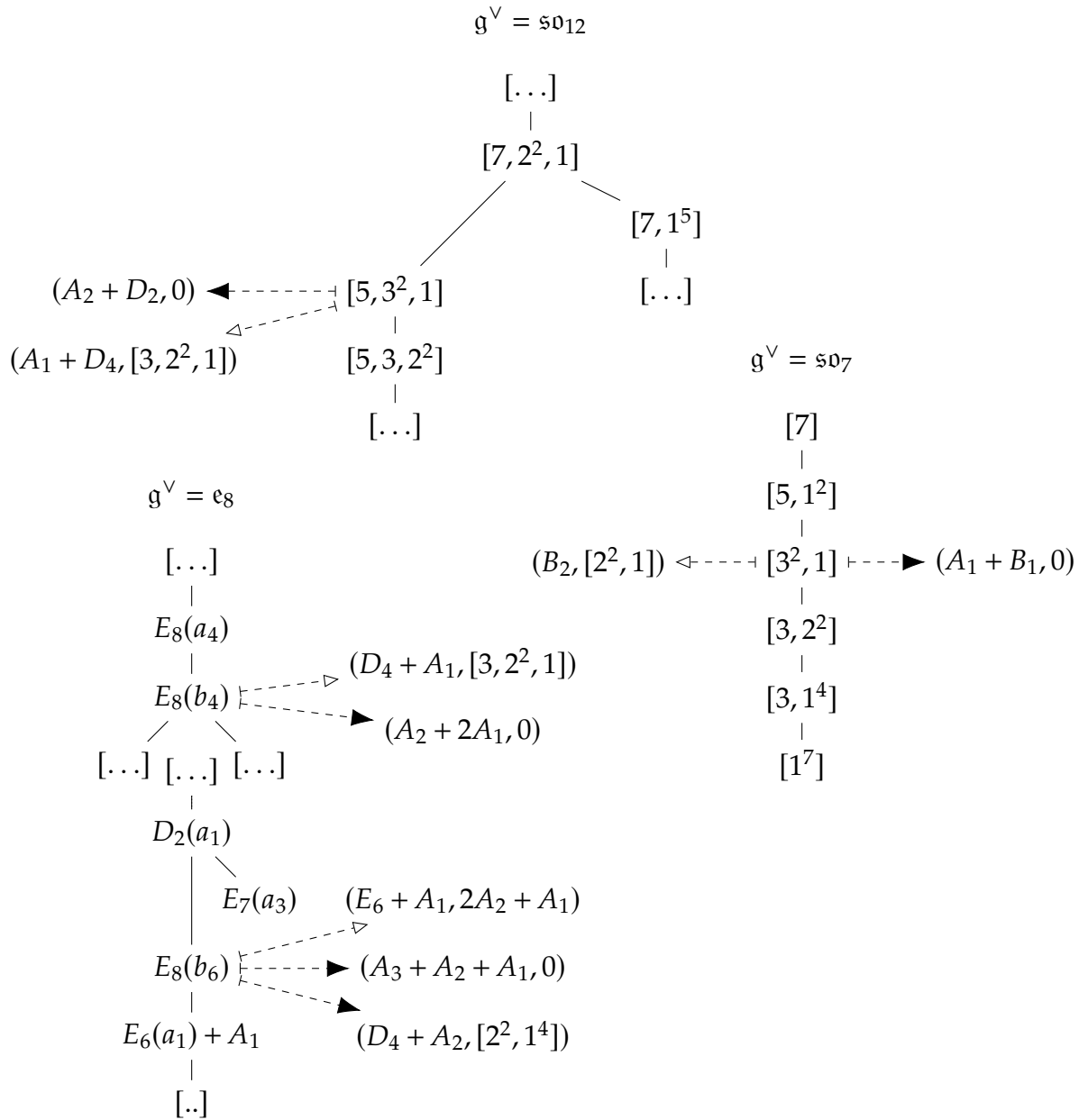
\begin{figure} [!h]
\begin{center}
\begin{tikzpicture}


 \node (so12) at (0,0) {$\mathfrak{g}^\vee=\mathfrak{so}_{12}$};
 
 \node (top) at (0,-1) {$[\ldots]$};
 
 \node (a) at (0,-2) {$[7,2^2,1]$};
 
 \node (b) at (-2,-4) {$[5,3^2,1]$};
 \node (b+) at (-6,-4) {$(A_2+D_2,0)$};
 \node (b++) at (-6,-5) {$(A_1+D_4,[3,2^2,1])$};
 
 \node (c) at (2,-3) {$[7,1^5]$};
 
 \node (d) at (-2,-5) {$[5,3,2^2]$};
 
  \node (dd) at (-2,-6) {$[\ldots]$};
   \node (cc) at (2,-4) {$[\ldots]$};

  \draw[preaction={draw=white, -,line width=10pt}] (top)--(a)--(b)--(d)--(dd);
  \draw[preaction={draw=white, -,line width=10pt}] (a)--(c)--(cc);
   \draw[|-{Latex[black,scale=2]}, dashed](b) -- (b+);
      \draw[>=open triangle 45,scale=2,|->, dashed](b) -- (b++);
     

 \node (so7) at (4,-5) {$\mathfrak{g}^\vee=\mathfrak{so}_7$};
 
 \node (so7top) at (4,-6) {$[7]$};
 
 \node (soA) at (4,-7) {$[5,1^2]$};
 
 \node (soB) at (4,-8) {$[3^2,1]$};
 \node (soB+) at (1,-8) {$(B_2,[2^2,1])$};
 \node (soB++) at (7,-8) {$(A_1+B_1,0)$};
 
 \node (soC) at (4,-9) {$[3,2^2]$};
 
  \node (soCC) at (4,-10) {$[3,1^4]$};
    
  \node (soD) at (4,-11) {$[1^7]$};

  \draw[preaction={draw=white, -,line width=10pt}] (so7top)--(soA)--(soB)--(soC)--(soCC)--(soD);
  
     \draw[>=open triangle 45,scale=2,|->, dashed](soB) -- (soB+);
   \draw[|-{Latex[black,scale=2]}, dashed] (soB) -- (soB++);


 \node (e8) at (-5,-7) {$\mathfrak{g}^\vee=\mathfrak{e}_8$};
 
 \node (e8top) at (-5,-8) {$[\ldots]$};
 
 \node (ea) at (-5,-9) {$E_8(a_4)$};
 
 \node (eb) at (-5,-10) {$E_8(b_4)$};
  \node (eb+) at (-1,-9.5) {$(D_4+A_1,[3,2^2,1])$};
  \node (eb++) at (-1,-10.5) {$(A_2+2A_1,0)$};
 
 \node (ec) at (-6,-11) {$[\ldots]$};
 
 \node (ecc) at (-4,-11) {$[\ldots]$};
 
 \node (ed) at (-5,-11.1) {$[\ldots]$};

 \node (ee) at (-5,-12) {$D_2(a_1)$};
 
 \node (ef) at (-4,-13) {$E_7(a_3)$};
 
 \node (eg) at (-5,-14) {$E_8(b_6)$};
 \node (eg+) at (-1,-13) {$(E_6+A_1,2A_2+A_1)$};
 \node (eg++) at (-1,-14) {$(A_3+A_2+A_1,0)$};
 \node (eg+++) at (-1,-15) {$(D_4+A_2,[2^2,1^4])$};
 
 \node (eh) at (-5,-15) {$E_6(a_1)+A_1$};
 \node (ehh) at (-5,-16) {$[..]$};

  \draw[preaction={draw=white, -,line width=10pt}] (e8top)--(ea)--(eb)--(ec);
  \draw[preaction={draw=white, -,line width=10pt}] (eb)--(ecc);
   \draw[preaction={draw=white, -,line width=10pt}] (ed)--(ee)--(eg)--(eh)--(ehh);
    \draw[preaction={draw=white, -,line width=10pt}] (ee)--(ef);
   \draw[densely dotted] (ed)--(ee);
  \draw[>=open triangle 45,scale=2,|->, dashed](eb) -- (eb+);
     \draw[|-{Latex[black,scale=2]}, dashed] (eb) -- (eb++);
    
      \draw[>=open triangle 45,scale=2,|->, dashed](eg) -- (eg+);
     \draw[|-{Latex[black,scale=2]}, dashed] (eg) -- (eg++); 
     \draw[|-{Latex[black,scale=2]}, dashed] (eg) -- (eg+++);

 \end{tikzpicture}
\end{center}
\caption{This diagram shows some examples of non-special sheets (denoted by white arrows) whose boundaries are special nilpotent orbits. We find that non-special sheets do not correspond to mass deformations of a $T^\rho[G]$ theory.}
\label{non-special}
\end{figure}

A special orbit that is not special rigid will have atleast one special sheet attached to it. But, it could also have non-special sheets attached to it. Here are some examples from the classical Lie algebras. We illustrate some of these examplein Fig \ref{non-special}

\begin{enumerate}
\item Consider the special orbit $\mathcal{O}_H = [5,3^2,1]$ in $\mathfrak{g}^\vee=so(12)$.

 We have $\mathfrak{g}=so(12)$ and $\dim(\mathcal{O}_H)=50$. This special orbit has two sheets attached to it and only one of it is a special sheet.
\begin{enumerate}
\item \textbf{Special Sheet} : $(A_2+D_2,0)$ 
 
 We claim that this corresponds to the mass deformation of a smoothable SCFT $T^{[3^3,1^3]}[\mathfrak{so}_{12}]$. First, one can check that the Bala-Carter Levi of $[3^3,1^3]$ is indeed $(A_2+D_2)$, as required by the Flavour condition.	 The Flavour symmetry associated to this defect is $F=$ and we have $d_{BV}([3^3,1^3]) = [5,3^2,1]$. This confirms that we have identified the mass deformed Coulomb branch correctly.

\item \textbf{Non-Special Sheet} : $(A_1+D_4,
   [3,2^2,1])$
    
    Let us assume that this non-special sheet \textit{could} correspond to a mass deformation of a $T^\rho[\mathfrak{so}_{12}]$ theory. Then, the Flavour condition requires that the Bala-Carter Levi associated to $\rho$ should be $(A_1+D_4)$. We have two possibilities
    \begin{equation}
    [7,2^2,1] \text{ , } [5,3,2^2]
\end{equation}  
    First, we consider the defect with Nahm orbit $[7,2^2,1]$. Note that $[7,2^2,1]$ is an orbit that is of principal Levi type\footnote{See \S\ref{definitionsappendix} for a recollection of definitions.}. So, it corresponds to a smoothable SCFT and the sheet associated to its mass deformation should be a Dixmier sheet. But, $(A_1+D_4,
   [3,2^2,1])$ is not a Dixmier sheet. Furthermore, we have that
   \begin{equation}
   d_{BV} ([7,2^2,1]]) = [3^2,1^6].
   \end{equation}
   But the nilpotent orbit at the boundary of the  $(A_1+D_4,
   [3,2^2,1])$ sheet is $[5,3^2,1]$. So, this rules out the possibility that $(A_1+D_4,
   [3,2^2,1])$ could parameterize the mass deformation of $T^{[7,2^2,1]}[\mathfrak{so}_{12}]$. 

    Next, we consider $T^{[5,3,2^2]}$. The Coulomb branch of this theory is the closure of the nilpotent orbit $[4^2,1^4]$ since we have 
\begin{equation}
d_{BV}([5,3,2^2]) = [4^2,1^4].
\end{equation}    
  This, again, is not the nilpotent orbit at the boundary of the $(A_1+D_4,
   [3,2^2,1])$ sheet. So, we conclude that this sheet does not correspond to a mass deformation of any $T^\rho[\mathfrak{so}_{12}]$ theory. 
\end{enumerate}   

\item Consider the special orbit $E_8(b_4)$ in $\mathfrak{g}^\vee = E_8$. 

We have $\mathfrak{g}^\vee$ and $\dim(E_8(b_4) = 230 $. This special orbit has two sheets attached to it. One of them is a special sheet while the other is not a special sheet.
 \begin{enumerate}
 \item \textbf{Special Sheet } : $(A_2 +2A_1,0)$
 
  This sheet parameterizes the mass deformed Coulomb branch of $T^{A_2+2A_1}[E_8]$.
  \item \textbf{Non-Special Sheet } : $(D_4+A_1,[3,2^2,1])$
  
  This sheet does not correspond to the mass deformation of any $T^\rho[E_8]$ theory.

 \end{enumerate}
\item As a further example in $E_8$, consider the special orbit $E_8(b_6)$.

We again have $\mathfrak{g}^\vee = E_8$ and $\dim(E_8(b_6)) = 220 $. This special orbit has three sheets attached to it. One of them is a Dixmier sheet (and hence, a special sheet), one is a special sheet that is not a Dixmier sheet and the other is a non-special sheet. We will see that the two special sheets correspond to mass deformations of some $T^\rho[E_8]$ theory while the non-special sheet does not.

 \begin{enumerate}
 \item \textbf{Special (and Dixmier) sheet} : $(A_3+A_2+A_1,0)$
 \item \textbf{Special (non-Dixmier) sheet} : $(D_4+A_1,[2^2,1^4])$
 \item \textbf{Non-special Sheet} : $(E_6+A_1,2A_2+A_1)$
 \end{enumerate}

\item Finally, let us also consider an example from non-simply laced Lie algebras. Take the twisted defect of the 6d theory with Hitchin orbit  $\mathcal{O}_H = [3^2,1]$ in $\mathfrak{g}^\vee=so(7)$.

We have $\mathfrak{g}=sp(6)$ and $\dim(\mathcal{O}_H)=14$. This special orbit has two sheets attached to it. One of the sheets is a special sheet and the other is a non-special sheet.

\begin{enumerate}

\item \textbf{ Special Sheet} : $(A_1+B_1,0)$
 
 This corresponds to the mass deformation of the smoothable SCFT $T^{[2^3]}[\mathfrak{sp}_6]$. \footnote{See \S\ref{tablegc3} for a summary of mass deformations of all $T^\rho[\mathfrak{sp}_6]$ theories.}

\item \textbf{ Non-Special Sheet} : $(B_2,[2^2,1])$
 
 This sheet does not correspond to the mass deformation of any $T^\rho[\mathfrak{sp}_6]$ theory.
 
 \end{enumerate}

\end{enumerate}

We collect all of the examples discussed in this section in Fig \ref{non-special}.

\clearpage

\subsection{Mass deformations and $A(\mathcal{O}_H)$ data}
\label{deformationsin4d}

 As outlined in the introductory section, our original motivation was to understand the effect of a mass deformation on the Coulomb branch of an arbitrary Class $\mathcal{S}$ theory. The description of the mass deformation of the three dimensional $T^\rho[G]$ theories is one step in the direction of a description of the mass deformation of the four dimensional theories. But, a few further steps are necessary to obtain a complete description of the mass deformed Coulomb branch of a generic Class $\mathcal{S}$ SCFT that is built using tame defects. 
  
 Here, we only mention the additional considerations that are purely \text{local} to the defect. In \cite{Chacaltana:2012zy}, it was realized that a precise description of the four dimensional Coulomb branch in the massless limit requires a study of the Sommers-Achar group associated to the $T^\rho[G]$ theory. This is a subgroup of the $\overline{A}(\mathcal{O}_H)$ group, a finite group that is a quotient of the component group $A(\mathcal{O}_H)$ associated to the Hitchin orbit $\mathcal{O}_H$. One may wonder how these Sommers-Achar groups interact with mass deformations and the theory of sheets. To understand this, we need to understand the role played by the Sommers-Achar group in the description of the 3d $T^\rho[G]$ theories. We postpone such a study for a later work. 
 
\subsection{Infinite families}

To complete this section, we note here examples of certain infinite families of Rigid and Malleable SCFTs. If one is studying a large class of CFTs in any dimension, a natural question to ask if these CFTs admit interesting ``large N'' limits. These are limits in which the number of degrees of freedom in the CFT grows in the some controlled fashion. It is typically the case that there are several large $N$ limits that one could consider and this is certainly the case here.  We present a few representative examples to show the existence of large N limits in which every SCFT in the sequence is either Rigid or Malleable. In the table, we also indicate the nilpotent orbit whose closure is the Coulomb branch of these theories.
 \begin{center}
 \begin{tabular}{c|c|c}
 Theories & Coulomb branch & Deformation Class \\ \hline
 $T^{[2N-3,3]}[SO_{2N}], N=4,\ldots$ & $[2^2,1^{2N-4}]$ in $\mathfrak{so}_{2n}$ &  Rigid \\
 $T^{[2N-5,5]}[SO_{2N}], N=6,\ldots$ & $[2^4,1^{2N-8}]$ in $\mathfrak{so}_{2n}$ &  Rigid \\
 $T^{[2N-5,3,1^2]}[SO_{2N}], N=5,\ldots$ & $[3^2,1^{2N-6}]$ in $\mathfrak{so}_{2n}$ &  Malleable \\
 $T^{[2N-7,3,1^2]}[SO_{2N}], N=7,\ldots$ & $[3^2,2^2,1^{2N-10}]$ in $\mathfrak{so}_{2n}$ &  Malleable \\
  \end{tabular}
  \end{center}
  
 The deformation type of these families can be determined in a way that is identical to the method outlined in \S\ref{tables}. The existence of the above families shows that the deformation class of the SCFT is a property that survives in a suitable large $N$ limit. If an AdS dual exists for these large $N$ limits, then it makes sense to ask how these properties of the SCFT are encoded in the dual geometry. So far, large $N$ limits appear to have been studied for $T^{\rho}[G]$ theories only for the case where $G$ is of type A and for very special choices of $\rho$ like $\rho=[1^N]$ \cite{Gulotta:2011si}.

\section{Quiver Gauge Theories for $T^\rho[G]$ ?}
\label{FurtherComments}

\subsection{Searching for UV Lagrangians}
\label{uvlagrangians}
In light of the deformations classes introduced in the previous section, it is natural to wonder under what circumstances can a UV Lagrangian detect the Rigid (or Malleable) nature of the IR SCFT that the theory flows to ? \footnote{We thank Shiraz Minwalla for a discussion on this question.} Note here that being Rigid (or Malleable) is a property of the IR SCFT. A priori, it is not clear whether a UV Lagrangian can always detect such a property of the IR SCFT. 

To probe this question in a systematic manner, one would have to begin by identifying the subset of $T^\rho[G]$ theories that do admit a Lagrangian description. For a 3d $\mathcal{N}=4$ SCFT, providing a Lagrangian description entails describing a UV Lagrangian theory which flows under RG flow to the IR SCFT. The simplest case to hope for would be that this RG flow is one in which the only relevant couplings are the gauge couplings of the UV gauge theory. A more involved scenario would be the case of a RG flow in which (a subset of the) masses and twisted masses of the gauge theory are also turned on. If a 3d $\mathcal{N}=4$ SCFT occurs as the end point of at least one RG flow from a UV Lagrangian (with or without masses/twisted masses being turned on),  we will say that the corresponding SCFT admits a Lagrangian description. 

When $G$ is of type $A$, it is known that such UV descriptions are provided by certain quiver gauge theories with unitary gauge groups \cite{Gaiotto:2008ak}. When $G$ is of type $B,C,D$, it is possible, for certain $\rho$,  to provide such a UV description using SO/Sp quiver gauge theories  \cite{Hanany:2016gbz} and/or using unitary quivers \cite{Cabrera:2018ldc}. And for $G$ exceptional, it is known that a quiver gauge theory description can be provided when the Coulomb branch is a small enough nilpotent orbit \cite{Hanany:2017ooe}. If we have a UV gauge theory which flows to a $T^\rho[G]$ SCFT when the masses and twisted masses are set to zero, then it follows that the Higgs branch of the SCFT has a description as a finite dimensional hyper-K\"{a}hler quotient. For the theories under question, the Higgs branches are Slodowy slices in the Nilpotent Cone of $G$. The realization of these slices as solutions of Nahm's equation (for the group $G$) only provides an infinite dimensional hyper-K\"{a}hler quotient construction. Outside of type A, the mathematical question of which Slodowy Slices admit finite dimensional hyper-K\"{a}hler quotients appears to have been settled only in special cases. See \cite{maffei2005quiver,henderson2014diagram} for some mathematical works in this direction. 

 Since we are studying 3d $\mathcal{N}=4$ theories, one can choose to study a corresponding question for the 3d Mirror theories in which nilpotent orbits are realized as Higgs branches and Slodowy Slices would be the Coulomb branches. In this duality frame, the constructions of \cite{Cremonesi:2014uva,Hanany:2016gbz} (see \cite{kobak1996classical} for related mathematical work) provide UV Lagrangian description. While such a description is useful for many purposes, it is not helpful to make transparent the continuous global symmetry acting as isometries on the Slodowy Slice.  But, we have noted that rigidity of a $T^\rho[G]$ theory is best detected in terms of the data on the Slodowy Slice side. So, even if some of the Rigid SCFTs can be detected using Higgs branch quiver constructions for nilpotent orbits, this would not exhaust the list of Rigid SCFTs that we are interested in. One would have to necessarily supplement this using data on the Slodowy Slice side.\footnote{We thank A. Hanany and N. Mekareeya for a conversation on this question} 

\subsection{Little String Quivers}

More recently, a proposal has been put forward in \cite{Haouzi:2016ohr,Haouzi:2016yyg} that every $T^\rho[G]$ theory has a UV description in terms of a Quiver gauge theory with only Unitary gauge groups.  The starting point for the construction of \cite{Haouzi:2016ohr,Haouzi:2016yyg} are defect operators arising in Little String theory. They argue that the world-volume theory on these defects, when wrapped on an $S^1$, are certain Dynkin Quiver Gauge theories, a name that refers to the fact that the gauge nodes of the quiver are always organized according to the Dynkin diagram of $G$.  Since the Little String theories flow to the $X[j]$ SCFTs in the IR, these defect theories should flow to the $T^\rho[G]$ SCFTs. In the rest of this section, we investigate this conjecture of \cite{Haouzi:2016ohr,Haouzi:2016yyg}.

We find that for certain defects, the Little String Quivers reproduce the expected dimension of the Higgs and Coulomb branches. In certain other cases, it is possible to identify a set of masses or FI parameters (twisted masses) in the UV theory that, when turned on, would lead to a flow to a IR SCFT with Higgs and Coulomb branches of the correct expected dimension. In the remaining cases, we were not able to find a way to reconcile the expected the dimension of the Coulomb branch of a $T^\rho[G]$ theory and the actual dimension of the Coulomb branch of the LSQs proposed in \cite{Haouzi:2016ohr,Haouzi:2016yyg}. We illustrate these by studying the LSQs for $G=D_4, E_6$.

For each LSQ, the quaternionic dimension of the Coulomb branch is the sum of the ranks of the $U(n)$ gauge groups in the quiver. The \emph{virtual} (quaternionic) dimension of the Higgs branch is $n_h-n_v$. In general the gauge symmetry of a quiver gauge theory might not be completely broken even at a generic point of the Higgs branch. In such a case, the virtual dimension would differ from the actual dimension of the Higgs branch by the rank of the unbroken (abelian) gauge group. For the LSQ quivers, each gauge node is balanced and there is always at least one fundamental hypermultiplet. Thus  the gauge symmetry can be Higgsed completely, and the virtual and actual dimensions of the Higgs branch coincide.

Following \cite{panyushev1994complexity}, we define the height of a nilpotent orbit by 
\begin{displaymath}
h(\mathcal{O})= \vec{w}(\mathcal{O})\cdot \vec{m},
\end{displaymath}
where $\vec{w}(\mathcal{O})$ is the weighted Dynkin diagram of $\mathcal{O}$ and $\vec{m}$ are the Dynkin labels of the labeled Dynkin diagram.

For $h(\mathcal{O}_H)\leq 3$, we find that the Higgs and Coulomb branch dimensions of the LSQ agree on the nose with those of $T^\rho[G]$. So the LSQ is a plausible realization of $T^\rho[G]$. In the $E_6$ theory, there are 2 such defects. These are the defects with Nahm orbits being $\mathcal{O}_N=E_6(a_1)$ and $D_5$. In the $E_7$ theory, there are 5 such defects : $\mathcal{O}_N= E_7(a_1),\,E_7(a_2),\,E_7(a_3),\,E_6$ and $E_6(a_1)$. Curiously, the nilpotent orbits with $h(\mathcal{O}_H)\leq 3$ are precisely the \textit{spherical} nilpotent orbits  \cite{panyushev1994complexity}. 

For $h(\mathcal{O}_H)\geq 4$, there is invariably a mismatch, so the LSQ \emph{cannot} be a realization of $T^\rho[G]$. Nevertheless, it \emph{could} serve as a UV fixed point which \emph{flows} to $T^\rho[G]$ under an RG flow induced by turning on some combination of hypermultiplet masses and Fayet-Iliopoulos parameters. Turning on hypermultiplet masses reduces the dimension of the Higgs branch and deforms the geometry of the Coulomb branch. Turning on FI parameters reduces the dimension of the Coulomb branch and deforms the geometry of the Higgs branch. In both cases, the branch dimensions are strictly non-increasing when one turns on these relevant pertubations (a result which can be shown rigorously by moving out along one or the other of the branches and applying the analysis of \S4.4 of \cite{Cordova:2016xhm}).

In a Lagrangian field theory (like these LSQs)\footnote{We are making a tacit assumption, here, which should be spelled out. We are using the phrase ``LSQ" to denote both the Lagrangian field theory and the SCFT it flows to (with the only relevant coupling being the gauge coupling, $e$) in the IR. When $|m_i/e|\ll 1$, turning on the mass parameters (or FI terms) in the Lagrangian field theory, or turning on the corresponding relevant perturbations of the SCFT, should flow to the same (new) IR SCFT. On the other hand, when $|m_i/e|\gg 1$, we can analyze the effect of turning on the (twisted) masses semiclassically in the Lagrangian field theory. We are assuming that, with $\mathcal{N}=4$ supersymmetry, there is no phase transition in extrapolating from $|m_i/e|\gg1$ to $|m_i/e|\ll1$.}, turning on an FI parameter decreases the Coulomb branch dimension by 1. Since there are $\mathrm{rank}(G)$ (= number of gauge nodes in the quiver) FI parameters available to turn on, this indicates that we can only flow to an IR SCFT whose Coulomb branch dimension, $d_{\text{Coul}}$, satisfies

\begin{equation}
0 \leq \dim_{\text{Coul}}(LSQ)-d_{\text{Coul}} \leq \mathrm{rank}(G)
\label{coulombbranchcondition}
\end{equation}
We find a smattering of additional LSQs, with $4\leq h(\mathcal{O}_H)\leq 8$ which could plausibly flow to the corresponding $T^\rho[G]$. For most orbits (indeed, for all orbits with $h(\mathcal{O}_H)\gt 8$ in the exceptional cases), either the Higgs branch dimension of $T^\rho[G]$ is greater than that of the LSQ, or the difference in Coulomb branch dimensions does not obey (\ref{coulombbranchcondition}), and there is no candidate for an RG flow from the LSQ to $T^\rho[G]$.

Before turning to examples in $D_4$ and $E_6$, we note here that the height of a nilpotent orbit has also played an important role in recent works  \cite{Hanany:2017ooe,Hanany:2018uzt}. In unpublished work, A. Hanany and G. Ferlito \cite{FerlitoHanany} have also considered the behaviour of Dynkin quivers at various heights. 

\subsubsection{Little String Quivers for $D_4$}

There are 11 non-trivial defects in total. We consider the Little String Quiver (LSQ) proposed for each defect and compute the values of the expected and actual Higgs and Coulomb branch dimensions. We also compute the height of the Hitchin nilpotent orbit for each defect.  We label each defect by the pair $(\mathcal{O}_N, \mathcal{O}_H)$ . We refer to pg. 56 of \cite{Haouzi:2016ohr} for the LSQ Diagrams.

{
\footnotesize
\renewcommand{\arraystretch}{2.25}

\begin{longtable}{|c|c|c|c|c|c|c|}
\hline
$\mathcal{O}_N$&$\mathcal{O}_H$&$h(\mathcal{O}_H)$&Little String Quiver&\mbox{\shortstack{$(d_{\text{Higgs}},d_{\text{Coul}})$\\LSQ}}&\mbox{\shortstack{$(d_{\text{Higgs}},d_{\text{Coul}})$\\$T^\rho[D_4]$}}&Reachable?\\
\hline
\endhead
$[1^8]$&$[7,1]$&6&\includegraphics[width=74pt]{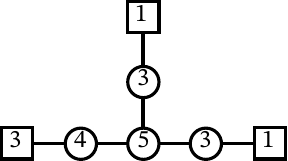}&$(9,15)$&$(12,12)$&\mbox{\shortstack{No.\\ $d_{\text{Higgs}}(T^\rho[D_4]) > d_{\text{Higgs}}(LSQ)$}}\\
\hline
$[2^2,1^4]$&$[5,3]$&6&\includegraphics[width=74pt]{quivers/D4quiver11111111}&$(6,12)$&$(7,11)$&\mbox{\shortstack{No.\\ $d_{\text{Higgs}}(T^\rho[D_4]) > d_{\text{Higgs}}(LSQ)$}}\\
\hline
$[2^4]^I$&$[4^2]^I$&6& \includegraphics[width=60pt]{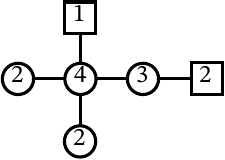}&$(5,11)$&$(6,10)$&\mbox{\shortstack{No.\\ $d_{\text{Higgs}}(T^\rho[D_4]) > d_{\text{Higgs}}(LSQ)$}}\\
\hline
$[2^4]^{II}$&$[4^2]^{II}$&6&same as above&$(5,11)$&$(6,10)$&\mbox{\shortstack{No.\\ $d_{\text{Higgs}}(T^\rho[D_4]) > d_{\text{Higgs}}(LSQ)$}}\\
\hline
$[3,1^5]$&$[5,1^3]$&6&same as above&$(5,11)$&$(6,10)$&\mbox{\shortstack{No.\\ $d_{\text{Higgs}}(T^\rho[D_4]) > d_{\text{Higgs}}(LSQ)$}}\\
\hline
$[3,2^2,1]$&$[3^2,1^2]$&4&\includegraphics[width=41pt]{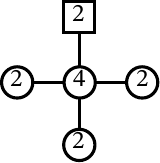}&$(4,10)$&$(4,9)$&\mbox{\shortstack{Yes, if we turn on\\ an FI term.}}\\
\hline
$[3^2,1^2]$&$[3^2,1^2]$&4&\includegraphics[width=79pt]{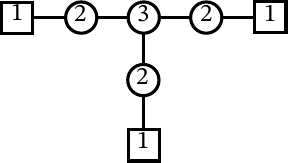}&$(3,9)$&$(3,9)$&\mbox{\shortstack{Yes. See, \emph{e.g.} \cite{Cabrera:2018ldc}.}}\\
\hline
$[5,1^3]$&$[3,1^5]$&2& \includegraphics[width=58pt]{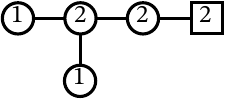}&$(2,6)$&$(2,6)$&\mbox{\shortstack{Yes. See, \emph{e.g.} \cite{Cabrera:2018ldc}.}}\\
\hline
$[4^2]^I$&$[2^4]^I$&2&same as above&$(2,6)$&$(2,6)$&\mbox{\shortstack{Yes. See, \emph{e.g.} \cite{Cabrera:2018ldc}.}}\\
\hline
$[4^2]^{II}$&$[2^4]^{II}$&2&same as above&$(2,6)$&$(2,6)$&\mbox{\shortstack{Yes. See, \emph{e.g.} \cite{Cabrera:2018ldc}.}}\\
\hline
$[5,3]$&$[2^2,1^4]$&2& \includegraphics[width=41pt]{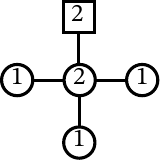}&$(1,5)$&$(1,5)$&\mbox{\shortstack{Yes. See, \emph{e.g.} \cite{Cabrera:2018ldc}.}}\\
\hline

\end{longtable}
}

\subsubsection{Little String Quivers for $E_6$}

We have 20 non-trivial defects in this case. We refer to Appendix A of \cite{Haouzi:2016yyg} for the LSQ data. We include the LSQs here for a few representative examples. 

{
\footnotesize
\renewcommand{\arraystretch}{2.25}

\begin{longtable}{|c|c|c|c|c|c|c|}
\hline
$\mathcal{O}_N$&$\mathcal{O}_H$&$h(\mathcal{O}_H)$&Little String Quiver&\mbox{\shortstack{$(d_{\text{Higgs}},d_{\text{Coul}})$\\LSQ}}&\mbox{\shortstack{$(d_{\text{Higgs}},d_{\text{Coul}})$\\$T^\rho[E_6]$}}&Reachable?\\
\hline
\endhead
$0$&$E_6$&22& \includegraphics[width=74pt]{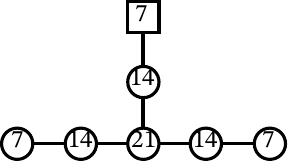}&$(49,77)$&$(36,36)$&\mbox{\shortstack{No.\\ $d_{\text{Coul}}(LSQ)-d_{\text{Coul}}(T^\rho[E_6]) > 6$}}\\
\hline
$A_1$&$E_6(a_1)$&16& \includegraphics[width=74pt]{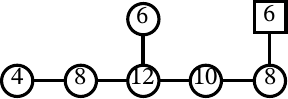} &$(24,48)$&$(25,35)$&\mbox{\shortstack{No. Both\\ $d_{\text{Higgs}}(T^\rho[E_6]) > d_{\text{Higgs}}(LSQ)$\\ and\\ $d_{\text{Coul}}(LSQ)-d_{\text{Coul}}(T^\rho[E_6]) > 6$}}\\
\hline
$2 A_1$&$D_5$&14&&$(13,43)$&$(20,34)$&\mbox{\shortstack{No. Both\\ $d_{\text{Higgs}}(T^\rho[E_6]) > d_{\text{Higgs}}(LSQ)$\\ and\\ $d_{\text{Coul}}(LSQ)-d_{\text{Coul}}(T^\rho[E_6]) > 6$}}\\
\hline
$3 A_1$&$E_6(a_3)$&10&&$(8,62)$&$(16,33)$&\mbox{\shortstack{No. Both\\ $d_{\text{Higgs}}(T^\rho[E_6]) > d_{\text{Higgs}}(LSQ)$\\ and\\ $d_{\text{Coul}}(LSQ)-d_{\text{Coul}}(T^\rho[E_6]) > 6$}}\\
\hline
$A_2$&$E_6(a_3)$&10&&$(17,49)$&$(15,33)$&\mbox{\shortstack{No.\\ $d_{\text{Coul}}(LSQ)-d_{\text{Coul}}(T^\rho[E_6]) > 6$}}\\
\hline
$A_2 + A_1$&$D_5(a_1)$&10& \includegraphics[width=74pt]{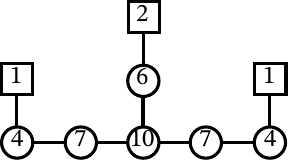} &$(10,38)$&$(13,32)$&\mbox{\shortstack{No.\\ $d_{\text{Higgs}}(T^\rho[E_6]) > d_{\text{Higgs}}(LSQ)$}}\\
\hline
$A_2 + 2 A_1$&$A_4 + A_1$&8&&$(10,38)$&$(12,31)$&\mbox{\shortstack{No. Both\\ $d_{\text{Higgs}}(T^\rho[E_6]) > d_{\text{Higgs}}(LSQ)$\\ and\\ $d_{\text{Coul}}(LSQ)-d_{\text{Coul}}(T^\rho[E_6]) > 6$}}\\
\hline
$2A_2$&$D_4$&10&&$(10,33)$&$(12,30)$&\mbox{\shortstack{No.\\ $d_{\text{Higgs}}(T^\rho[E_6]) > d_{\text{Higgs}}(LSQ)$}}\\
\hline
$A_3$&$A_4$&8& \includegraphics[width=74pt]{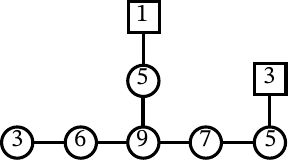}&$(10,35)$&$(10,30)$&\mbox{\shortstack{Yes, if we turn on\\ FI terms.}}\\
\hline
$2 A_2 + A_1$&$D_4(a_1)$&6&  \includegraphics[width=74pt]{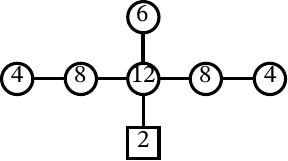} &$(12,42)$&$(9,29)$&\mbox{\shortstack{No. \\ $d_{\text{Coul}}(LSQ)-d_{\text{Coul}}(T^\rho[E_6]) > 6$}}\\
\hline
$A_3 + A_1$&$D_4(a_1)$&6& \includegraphics[width=74pt]{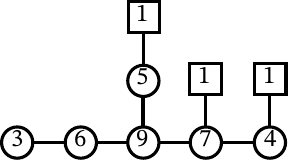} &$(8,34)$&$(8,29)$&\mbox{\shortstack{Yes, if we turn on\\ FI terms.}}\\
\hline
$D_4(a_1)$&$D_4(a_1)$&6&&$(9,37)$&$(7,29)$&\mbox{\shortstack{No. \\ $d_{\text{Coul}}(LSQ)-d_{\text{Coul}}(T^\rho[E_6]) > 6$}}\\
\hline
$A_4$&$A_3$&6&&$(7,31)$&$(6,24)$&\mbox{\shortstack{No. \\ $d_{\text{Coul}}(LSQ)-d_{\text{Coul}}(T^\rho[E_6]) > 6$}}\\
\hline
$D_4$&$2 A_2$&4& \includegraphics[width=74pt]{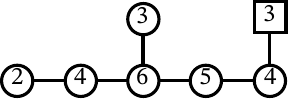}&$(6,24)$&$(6,26)$&\mbox{\shortstack{No. \\ $d_{\text{Coul}}(LSQ)-d_{\text{Coul}}(T^\rho[E_6]) < 0$}}\\
\hline
$A_4+A_1$&$A_2+2A_1$&4& \includegraphics[width=74pt]{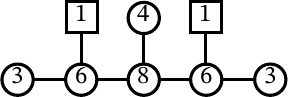}&$(6,30)$&$(5,25)$&\mbox{\shortstack{Yes, if we turn on\\ both FI terms and a mass.}}\\
\hline
$D_5(a_1)$&$A_2+A_1$&4& \includegraphics[width=74pt]{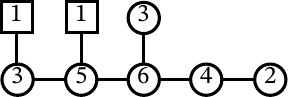}&$(4,23)$&$(4,23)$&\mbox{\shortstack{Yes. The moduli\\ space dimensions match.}}\\
\hline
$A_5$&$A_2$&4&\includegraphics[width=74pt]{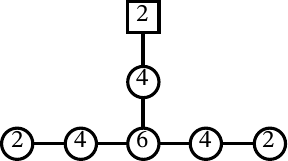} &$(2,22)$&$(4,21)$&\mbox{\shortstack{No.\\ $d_{\text{Higgs}}(T^\rho[E_6]) > d_{\text{Higgs}}(LSQ)$}}\\
\hline
$E_6(a_3)$&$A_2$  \footnote{For this defect, the authors of \cite{Haouzi:2016yyg} assign the Hitchin orbit to be ``$3 A_1$'', a  \emph{non-special} orbit. They do not clarify why $3 A_1$ and not $A_2$ is the Hitchin Orbit for this case. Incidentally, the LSQ has the right Coulomb Branch dimension for $A_2$.}&4& \includegraphics[width=74pt]{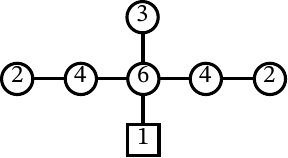}&$(3,21)$&$(3,21)$&\mbox{\shortstack{Yes. The moduli\\ space dimensions match.}}\\
\hline
$D_5$&$2 A_1$&2&\includegraphics[width=74pt]{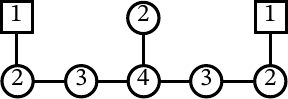}&$(2,16)$&$(2,16)$&\mbox{\shortstack{Yes. Constructed\\ in  \cite{Hanany:2017ooe}.}}\\
\hline
$E_6(a_1)$&$A_1$&2&\includegraphics[width=74pt]{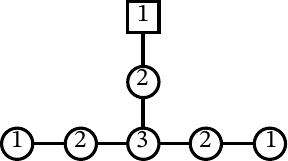}&$(1,11)$&$(1,11)$&\mbox{\shortstack{Yes. Constructed\\ in  \cite{Hanany:2017ooe}.}}\\
\hline
\end{longtable}
}

\section{Connections to Symplectic Duality}
\label{SymplecticDuality}

In this section, we explore two consequences of our results for the interface between Symplectic Singularities and 3d $\mathcal{N}=4$ SCFTs and to the recent work on ``Symplectic Duality" \cite{braden2014quantizations,nakajima2015questions}. 

The notion of Symplectic Duality was defined in \cite{braden2014quantizations} and according to this definition, two symplectic resolutions are dual to each other if a series of highly non-trivial relations hold between certain natural objects related to the two resolutions (See \S 10 of \cite{braden2014quantizations}). At the highest level, Symplectic Duality is a Koszul duality between certain geometric and algebraic categories associated to the two Symplectic Resolutions together with a switching of roles of two natural functors that act on these categories (the twisting and shuffling functors). These categories are constructed in a way analogous to the construction of the geometric and algebraic Bernstein-Gelfand-Gelfand (BGG) Category $\mathcal{O}$ corresponding to representations of the $\mathcal{U}(\mathfrak{g})$, the universal enveloping algebra of $\mathfrak{g}$. For the case where the categories are indeed the original BGG categories, the corresponding symplectic resolution is the Springer resolution of the nilpotent cone of $\mathfrak{g}$. 

In  \cite{braden2014quantizations}, the statement of Symplectic Duality at the level of categories is then shown to have important consequences at the level of cohomology of these resolutions. This duality at the level of cohomology further leads to an order-reversing duality between the strata (symplectic leaves) of the singular base manifolds of the two symplectic resolutions. A simple instance of this order-reversing duality is that the top stratum on one side of the duality is always exchanged with the lowest stratum on the other side. On the other hand, the duality between intermediate strata can be quite involved. 

But, it can sometimes turn out that the intermediate strata can themselves be realized as top/lowest strata of certain smaller Symplectic Resolutions. For the case of the Springer resolutions, these smaller resolutions are the parabolic Springer resolutions\footnote{These resolutions are often called ``partial Springer resolutions". Since we have already used the adjective ``partial'' in a different context in this paper, we have avoided calling these partial Springer resolutions and instead prefer the term parabolic Springer resolution.} of the form $T^\star(G_\mathbb{C}/P) \rightarrow \overline{\mathcal{O}}$ for certain choice of a parabolic $P$. The existence of these smaller resolutions is helpful since one can then hope to devise an inductive procedure of deducing the order-reversing duality relevant for the bigger resolution by using the duality between top stratum and lowest stratum for the smaller resolutions. When $G$ is a Lie group of Cartan type $A$, every nilpotent orbit can be realized as either the top or lowest stratum of a suitable parabolic Springer resolution. This corresponds to the fact that every nilpotent orbit of type $A$ is both principal Levi type (restricts to the principal orbit in a Levi) and Richardson (can be induced from the zero orbit in a Levi). So, the entire order-reversing duality map can be deduced just from the knowledge that the top stratum and lowest stratum are exchanged under the duality.

 However, this property does not hold outside of type A. For example, there exist strata of the Nilpotent cone that can not be realized as the open stratum of (the base in) some parabolic Springer resolution. Correspondingly, we have nilpotent orbits that are not Richardson. So, to deduce the action of the duality on these strata, we need some additional input. For us, this additional input is in the form of the condition that the mass deformations of the corresponding 3d $\mathcal{N}=4$ SCFT have consistent interpretations as a) VEVs for scalars in background vector multiplets for the Flavour symmetry group and b) they act as deformations of the Coulomb branch in a way that resolves (perhaps only partially) the singular geometry. It is this additional constraint that led us to consider a refinement of the theory of sheets in \S\ref{taxonomy}.
 
 This dual role played by the mass parameters also plays an important role in the switching of the twisting and shuffling functors under Symplectic Duality. Here, we will specialize to the case of $T^\rho[G]$ theories. We will continue to denote their Flavour symmetry by $F$. The twisting and shuffling functors are close relatives of the $W(F)$ action on the space of mass parameters. 
 
 From the perspective of the Higgs branch (the Slodowy Slice), the action of $W(F)$ on masses is the action of the $W(F)$ on the parameters living in the Cartan subalgebra of group of continuous hyper-K\"{a}hler isometries of the Slodowy Slice. This is the de-categorified version of the action of the shuffling functor \cite{braden2014quantizations}.
 
 From the perspective of the Coulomb branch, the $W(F)$ action has a different interpretation. In the framework of \cite{braden2012quantizations}, the family of Coulomb branches (parameterized by the value of the complex mass parameters) obtained by mass deformations can be identified with a canonical Poisson deformation associated to every Symplectic resolution \footnote{We emphasize here that while the Symplectic Resolution depends on the choice of a parabolic $P$, the mass deformation depends only on the choice of a Levi.}. The Flavour symmetry group $F$ does \textit{not} act on the Coulomb branch. Nevertheless, we have an action of $W(F)$ on the space of mass deformations.  In other words, it is an action on the family of Coulomb branches that are obtained after mass deformation \footnote{One can make a corresponding statement about twisted masses/F.I parameters as well but for the $T^\rho[G]$ theories, mass deformations offer the more interesting constraints. This is due to the fact that the space of mass deformations varies in a non-trivial way for different choice of $\rho$ (for fixed $G$). This is in start contrast to the number of twisted masses/F.I parameters which always equals $\mathrm{rank}(\mathfrak{g})$ (independent of the choice of $\rho$).}. We derived this action in terms of sheet data in \S\ref{finitegrouponmasses}. This is the de-categorified version of the twisting functor. In \cite{bullimore2016boundaries}, the authors have argued that the twisting and shuffling functors can themselves be identified with actions of the braid group associated to $W(F)$ on certain boundary conditions in the 3d $\mathcal{N}=4$ theory reduced to two dimensions.

\subsection{Conjectures for Symplectic Duality}

Our work in this paper makes a direct connection only with the consequences of Symplectic Duality at the level of the order reversing duality on strata. Based on the results of this paper and the observation that a large class of instances of Symplectic Duality can be deduced by observing that the two relevant Symplectic Resolutions are Coulomb and Higgs branches of a specific 3d $\mathcal{N}=4$ SCFT  (see \cite{braden2014quantizations}\footnote{Here, the initial observation in this direction is attributed to Gukov-Witten} and \cite{bullimore2016boundaries}), we are led to make the following observations about Symplectic duality for the Springer resolution. 

\subsubsection{Symplectic duality is a duality between slices and sheets}
\label{slicessheets}
The first conjecture concerns the role of the theory of sheets. Our study of mass deformations clearly demonstrates that the objects that get naturally paired in $T^\rho[G]$ theories are slices in $\mathfrak{g}$ and sheets in $\mathfrak{g}^\vee$. Correspondingly, the consequences of Symplectic duality for the $T^\rho[G]$ theories are better thought of as relating slices and sheets and not just slices and orbit closures (as it was originally articulated in \cite{braden2014quantizations}). The inclusion of sheet data on the $\mathfrak{g}^\vee$ side is a \textit{necessary} refinement outside of type A since a given nilpotent orbit (even a Richardson one) can be part of multiple sheets. The distinct sheets associated a fixed nilpotent orbit correspond to the different families  of non-nilpotent orbits that are of the same dimension as the nilpotent orbit. Every such family contains infinitely many non-nilpotent orbits. Their realization as the moduli space of Nahm's equations endows any fixed orbit with a fixed hyper-K\"{a}hler structure and a unique metric. Alternatively, one can take the non-nilpotent parts sheet to parameterize a family of holomorphic symplectic forms on a K\"{a}hler manifold $(M,\omega_I)$. In this alternative interpretation, changing the eigenvalues of the semi-simple part of the non-nilpotent elements in a sheet is interpreted as changing the holomorphic symplectic form $\Omega_I$ on $(M,\omega_I)$.  For the sheet containing regular semi-simple elements (the principal Dixmier sheet), this observation is essentially due to \cite{kronheimer1990hyper}. The manifold $M$ is of the form of a homogeneous space $G_C/L_C$ when the non-nilpotent orbits in a sheet are all semi-simple. When they are of a mixed nature, the manifold $M$ is of the form $G_{C}/Z(\tilde{a})$ where $Z(\tilde{a})$ is the centralizer of one of the non-nilpotent elements and $M$ can now have non-trivial closure and non-trivial topology. The local nature of the singularity of $G_{C}/Z(\tilde{a})$ is what we have termed the \textit{residual singularity} in \S\ref{localhitchin} and is the closure of a nilpotent orbit in a proper Levi subalgebra. 

From the theory of sheets (that we have reviewed in \S\ref{sheets}), the following facts can be deduced about symplectic resolutions for nilpotent orbits. 

\begin{enumerate}[label=(\alph*)]
\item Every symplectic resolution corresponds to a special sheet (in fact a Dixmier sheet) attached to a Richardson Orbit.
\item A partial symplectic resolution of a special orbit could correspond to a special sheet or a non-special sheet. So, it makes sense to distinguish between \textit{special symplectic partial resolutions} vs \textit{non-special symplectic partial resolutions}.
\item A partial symplectic resolution of a non-special orbit corresponds to a non-special sheet. 
\end{enumerate}

We expect that the distinction between a special sheet attached to a special orbit and a non-special sheet attached to a special orbit should also be relevant for the study of Symplectic Duality. In particular, we conjecture that a non-special sheet attached to a special orbit is \textit{not} Symplectic dual to any slice on the $\mathfrak{g}^\vee$ side. Equivalently, we conjecture that only special sheets have Symplectic dual slices.

\subsubsection{Rigid SCFT is the appropriate notion of Rigidity for Symplectic Duality}
\label{correctrigidity}

Our second conjecture concerns the notion of \textit{rigidity}. We conjecture that the appropriate notion of rigidity for the study of Symplectic Duality is the one that corresponds to a \textit{Rigid SCFT} and not the one that corresponds to a \textit{Rigid Orbit}. The latter notion can be recovered from the former as a special case. This conjecture can be thought of as a refinement of the conjectures following from \cite{braden2014quantizations} which had already suggested an important role for rigid orbits.\footnote{We thank N. Proudfoot and B. Webster for discussions on this.} Below, we discuss why this refinement makes a difference.

By the observation in \S\ref{slicessheets} above, one can think of Symplectic duality for the Springer resolution as acting on a simultaneous stratification of $(\mathfrak{g},\mathfrak{g}^\vee)$ by $(\text{slices}, \text{sheets})$ in which only the special sheets (on the $\mathfrak{g}^\vee$ side) occur as Symplectic duals for slices (on the $\mathfrak{g}$ side). If one were studying the usual theory of sheets on $\mathfrak{g}^\vee$, a simple count of the special sheets in the theory would, in general, be smaller than the number of nilpotent orbits (and hence slices) in $\mathfrak{g}$. But, with the refined notion of a sheet that is based on the notion of a rigid SCFT, the number of special (refined) sheets in $\mathfrak{g}^\vee$ equals the number of nilpotent orbits in $\mathfrak{g}$ by construction. 

Now, for the purposes of our paper, we only needed to study sheets attached to special orbits and the notion of Rigidity (refined or otherwise) of special orbits. The usual notion of Rigidity applies equally well to non-special orbits. So, it is natural to wonder if there is a corresponding refined notion of rigidity that applies to non-special orbits as well. 

\subsubsection{Relation to the geometric theory of character sheaves}

It would be interesting to connect our present discussion to the theory of character sheaves for the Lie algebra $\mathfrak{g}$. This theory was originally developed by G. Lusztig in a series of papers \cite{lusztig1984intersection,lusztig1985character}. The theory has since been given a more geometric flavour in \cite{ginzburg1993induction,mirkovic2004character}. In particular, one would like to relate the duality  between slices and refined sheets arising in our work with the duality between \textit{distinguished Lusztig strata} in the Lie algebra $\mathfrak{g}$ (defined in \cite{lusztig1984intersection}) and the dual stratification on the $\mathfrak{g}^\vee$ side arising from the parameterization of character sheaves using special nilpotent orbits in $\mathfrak{g}^\vee$. These stratifications enter geometric approaches to the theory of character sheaves (for instance, see \cite{mirkovic2004character,evens1999characteristic}). Barbasch-Vogan duality $d_{BV}$ plays an important role in both situations and hence, it seems reasonable to believe that a precise connection can be established. A setting in which such a connection can be explored is the gauge theory approach to the geometric Langlands program which relies on the S-duality of 4d $\mathcal{N}=4$ SYM as its starting point \cite{Kapustin:2006pk}. When this setup is specialized to the local case, it is expected to lead to the local geometric Langlands correspondence which can be thought of as a `categorified' version of Lusztig's theory of character sheaves \cite{bezrukavnikov2016two,ben2009character}. Of particular interest would be the identification of cuspidal character sheaves in the gauge theory setup. These are, in a sense, the fundamental objects in the theory. This is because the non-cuspidal character sheaves for a Lie algebra $\mathfrak{g}$ can be obtained by induction from cuspidal character sheaves for certain proper Levi subalgebras $\mathfrak{l} \subset \mathfrak{g}$ \cite{ginzburg1993induction}. So, a classification of cuspidal sheaves for $\mathfrak{g}$ and for all proper Levi subalgebras $\mathfrak{l}$ amounts to a classification of all character sheaves for $\mathfrak{g}$. In the approach to geometric Langlands based on 4d $\mathcal{N}=4$ SYM, boundary conditions in the four dimensional theory and their behaviour under S-duality plays a crucial role. Of particular interest are the boundary conditions involving the $T^\rho[G]$ theories \cite{Gaiotto:2016hvd}. When the $T^\rho[G]$ boundary conditions are considered in conjunction with the available symmetry breaking patterns in the bulk 4d theory \cite{Balasubramanian:2016zzr}, we get a large class of possible boundary conditions in the 4d theory. From among these boundary conditions, it should be possible to separate the ones corresponding to the cuspidal objects and the non-cuspidal (or Eisenstein) objects of the local theory. We expect the results of this paper to be helpful in carrying out such a separation.

\subsection{Order Reversing Dualities and Symplectic Duality}
\label{comparingdefinitions}

In this section, we wish to clarify a subtle point which is important in the
study of order-reversing dualities and their relationship to Symplectic
Duality. For background material on order reversing dualities, see \cite{Chacaltana:2012zy,Balasubramanian:2014jca}.

  This subtlety appears only when $\mathfrak{g}$ is a semi-simple Lie algebra that
contains atleast one non-simply laced Lie algebra as one of its simple
factors. So, to simplify the discussion, let us go ahead assume $\mathfrak{g}$ is
simple and non-simply laced.  Nilpotent orbits and Slodowy Slices in
such non-simply laced Lie algebras arise as vacuum moduli spaces
associated to twisted defects of the six dimensional theory \cite{Chacaltana:2012zy}.  As
we reviewed in an earlier section, the Higgs branch associated to such a
defect is a Slodowy Slice in $\mathfrak{g}$ and the Coulomb Branch is a Nilpotent
Orbit in $\mathfrak{g}^\vee$. The relationship between them is an order reversing
duality that passes to the Langlands dual Lie algebra. One can see that
this is the case by relating these defects to S-duality of boundary
conditions in $\mathcal{N}=4$ SYM and  this S-duality exchanges the gauge theories
with $G$ and $G^\vee$ gauge groups. There is, however, a different order
reversing duality map that stays \textit{within} the Lie Algebra $\mathfrak{g}$.  This map is
also relevant for the physics of these defects in the following way. If
one were to consider an alternate dimensional reduction from six
dimensions to four dimensions in which these defects reduce to surface
operators of the $\mathcal{N}=4$ theory, then there is a S-duality map between Surface
Operators of $G$ and Surface Operators of $G^\vee$ theory.  When restricted
to Surface Operators corresponding to Special Nilpotent orbits, this map
squares to the Identity and is order (and dimension) preserving.  When
this map is composed with the Order reversing duality  that relates the
Higgs and Coulomb branches of $T^\rho[G]$, then we get an order-reversing map from Slodowy
Slices of $\mathfrak{g}$ to Nilpotent Orbits of $\mathfrak{g}$. \footnote{The case where $\mathfrak{g}=F_4$ is an
interesting special case. Although the Lie algebra is ``self-dual", the
two order-reversing dualities are different \cite{barbasch1985unipotent}}. 

 When one studies the BGG Category $\mathcal{O}$ associated to the two Springer resolutions, then Category $\mathcal{O}$ is both Koszul self-dual \textit{and} Koszul dual to the
Langlands dual Category $\mathcal{O}$.  The definition of Symplectic Duality adopted in \cite{braden2014quantizations} includes this Koszul duality as an important requirement. One of the
important examples of Symplectic Duality is the relationship between
Category $\mathcal{O}$ for $T^\star(G_\mathbb{C}/B)$ and Category $\mathcal{O}$ for $T^\star(G^\vee/B^\vee)$ and this is proved
in \cite{braden2014quantizations} using the classic result on Koszul duality due to \cite{beilinson1996koszul} and the results of
\cite{mazorchuk2009quadratic}. Using these same results, it is also possible to show that $T^\star(G_\mathbb{C}/B)$
is also its own Symplectic Dual in the sense of \cite{braden2014quantizations} in that it satisfies
their list of requirements for being Symplectic Self-Dual.

In the program initiated in \cite{bullimore2016boundaries}, boundary conditions in 3d $\mathcal{N}=4$ theories in the presence of an analog of the $\Omega$-background are used to study the categorical statement of Symplectic Duality. In this setup, the construction of a pair of
Symplectic Dual resolutions is through the study of vacuum moduli spaces of 3d $\mathcal{N}=4$
SCFT that has the corresponding geometries as Higgs and Coulomb branches. This definition is known to match the definition of \cite{braden2014quantizations} in
several non-trivial examples \cite{bullimore2016boundaries}. But, in the specific context of Langlands
duality, equality between the definition adopted by \cite{bullimore2016boundaries} and original
definition of \cite{braden2014quantizations} is not automatic. For example, at the present moment,
we are not aware of the existence of any three dimensional SCFTs for
which the Nilpotent cone of $\mathfrak{g}$ (recall that $\mathfrak{g}$ is strictly non-simply laced in this
discussion) occurs as both the Higgs and Coulomb Branch.  The
existence or non-existence of such theories is relevant for a precise
comparison between the defining requirements of Symplectic Duality as
outlined in \cite{braden2014quantizations} and definition adopted in \cite{bullimore2016boundaries} where Symplectic
Duality is taken to relate Higgs and Coulomb branches of a 3d $\mathcal{N}=4$
SCFT.

Our observations in the beginning of the section regarding the relevance of the refined theory of sheets and special symplectic resolutions was made in the context of those instances of symplectic duality where the two relevant geometries do arise as Coulomb and Higgs branches of a single 3d $\mathcal{N}=4$ SCFT. But, they can be extended to  include the instances of symplectic self-duality by a combining the results of this paper with an extension of \cite{Gukov:2006jk,Gukov:2008sn} in the spirit of this paper.  We also anticipate that the geometry of the mass deformed Coulomb branches could be related to the deformations studied in \cite{namikawa2009induced,fu2010q} and \cite{losev2016deformations}. 

\section{Acknowledgements}

We would like to thank Pramod Achar, Philip Argyres, Florian Beck, David Ben-Zvi, Mathew Bullimore, Giovanna Carnovale, Oscar Chacaltana, Emanuel Diaconescu, Alexander Elashvili, Ami Hanany, Justin Hilburn, Madalena Lemos, Mario Martone, Noppadol Mekareeya, Shiraz Minwalla, Greg Moore, Andy Neitzke, Wolfger Peelaers, Nick Proudfoot, Eric Sommers and Ben Webster for discussions at various stages. We especially thank Tudor Dimofte and Yuji Tachikawa for detailed comments on an earlier draft of this paper. We would also like to thank William M. McGovern and David Vogan for helpful correspondence.

A.B would also like to thank ICTS-TIFR (Bengaluru), TIFR (Mumbai), University of Amsterdam and the organizers of the DESY Theory workshop for opportunities to talk about this work when it was in progress. A.B and J.D would also like to thank the organizers of the 2017 Pollica workshop on ``\textit{Dualities and superconformal field theories}'' for their hospitality and the participants of the workshop for many stimulating questions and discussions. We acknowledge the support received from the ERC STG grant 306260 for participation in the 2017 Pollica Workshop. Finally, we thank ICTS-TIFR (Bengaluru) for hospitality during the ``\textit{Quantum Fields, Geometry and Representation Theory 2018}" program (code: ICTS/qftgrt/2018/07) where this paper was completed.

A.B is currently supported by a postdoctoral fellowship of the NHETC, Rutgers and DOE grant DE-SC0010008. While at DESY(Theory) and University of Hamburg, A.B was supported by a postdoctoral fellowship under the SFB 676 collaborative research group of DFG.

The work of J.D.~was supported by the National Science Foundation under Grant Number PHY-1620610.

\appendix

\section{Notions for Adjoint Orbits}
\label{OrbitsAppendix}
In this Appendix, we summarize certain aspects of the theory of adjoint orbits in a complex Lie algebra that play an important role in the paper. See \cite{collingwood1993nilpotent} for a textbook treatment of many of these topics. Other classic references include \cite{spaltenstein2006classes,carterfinite}.

\subsection{Bala-Carter Theory}
\label{balacarterappendix}

In order classify all nilpotent orbits in a complex Lie algebra $\mathfrak{g}$, one can could study how nilpotent orbits restrict to Parabolic subalgebras of $\mathfrak{g}$. The fundamental insight of Bala-Carter \cite{bala1976classes1,bala1976classes2} was that if one were to classify distinguished nilpotent orbits occurring in Levi factors of parabolic subalgebras, then one obtains a classification of all nilpotent orbits. In the paper, the Bala-Carter Levi associated to the Nahm orbit plays an important role. 

In the case of nilpotent orbits in Exceptional Lie algebras, we have used their Bala-Carter labels to identify them. The BC Levi is obvious from the BC Label. For example, the BC Levi for a nilpotent orbit that is denoted as ``$E_6(a_1)$" is $E_6$. In the case of nilpotent orbits in classical Lie algebras, we have identified them using the standard partition type labels \cite{collingwood1993nilpotent}. The BC Levi is not obvious from the partition label. But, the Bala-Carter Levi associated to a partition label can be calculated using a simple combinatorial algorithm. For completeness, we include the algorithm here for all the classical types.

\subsubsection*{Type A}

Nilpotent orbits in the Lie algebra $A_N$ are classified by part ions of $N+1$. Let $[n_i]$ be a partition of $n+1$. Then, the BC Levi associated to $[n_i]$ is just $A_{n_1-1} + A_{n_2-1} \ldots $

\subsubsection*{Types B/D}

Nilpotent orbits in the Lie algebra $B_N$/$D_N$ are classified by D-type partitions of $2N+1$/$2N$. These are partitions in which even parts occur an even number of times. Let $[n_i]$ be such a partition. If all $n_i$ are distinct and odd, then the orbit is distinguished and the BC Levi is $B_N/D_N$.  If some of the parts repeat, we have the following inductive procedure. For every pair in $n_i$ that is equal $k$, add a factor of $A_{k-1}$ to its BC Levi and form a reduced partition with the repeating pair removed. Proceed until the partition is empty. If the final remaining part is a $[3]$, then add a factor $\tilde{A}_1$ to the BC Levi.

\subsubsection*{Type C}

Nilpotent orbits in the Lie algebra $C_N$ are classified by C-type partitions of $2N$. These are partitions in which every odd part occurs an even number of times. If all the $[n_i]$ are distinct and even, then the corresponding orbit is distinguished. Its BC Levi is just $C_N$. If some of the parts repeat, one proceeds inductively as in the case above. For every pair in $n_i$ that is equal to $k$, add a factor of $\tilde{A}_{k-1}$ to the BC Levi and form a reduced partition with the repeating pair removed. Proceed until the partition is empty. If the final partition is $[2]$, then add a factor $A_1$ to the BC Levi.

\subsection{Induction of Nilpotent Orbits}
\label{inductionappendix}
Here, we briefly recall the idea of the idea of \textit{Orbit Induction} and summarize the how the procedure works in the classical Lie algebras. The idea is originally due to Lusztig-Spaltenstein \cite{lusztig1979induced}. We refer to \cite{collingwood1993nilpotent} and \cite{Chacaltana:2012zy} for further discussions. The basic idea is the following. Let $\mathfrak{p}$ be a parabolic subalgebra of $\mathfrak{g}$ and let $\mathfrak{l,n}$ be the Levi factor and Nilradical of $\mathfrak{p}$. Now, we pick a nilpotent orbit $\mathcal{O}_\mathfrak{l}$ of $\mathfrak{l}$. Now, pick representative of $\mathcal{O}_{\mathfrak{l}} + n$. This has a natural inclusion into the Lie algebra $\mathfrak{g}$. The adjoint orbit through this element will have a unique open dense orbit. This is said to the orbit induced from $(\mathfrak{l}, \mathcal{O}_\mathfrak{l})$. Although a parabolic subalgebra is chosen to describe the induction, the induced orbit ultimately depends only on the Levi factor $\mathfrak{l}$. 

Orbit induction furthermore obeys the following transitive property. If an orbit $\mathcal{O}$ is induced from $(\mathfrak{l}_1,\mathcal{O}_1)$ for a Levi $\mathfrak{l}_1 \subset \mathfrak{g}$ and the orbit $\mathcal{O}_1$ is in turn induced from $(\mathfrak{l}_2,\mathcal{O}_2)$ for a Levi $\mathfrak{l}_2 \subset \mathfrak{l}_1$, then it is possible to induce the orbit $\mathcal{O}$ from $(\mathfrak{l}_2,\mathcal{O}_2)$. Owing to this transitive property, the classification problem for orbit inductions reduces to a classification problem for orbit inductions from rigid nilpotent orbits in maximal Levi subalgebras. This is because rigid nilpotent orbits are precisely those that are not induced from any proper Levi subalgebra and maximal Levi subalgebras are those Levi subalgebras that do not arise as proper Levi subalgebras of some other Levi subalgebra.

 Orbit induction for Exceptional Lie algebras has to studied on a case-by-case basis. The results are originally due to A. Elashvili and can be found in summarized form in \cite{elashvili2009induced}. Below, we summarize the orbit induction procedure in the classical cases. 

\subsubsection{Induction in Classical Lie Algebras}

\subsubsection*{Type A}

Let us denote by $\Delta = \{e_1 -e_2, e_2-e_3,  \ldots e_{n-1}- e_n \}$ the set of simple roots of the Lie algebra $A_{n-1}$. Let $[p_i]$ be a partition of $n$. One can associate a canonical Levi subalgebra of $\mathfrak{g}$ associated to such a partition. One can do this in the following way. First, we pick a parabolic subalgebra corresponding to the following subset of the set of simple roots
 \begin{equation}
 \Delta_{p_i} = \{ e_1 - e_2, e_2-e_3, \ldots e_{p_1-1} - e_{p_1}, e_{p_1+1} - e_{p_1}, \ldots , e_{p_1+p_2-1}- e_{p_1+p_2}, e_{p_1+p_2+1} - e_{p_1+p_2}, \ldots  \}
 \label{levisubset}
 \end{equation}
 
 When the partition is $[1^n]$, the associated parabolic subalgebra is the Borel subalgebra $\mathfrak{b}$. When the partition is $[n]$, the associated parabolic subalgebra is the full Lie algebra $\mathfrak{g}$. The Levi factor of the parabolic subalgebra associated to $\Delta_p$ is $\mathfrak{l}_{[p_i]} = A_{p_1-1} + A_{p_2-1}+ \ldots  $. Now, let us pick the zero nilpotent orbit in the Levi $\mathfrak{l}_{[p_i]}$. The orbit obtained by induction from this orbit is
 \begin{equation}
 Ind_{\mathfrak{l}_{[p_i]}}^{\mathfrak{g}} = \mathcal{O}_{[p_i]^T}
 \end{equation}
where $[p_i]^T$ is the transpose partition and $\mathcal{O}_{[p_i]^T}$ is the nilpotent orbit corresponding to the partition $[p_i]^T$.

\subsubsection*{Any classical Lie algebra}

To describe induction in the other classical types, we need to make use, as an intermediate step, an operation that takes a partition that is not of $B/C/D$ type and outputs a partition that is of $B/C/D$ type. This is called $X$-collapse for $X=B,C,D$. We refer to \cite{collingwood1993nilpotent,Chacaltana:2012zy} for a description of the $X$-collapse operation. 

We take as a starting point a Levi subalgebra $\mathfrak{l}$ and a nilpotent orbit $\mathcal{O}_l$ in the Levi subalgebra. We would obtain the partition label for $Ind_{\mathfrak{l}}^{\mathfrak{g}}(\mathcal{O}_{\mathfrak{l}})$. Let us denote by $t$ the rank of the semi-simple part of $\mathfrak{l}$. Let $r$ be the rank of $\mathfrak{g}$. The center of $\mathfrak{l}$ is $r-t$ dimensional. The semi-simple part of $\mathfrak{l}$ would be of the form $ \mathfrak{l}_{ss}=A_{k_1} + A_{k_2} + \ldots + A_{k_l} + X_{k'} $, where $X_{k'}$ is of the same Cartan type as $\mathfrak{g}$. Let $d_{k'}$ be the dimension of the standard representation of $X_{d_{k'}}$. That is, $d_{k'} = 2k'$ if $X_{k'}$ is of type $C,D$ and $d_{k'} = 2k'+1$ if  $X_{k'}$ is of type $B$.

Now, any nilpotent orbit of $\mathfrak{l}$ is uniquely identified by its restriction to each of the simple factors in the semi-simple part of $\mathfrak{l}$. In other words, a nilpotent orbit of $\mathfrak{l}_{ss}$ can be given a label that corresponds to a series of partitions of $k_1+1, k_2+1, \ldots, k_{l+1}, 2k'$ . Let us denote these partitions by $[p_{k_1,i}], [p_{k_2,i}],\ldots, [p_{k_l,i}], [p_{k',i}]$. Now, treat each of these partitions to be of length $N$ (by adding zeros as parts if necessary) where $N=2n$ for $\mathfrak{g}=D_n/C_n$ and $N=2n+1$ for $\mathfrak{g}=B_n$. Now, the partition label $[p_i]$ of $Ind_{\mathfrak{l}}^{\mathfrak{g}}(\mathcal{O}_{\mathfrak{l}})$ is given by

\begin{equation}
\begin{split}
[\tilde{p}_i] &= [p_{0,i}] + 2[p_{k_1,i}] + 2[p_{k_2,i}]+ \ldots + 2[p_{k_l,i}] + [p_{k',i}]\\
[p_i] &= [\tilde{p}_i]_X
\end{split}
\end{equation}
where $[p_{0,i}]$ is a length $N$ partition of the form $[N-d_{k'}-2(k_1+k_2+ \ldots k_l),0,0,0,\ldots, 0]$ and $[\cdot]_X$ denotes the X-collapse operation. 

The above algorithm is equivalent to the one obtained by Kempken \cite{kempken1983induced} and Spaltenstein \cite{spaltenstein2006classes}. When dealing with very even orbits in Lie algebras of type $D_{2n}$, there are some further subtleties. We refer to \cite{collingwood1993nilpotent} for these cases.

To make the discussion more symmetric, we note that the induction algorithm for $\mathfrak{g}=A_n$ can also stated in a form that is similar to the one for types $B/C/D$. Here, we take $N=n+1$. The semi-simple part of a Levi subalgebra $\mathfrak{l}$ of $A_n$ is always of the form $A_{k_1} + A_{k_2} + \ldots + A_{k_l}$. Using notation as above, the partition label of $Ind_{\mathfrak{l}}^{\mathfrak{g}}(\mathcal{O}_{\mathfrak{l}})$ is $[p_i] = [p_{0,i}] + [p_{k_1,i}] + [p_{k_2,i}]+ \ldots + [p_{k_l,i}]$, where $[p_{0,i}]$ is a length $N$ partition of the form $[N-(k_1+k_2+ \ldots k_l),0,0,0,\ldots, 0]$. One can check that this equals $[r_i]^T$ where $[r_i]$ is the partition corresponding to the Levi $\mathfrak{l}$ in the sense that the set simple roots corresponding to $\mathfrak{l}$ is $\Delta_{r_i}$ (see Eq. \ref{levisubset} for definition of $\Delta_{r_i}$).

\subsubsection{Kempken-Spaltenstein Criterion}

There is a combinatorial condition on the partition label of a classical nilpotent orbit in order for it to correspond to a rigid nilpotent orbit. A rigid nilpotent orbit is one which is not induced from any proper Levi subalgebra. This condition was first obtained by Kempken \cite{kempken1983induced}\footnote{In this work, rigid nilpotent orbits were called \textit{original} orbits.} and Spaltenstein \cite{spaltenstein2006classes} and they are summarized with proof in \cite{collingwood1993nilpotent}. Let us denote by $\mathcal{P}_{B/C/D}$ the set of $B/C/D$ type partitions. As per the usual convention, the parts $p_i$ are ordered to be in descending order.

A partition $[p_i] \in \mathcal{P}_{B/D}$ corresponds to a rigid nilpotent orbit if and only if
\begin{equation}
\begin{split}
p_{i} - p_{i+1} &= 0,1 \\
p_n &= 1 \\
\mid \{j \mid p_j = i\} \mid &\neq 2 ,\text{ for any odd number $i$. }
\end{split}
\end{equation}

A partition $[p_i] \in \mathcal{P}_{C}$ corresponds to a rigid nilpotent orbit if and only if 
\begin{equation}
\begin{split}
p_{i} - p_{i+1} &= 0,1 \\
p_n &= 1 \\ 
\mid \{j \mid p_j = i\} \mid &\neq 2 ,\text{ for any even number $i$. }
\end{split}
\end{equation}

\subsection{Definitions}
\label{definitionsappendix}
We collect here some definitions that play an important role in the body of the paper.

\subsubsection{Parabolic subalgebra}
\label{parabolicsubalgebra}
A parabolic subalgebra $\mathfrak{p}$ of $\mathfrak{g}$ is defined to be a subalgebra that contains a Borel subalgebra $\mathfrak{b}$ of $\mathfrak{g}$. A canonical way to construct a parabolic subalgebra is the following procedure. We start with a real semi-simple element $x \in \mathfrak{g} $. Then, the set of all elements $\alpha \in \mathfrak{g}$ that obey $[x,\alpha] = \lambda \alpha, \lambda \geq 0 $ form a parabolic subalgebra $\mathfrak{p}$. The nilradical of $\mathfrak{p}$ is the further subset of elements that obey $[x,\alpha] = \lambda \alpha, \lambda > 0 $.

\subsubsection{Levi subalgebra}

Every parabolic subalgebra $\mathfrak{p}$ has a Levi decomposition $\mathfrak{p}=\mathfrak{l} + \mathfrak{n}$, where $\mathfrak{n}$ is the nilpotent radical of $\mathfrak{p}$ and $\mathfrak{l}$ is a reductive Lie algebra that is called the \textit{Levi factor} of $\mathfrak{p}$. All subalgebras of $\mathfrak{g}$ that occur as Levi factors are called Levi subalgebras. The canonical procedure to construct parabolic subalgebra can be adapted to construct all Levi subalgebras. Using the notation as in \ref{parabolicsubalgebra}, the Levi factor of a parabolic subalgebra is nothing but the set of elements $\alpha \in \mathfrak{g}$ that obey $[x,\mathfrak{\alpha}]=0$. In other words, Levi subalgebras arise as Lie algebra centralizers of semi-simple elements.

\subsubsection{Rigid Nilpotent Orbit}

These are nilpotent orbits of $\mathfrak{g}$ that are not induced (in the sense of Lusztig-Spaltenstein) from any proper Levi subalgebra $\mathfrak{l} \subset \mathfrak{g}$. For Lie algebras of Cartan type A, every non-trivial nilpotent orbit can be induced from a proper Levi subalgebra. Consequently, there are no rigid nilpotent orbits. For Lie algebras of other types, non-trivial Rigid nilpotent orbits always exist. For example, the minimal nilpotent orbit (the unique nilpotent orbit of the smallest dimension) is always rigid outside of type A.

\subsubsection{Special Nilpotent Orbit}

These are those nilpotent orbits that are in the image of the Spaltenstein/Barbasch-Vogan duality maps between nilpotent orbits \cite{Chacaltana:2012zy}. An equivalent definition is the one due to Lusztig using the Springer correspondence (see a review in \cite{Balasubramanian:2014jca} adapted to the physics context). 

 For nilpotent orbits in classical Lie algebras, there is a combinatorial condition on the partition labels $[p_i]$ corresponding to special orbits. Let $[p_i]$ be a $B/C/D$ type partition. Then, $[p_i]$ is a  special iff its transpose $[p_i]^T$ is a $B/C/C$ partition (respectively).

\subsubsection{Principal/Regular Nilpotent Orbit}

This is the unique nilpotent orbit with the smallest possible centralizer. The dimension of the smallest possible centralizer is always $\mathrm{rank}(\mathfrak{g})$. Hence, the dimension of the principal nilpotent orbit is $\dim(\mathfrak{g}) - \mathrm{rank}(\mathfrak{g})$. The principal nilpotent orbit is also called the regular nilpotent orbit.

\subsubsection{Regular Semi-Simple Orbit}

Any semi-simple orbit with the smallest possible centralizer is called a regular semi-simple orbit. The dimension of every regular semi-simple orbit is $\dim(\mathfrak{g}) - \mathrm{rank}(\mathfrak{g})$. There an infinite number of regular semi-simple orbits but there is a unique sheet that contains them all. We call this the principal/regular sheet of the Lie algebra.

\subsubsection{Distinguished Nilpotent Orbit}

If a nilpotent orbit of $\mathfrak{g}$ is such that the only Levi subalgebra that contains it is $\mathfrak{g}$ itself, then it is called a distinguished nilpotent orbit. For Lie algebras of type A, the principal nilpotent orbit is the only distinguished nilpotent orbit. For other types, distinguished orbits that are different from the principal orbit exist.

\subsubsection{Nilpotent Orbit of Principal Levi type}
\label{pltype}

The Bala-Carter theory classifies nilpotent orbits by classifying distinguished nilpotent orbits in Parabolic subalgebras. The classification ultimately depends only on the Levi factors of these parabolic subalgebras. So, the BC labels for nilpotent orbits involve a Levi subalgebra $\mathfrak{l}$ of $\mathfrak{g}$ and a distinguished nilpotent orbit $\mathcal{O}_{dist}$ of the Levi subalgebra. When $\mathcal{O}_{dist}$ is the principal/regular orbit, we denote the corresponding nilpotent to be a nilpotent orbit of principal Levi type.

\subsubsection{Richardson Nilpotent Orbit}

When a nilpotent orbit is non-trivially induced, one can describe each such instance of induction by identifying the Levi subalgebra $\mathfrak{l}$ and a rigid nilpotent orbit $\mathcal{O}_{rigid}$ in $\mathfrak{l}$. When this rigid nilpotent orbit happens to just be the zero orbit of $\mathfrak{l}$, the original nilpotent orbit is called a Richardson orbit. Richardson orbits are always special.

\subsubsection{Sheet}

Let $\mathcal{U}$ be the union of adjoint orbits in the Lie algebra that are of a fixed dimension. Then, the irreducible components of $\mathcal{U}$ are called sheets. Every sheet has a unique nilpotent orbit at its boundary. In type A, every nilpotent orbit belongs to a unique sheet. But, outside of type A, this property can fail to hold for certain nilpotent orbits.

\subsubsection{Dixmier Sheet}

A sheet which contains semi-simple orbits is known as a Dixmier sheet. In type A, every sheet is a Dixmier sheet. This is no longer true outside of type A. The boundary of every Dixmier sheet is a Richardson nilpotent orbit. The existence of non-Dixmier sheets is equivalent to the existence of non-Richardson nilpotent orbits.

\subsubsection{Special Sheet}

The non-nilpotent elements of a sheet could, in general, have a Jordan decomposition of the form $a_n + a_{ss},[a_{ss},a_n]=0$, for $a_n$ being a representative of a nilpotent orbit of $\mathfrak{l} = Z_{\mathfrak{g}}(a_{ss})$. If the nilpotent orbit $\mathcal{O}_{a_n}$ is a special nilpotent orbit in $\mathfrak{l}$, then denote the corresponding sheet $(\mathfrak{l},\mathcal{O}_{a_n})$ to be a special sheet of $\mathfrak{g}$. The boundary of a special sheet is necessarily a special nilpotent orbit of $\mathfrak{g}$.

\subsubsection{Non-Special Sheet}

This is defined in a way that is similar to the definition of a special sheet. There are sheets in which the the nilpotent orbit $\mathcal{O}_{a_n}$ is a non-special nilpotent orbit of $\mathfrak{l}$. The boundary of a non-special sheet could be a special orbit or a non-special nilpotent orbit of $\mathfrak{g}$.

\subsubsection{Refined Sheet}

A refined sheet is a subspace of a sheet that obeys a further restriction. This restriction arises from the physical requirement that we want the eigenvalues of the semi-simple part of the non-nilpotent elements in such a sheet to correspond to mass parameters of the 3d SCFTs that we study in this paper. This additional restriction is what we have called the \textit{Flavour condition} in the paper.

\section{Tables for Twisted Defects}
\label{MoreTaxnonomy}

We included here tables detailing mass deformations for twisted defects in certain low rank simple Lie algebras. For twisted defects that are classified by nilpotent orbits in $B_n,C_n$ algebras, there exist general formulas for the dimension of the Coulomb branch and the flavour symmetry analogous to (\ref{dimensionD}) and (\ref{flavourD}). Let $[p_i]$ the partition label for the Nahm orbit and let $[r_i]$ be its transpose and $t_i$ be the multiplicity of the number $i$ in the partition $[p_i]$. 

In terms of this data, the flavour symmetry of the $T^{[p_i]}[G]$ theory is given by
\begin{equation}
\begin{aligned}
\mathfrak{f} &= \prod_{i \in odd} so(t_i) \times \prod_{i \in even} sp(t_i/2),& \mathfrak{g}&=so(2n+1), \\
\mathfrak{f} &= \prod_{i \in even} so(t_i) \times \prod_{i \in odd} sp(t_i/2),&\mathfrak{g}&=sp(2n).
\end{aligned}
\end{equation}

Now, let the Coulomb branch be the orbit with label $[q_i]$ in $\mathfrak{g}^\vee$. Let $[s_i]$ be the transpose of $[q_i]$ and let $m_i$ be the multiplicity of the number $i$ in the partition $[q_i]$. The dimension of the Coulomb branch is given by
\begin{equation}
\begin{aligned}
\dim(\mathcal{O}_{[q_i]}) &= \dim(\mathfrak{so}_{2n+1}) - \frac{1}{2} \bigg( \sum_i s_i^2 - \sum_{i \in odd} m_i \bigg),&\mathfrak{g}^\vee&=so(2n+1) \\
\dim(\mathcal{O}_{[q_i]}) &=\dim(\mathfrak{sp}_{2n}) - \frac{1}{2} \bigg( \sum_i s_i^2 + \sum_{i \in odd} m_i \bigg),& \mathfrak{g}^\vee&=sp(2n)
\end{aligned}
\end{equation}

In the case of $G_2,F_4$, the flavour symmetries and the dimensions of the Coulomb branches can be obtained from \cite{carterfinite,collingwood1993nilpotent} and have been summarized in \cite{Chacaltana:2012zy}. We use used these and the formula (\ref{ssdimension}) in compiling the tables below.

\subsection{$\mathfrak{g}=B_3, \mathfrak{g}^\vee=C_3$}

\begin{center}

\begin{tabular}{c|c|c|c|c|c|c}
$O_N$&$O_H$&$\dim(\mathcal{O}_H)$&$\dim(\mathcal{O}_{a_{ss}}^{g^\vee})$&$F$&$(\mathfrak{l}^\vee,\mathcal{O}_H^{l^\vee})$&$\dim(Z(\mathfrak{l}^\vee))$\\
\toprule[0.75mm]
$[1^7$&$[6]$&18&18&$B_3$&$(0,0)$&3\\ \hline 
$_1[2^2,1^3]$&$[4,2]$&16&16&$C_1\times B_1$&$(A_1,0)$&2\\
$_2[3,1^4]$&$[4,2]$&16&16&$D_2$&$(C_1,0)$&2\\ \hline 
$[3,2^2]$&$[3^2]$&12&12&$C_1$&$(A_1+C_1,0)$&1\\
$[3^2,1]$&$[2^3]$&12&12&$B_1$&$(A_2,0)$&1\\
$[5,1^2]$&$[2^2,1^2]$&10&10&$B_1$&$(C_2,0)$&1\\
$[7]$&$[1^6]$&0&-&-&$(C_3,0)$&0\\ \toprule[0.75mm]
\end{tabular}

\end{center}
\begin{center}

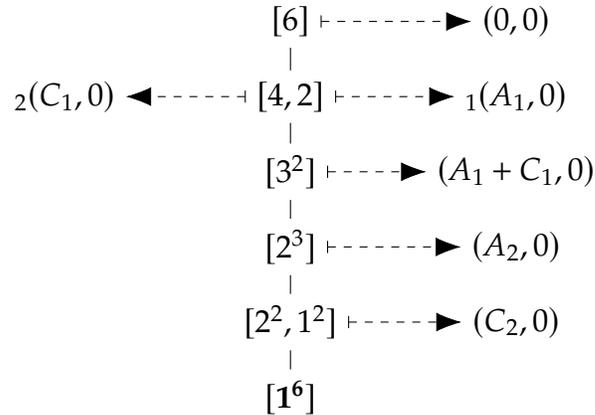
\begin{figure} [!h]
\begin{center}
\begin{tikzpicture}
 \node (a) at (0,0) {$[6]$};
 \node (a+) at (3,0) {$(0,0)$};
 
 \node (b) at (0,-1) {$[4,2]$};
 \node (b+) at (3,-1) {$_1(A_1,0)$};
  \node (b++) at (-3,-1) {$_2(C_1,0)$};
 
 \node (c) at (0,-2) {$[3^2]$};
 \node (c+) at (3,-2) {$(A_1+C_1,0)$};
 
 \node (d) at (0,-3) {$[2^3]$};
  \node (d+) at (3,-3) {$(A_2,0)$};
 
  \node (e) at (0,-4) {$[2^2,1^2]$};
 \node (e+) at (3,-4) {$(C_2,0)$};  
  
   \node (f) at (0,-5) {$\mathbf{[1^6]}$};
  \draw[preaction={draw=white, -,line width=10pt}] (a)--(b)--(c)--(d)--(e)--(f);
  \draw[|-{Latex[black,scale=2]}, dashed](a) -- (a+); 
 \draw[|-{Latex[black,scale=2]}, dashed](b) -- (b+);  
  \draw[|-{Latex[black,scale=2]}, dashed](b) -- (b++);
  \draw[|-{Latex[black,scale=2]}, dashed](c) -- (c+);
  \draw[|-{Latex[black,scale=2]}, dashed](d) -- (d+);
  \draw[|-{Latex[black,scale=2]}, dashed](e) -- (e+);
    
 \end{tikzpicture}
\end{center}
\caption{This diagram shows the special sheets for the Lie algebra $C_3$.}
\label{hasseC3}
\end{figure}

\end{center}

\subsection{$\mathfrak{g}=C_3, \mathfrak{g}^\vee=B_3$}
\label{tablegc3}
\begin{center}

\begin{tabular}{c|c|c|c|c|c|c}
$O_N$&$O_H$&$\dim(\mathcal{O}_H)$&$\dim(\mathcal{O}_{a_{ss}}^{g^\vee})$&$F$&$(\mathfrak{l}^\vee,\mathcal{O}_H^{l^\vee})$&$\dim(Z(\mathfrak{l}^\vee))$\\
\toprule[0.75mm]
$[1^6]$&$[7]$&18&18&$C_3$&$(0,0)$&3\\ \hline 
$_1[2,1^4]$&$[5,1^2]$&16&16&$C_2$&$(B_1,0)$&2\\
$_2[2^2,1^2]$&$[5,1^2]$&16&16&$C_1 \times D_1$&$(A_1,0)$&2\\ \hline 
$[2^3]$&$[3^2,1]$&14&14&$B_1$ &$(A_1+B_1,0)$&1\\
$[3^2]$&$[3,2^2]$&12&12&$C_1$ &$(A_2,0)$&1\\ \hline 
$_1[4^2,1^2]$&$[3,1^4]$&10&10&$C_1$ &$(B_2,0)$&1\\
$_2[4,2]$&$[3,1^4]$&10&-& $0$ &$(B_3,[3,1^4])$&0\\ \hline 
$[6]$&$[1^7]$&0&-& $0$&$(B_3,0)$&0\\ \toprule[0.75mm]
\end{tabular}
\end{center}

\begin{center}

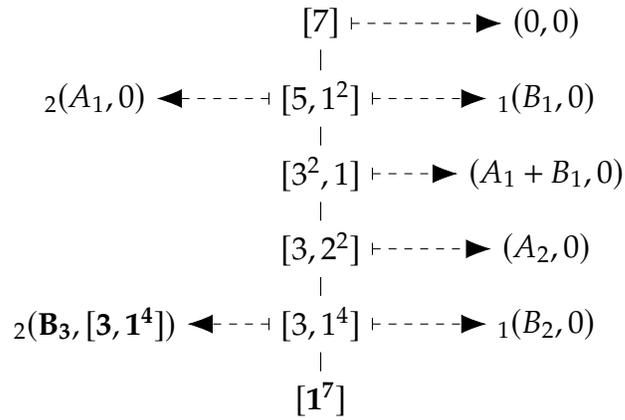
\begin{figure} [!h]
\begin{center}
\begin{tikzpicture}
 \node (a) at (0,0) {$[7]$};
 \node (a+) at (3,0) {$(0,0)$};
 
 \node (b) at (0,-1) {$[5,1^2]$};
 \node (b+) at (3,-1) {$_1(B_1,0)$};
 \node (b++) at (-3,-1) {$_2(A_1,0)$};
 
 \node (c) at (0,-2) {$[3^2,1]$};
  \node (c+) at (3,-2) {$(A_1+B_1,0)$};
 
 \node (d) at (0,-3) {$[3,2^2]$};
  \node (d+) at (3,-3) {$(A_2,0)$};
 
  \node (e) at (0,-4) {$[3,1^4]$};
    \node (e+) at (3,-4) {$_1(B_2,0)$};
   \node (e++) at (-3,-4) {$_2\mathbf{(B_3,[3,1^4])}$};
   
   \node (f) at (0,-5) {$\mathbf{[1^7]}$};
  \draw[preaction={draw=white, -,line width=10pt}] (a)--(b)--(c)--(d)--(e)--(f);
 \draw[|-{Latex[black,scale=2]}, dashed](a) -- (a+); 
 \draw[|-{Latex[black,scale=2]}, dashed](b) -- (b+);  
  \draw[|-{Latex[black,scale=2]}, dashed](b) -- (b++);
  \draw[|-{Latex[black,scale=2]}, dashed](c) -- (c+);
  \draw[|-{Latex[black,scale=2]}, dashed](d) -- (d+);
  \draw[|-{Latex[black,scale=2]}, dashed](e) -- (e+);
     \draw[|-{Latex[black,scale=2]}, dashed](e) -- (e++); 
 \end{tikzpicture}
\end{center}
\caption{This diagram shows the special sheets for the Lie algebra $B_3$.}
\label{hasseB3}
\end{figure}
\end{center}

\subsection{$\mathfrak{g}=B_4,\mathfrak{g}^\vee=C_4$}
\label{tablegc4}
\begin{center}

\begin{tabular}{c|c|c|c|c|c|c}
$O_N$&$O_H$&$\dim(\mathcal{O}_H)$&$\dim(\mathcal{O}_{a_{ss}}^{g^\vee})$&$F$&$(\mathfrak{l}^\vee,\mathcal{O}_H^{l^\vee})$&$\dim(Z(\mathfrak{l}^\vee))$\\
\toprule[0.75mm] 
$[1^9]$&$[8]$&32&32&$B_4$&$(0,0)$&4\\ \hline 
$[2^2,1^5]$&$[6,2]$&30&30&$B_2\times C_1$&$(A_1,0)$&3\\
$[3,1^6]$&$[6,2]$&30&30&$D_3$&$(C_1,0)$&3\\ \hline 
$_1[2^4,1]$&$[4^2]$&28&28&$C_2$&$(A_2,0)$&2\\
$_2[3,2^2,1^2]$&$[4^2]$&28&28&$C_1 \times U(1)$&$(A_1+C_1,0)$&2\\ \hline 
$[3^2,1^3]$&$[4,2^2]$&26&26&$B_1\times U(1)$&$(A_2,0)$&2\\
$[3^3]$&$[3^2,2]$&24&24&$B_1$&$(A_2+C_1,0)$&1\\
$[5,1^4]$&$[4,2,1^2]$&24&24&$D_2$&$(C_2,0)$&2\\
$[5,2^2]$&$[3^2,1^2]$&22&22&$C_1$&$(A_1+C_2,0)$&1\\ \hline 
$_1[4^2,1]$&$[2^4]$&20&20&$C_1$&$(A_3,0)$&1\\
$_2[5,3,1]$&$[2^4]$&20&-&0&$(C_4,[2^4])$&0\\ \hline 
$[7,1^2]$&$[2^2,1^4]$&14&14&$U(1)$&$(C_3,0)$&1\\
$[9]$&$[1^8]$&0&-&0&$(C_4,0)$&0\\ \toprule[0.75mm]
\end{tabular}
\end{center}

\begin{center}

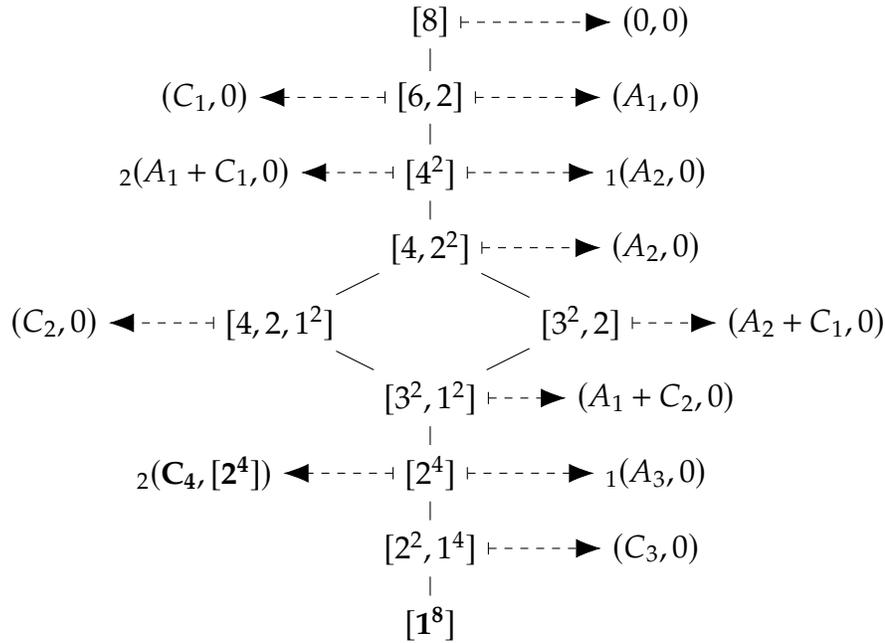
\begin{figure} [!h]
\begin{center}
\begin{tikzpicture}
 \node (a) at (0,0) {$[8]$};
 \node (a+) at (3,0) {$(0,0)$};
    
 \node (b) at (0,-1) {$[6,2]$};
   \node (b+) at (3,-1) {$(A_1,0)$};
    \node (b++) at (-3,-1) {$(C_1,0)$};
    
 \node (c) at (0,-2) {$[4^2]$};
 \node (c+) at (3,-2) {$_1(A_2,0)$};
 \node (c++) at (-3,-2) {$_2(A_1+C_1,0)$};
 
 \node (d) at (0,-3) {$[4,2^2]$};
  \node (d+) at (3,-3) {$(A_2,0)$};
 
  \node (e) at (-2,-4) {$[4,2,1^2]$};
  \node (e+) at (-5,-4) {$(C_2,0)$};
  
    \node (ee) at (2,-4) {$[3^2,2]$};
    \node (ee+) at (5,-4) {$(A_2+C_1,0)$};
    
   \node (f) at (0,-5) {$[3^2,1^2]$};
    \node (f+) at (3,-5) {$(A_1+C_2,0)$};
    
    \node (g) at (0,-6) {$[2^4]$};
    \node (g+) at (3,-6) {$_1(A_3,0)$};
     \node (g++) at (-3,-6) {$_2\mathbf{(C_4,[2^4])}$};
    
     \node (h) at (0,-7) {$[2^2,1^4]$};
     \node (h+) at (3,-7) {$(C_3,0)$};
     
     \node (i) at (0,-8) {$\mathbf{[1^8]}$};
  \draw[preaction={draw=white, -,line width=10pt}] (a)--(b)--(c)--(d)--(e)--(f)--(g)--(h)--(i);
  \draw[preaction={draw=white, -,line width=10pt}] (d)--(ee)--(f);
 \draw[|-{Latex[black,scale=2]}, dashed](a) -- (a+); 
 \draw[|-{Latex[black,scale=2]}, dashed](b) -- (b+);  
  \draw[|-{Latex[black,scale=2]}, dashed](b) -- (b++);
  \draw[|-{Latex[black,scale=2]}, dashed](c) -- (c+);
    \draw[|-{Latex[black,scale=2]}, dashed](c) -- (c++);
  \draw[|-{Latex[black,scale=2]}, dashed](d) -- (d+);
  \draw[|-{Latex[black,scale=2]}, dashed](e) -- (e+);
  \draw[|-{Latex[black,scale=2]}, dashed](ee) -- (ee+); 
   \draw[|-{Latex[black,scale=2]}, dashed](f) -- (f+);
    \draw[|-{Latex[black,scale=2]}, dashed](g) -- (g+);
     \draw[|-{Latex[black,scale=2]}, dashed](g) -- (g++);
     \draw[|-{Latex[black,scale=2]}, dashed](h) -- (h+); 
 \end{tikzpicture}
\end{center}
\caption{This diagram shows the special sheets for the Lie algebra $C_4$.}
\label{hasseC4}
\end{figure}
\end{center}

\subsection{$\mathfrak{g}=C_4,\mathfrak{g}^\vee=B_4$ }

\begin{center}

\begin{tabular}{c|c|c|c|c|c|c}
$O_N$&$O_H$&$\dim(\mathcal{O}_H)$&$\dim(\mathcal{O}_{a_{ss}}^{g^\vee})$&$F$&$(\mathfrak{l}^\vee,\mathcal{O}_H^{l^\vee})$&$\dim(Z(\mathfrak{l}^\vee))$\\
\toprule[0.75mm]
$[1^8]$&$[9]$&32&32&$C_4$&$(0,0)$&4\\ \hline 
$_1[2,1^6]$&$[7,1^2]$&30&30&$C_3$&$(B_1,0)$&3\\
$_2[2^2,1^4]$&$[7,1^2]$&30&30&$C_2 \times U(1)$&$(A_1,0)$&3\\ \hline 
$[2^3,1^2]$&$[5,3,1]$&28&28&$C_1\times B_1$&$(A_1+B_1,0)$&2\\
$[2^4]$&$[5,3,1]$&28&28&$D_2$&$(2A_1,0)$&2\\ \hline 
$[3^2,1^2]$&$[5,2^2]$&26&26&$C_1\times C_1$&$(A_2,0)$&2\\
$[3^2,2]$&$[3^3]$&24&24& $C_1$&$(A_2+B_1,0)$&1\\
$[4,2,1^2]$&$[5,1^4]$&24&24&$C_1$&$(B_3,[3,1^4])$&2\\
$[4,2^2]$&$[3^2,1^3]$&22&22&$U(1)$&$(A_1+B_2,0)$&1\\
$[4^2]$&$[3,2^2,1^2]$&20&20&$U(1)$&$(A_3,0)$&1\\ \hline 
$_1[6,1^2]$&$[3,1^6]$&14&14&$C_1$&$(B_3,0)$&1\\
$_2[6,2]$&$[3,1^6]$&14&-&$0$&$(B_4,[3,1^6])$&0\\ \hline 
$[8]$&$[1^9]$&0&-&$0$&$(B_4,0)$&0\\ \toprule[0.75mm]
\end{tabular}

\end{center}

\begin{center}

\begin{figure} [!h]
\begin{center}
\begin{tikzpicture}
 \node (a) at (0,0) {$[9]$};
 \node (a+) at (3,0) {$(0,0)$};
 
 \node (b) at (0,-1) {$[7,1^2]$};
  \node (b+) at (3,-1) {$_1(B_1,0)$};
  \node (b++) at (-3,-1) {$_2(A_1,0)$}; 
 
 \node (c) at (0,-2) {$[5,3,1]$};
  \node (c+) at (3,-2) {$(A_1+B_1,0)$};
   \node (c++) at (-3,-2) {$(2A_1,0)$};
 
 \node (d) at (0,-3) {$[5,2^2]$};
 \node (d+) at (3,-3) {$(A_2,0)$};
 
  \node (e) at (-2,-4) {$[5,1^4]$};
   \node (e+) at (-5,-4) {$(B_2,0)$};
   
    \node (ee) at (2,-4) {$[3^3]$};
    \node (ee+) at (5,-4) {$(A_2+B_1)$};
    
   \node (f) at (0,-5) {$[3^2,1^3]$};
    \node (f+) at (3,-5) {$(A_1+B_2,0)$};
   
    \node (g) at (0,-6) {$[3,2^2,1^2]$};
      \node (g+) at (3,-6) {$(A_3,0)$};
      
     \node (h) at (0,-7) {$[3,1^6]$};
        \node (h+) at (3,-7) {$_1[(B_3,0)$};
          \node (h++) at (-3,-7) {$_2\mathbf{(B_4,[3,1^6]})$};
     
        \node (i) at (0,-8) {$\mathbf{[1^9]}$};
  \draw[preaction={draw=white, -,line width=10pt}] (a)--(b)--(c)--(d)--(e)--(f)--(g)--(h)--(i);
  \draw[preaction={draw=white, -,line width=10pt}] (d)--(ee)--(f);
  
 \draw[|-{Latex[black,scale=2]}, dashed](a) -- (a+); 
 \draw[|-{Latex[black,scale=2]}, dashed](b) -- (b+);  
  \draw[|-{Latex[black,scale=2]}, dashed](b) -- (b++);
  \draw[|-{Latex[black,scale=2]}, dashed](c) -- (c+);
    \draw[|-{Latex[black,scale=2]}, dashed](c) -- (c++);
  \draw[|-{Latex[black,scale=2]}, dashed](d) -- (d+);
  \draw[|-{Latex[black,scale=2]}, dashed](e) -- (e+);
  \draw[|-{Latex[black,scale=2]}, dashed](ee) -- (ee+); 
   \draw[|-{Latex[black,scale=2]}, dashed](f) -- (f+);
    \draw[|-{Latex[black,scale=2]}, dashed](g) -- (g+);
     \draw[|-{Latex[black,scale=2]}, dashed](h) -- (h+); 
      \draw[|-{Latex[black,scale=2]}, dashed](h) -- (h++); 
    
 \end{tikzpicture}
\end{center}
\caption{This diagram shows the special sheets for the Lie algebra $B_4$.}
\label{hasseB4}
\end{figure}
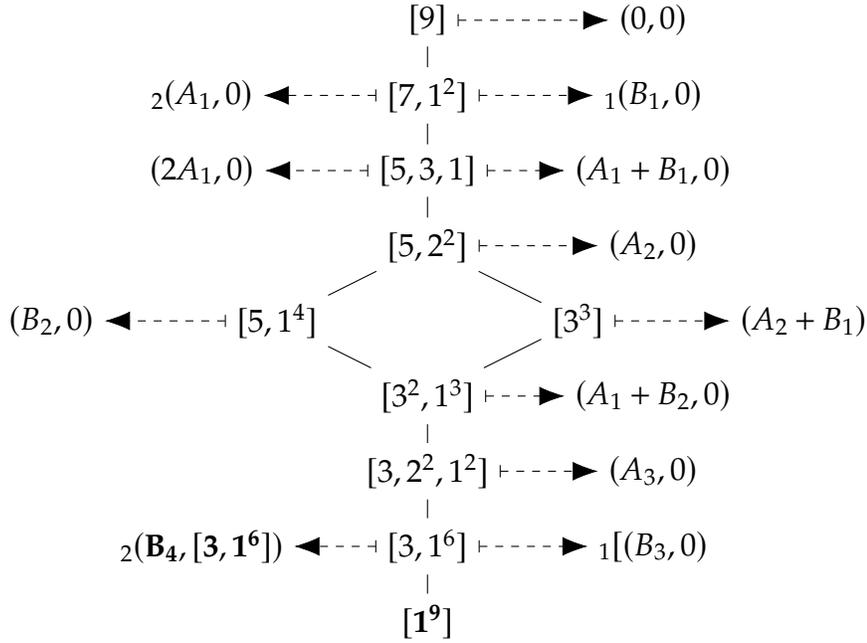
\end{center}

\subsection{$\mathfrak{g}=\mathfrak{g}^\vee=G_2$}
\begin{center}
\begin{tabular}{c|c|c|c|c|c|c}
$O_N$&$O_H$&$\dim(\mathcal{O}_H)$&$\dim(\mathcal{O}_{a_{ss}}^{g^\vee})$&$F$&$(\mathfrak{l}^\vee,\mathcal{O}_H^{l^\vee})$&$\dim(Z(\mathfrak{l}^\vee))$\\
\toprule[0.75mm] 
$1$&$G_2$&12&12&$G_2$&$(0,0)$&2\\ \hline 
$_1A_1$&$G_2(a_1)$&10&10&$A_1$&$(\tilde{A_1},0)$&2\\
$_2\tilde{A_1}$&$G_2(a_1)$&10&10&$A_1$&$(A_1,0)$&1\\
$_3G_2(a_1)$&$G_2(a_1)$&10&-&&$(G_2,G_2(a_1))$&0\\ \hline 
$G_2$&1&0&-&&$(G_2,0)$&0\\ \toprule[0.75mm]
\end{tabular}
\end{center}

\begin{figure} [!h]
\begin{center}
\begin{tikzpicture}
 \node (a) at (0,0) {$G_2$};
 
 \node (b) at (0,-2) {$G_2(a_1)$};
  \node (b+) at (-3,-3) {$_1(A_1,0)$};
  \node (b++) at (3,-3) {$_2(\tilde{A_1},0)$};
   \node (b+++) at (3,-4) {$_3\mathbf{(G_2,G_2(a_1))}$};
 \node (c) at (0,-4) {$\mathbf{1}$};

  \draw[preaction={draw=white, -,line width=10pt}] (a)--(b)--(c);
  
   \draw[|-{Latex[black,scale=2]}, dashed](b) -- (b++);
     \draw[|-{Latex[black,scale=2]}, dashed] (b) -- (b+);   
     \draw[|-{Latex[black,scale=2]}, dashed] (b) -- (b+++); 
 \end{tikzpicture}
\end{center}
\caption{This diagram shows the special sheets for the Lie algebra $G_2$.}
\label{hasseG2}
\end{figure}
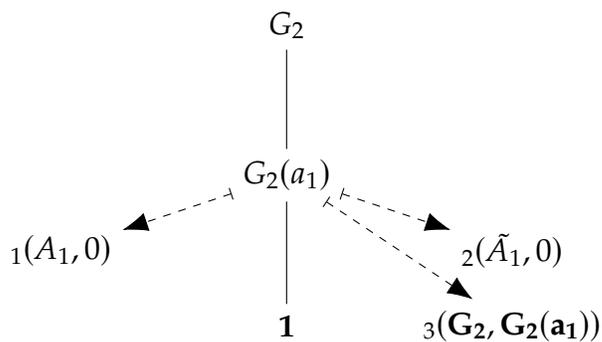

\subsection{$\mathfrak{g}=\mathfrak{g}^\vee=F_4$}
\begin{center}
\begin{tabular}{c|c|c|c|c|c|c}
$O_N$&$O_H$&$\dim(\mathcal{O}_H)$&$\dim(\mathcal{O}_{a_{ss}}^{g^\vee})$&$F$&$(\mathfrak{l}^\vee,\mathcal{O}_H^{l^\vee})$&$\dim(Z(\mathfrak{l}^\vee))$\\
\toprule[0.75mm] 
$0$&$F_4$&48&48&$F_4$&$(0,0)$&4\\ \hline 
$_1A_1$&$F_4(a_1)$&46&46&$C_3$&$(\tilde{A_1},0)$&3\\
$_2\tilde{A_1}$&$F_4(a_1)$&46&46&$A_3$&$(A_1,0)$&3\\ \hline 
$A_2$&$B_3$&42&42&$A_2$&$(\tilde{A_2},0)$&2\\
$\tilde{A_2}$&$C_3$&42&42&$G_2$&$(A_2,0)$&2\\ \hline 
$_1(A_2 + \tilde{A_1})$&$F_4(a_3)$&40&40&$A_1$&$(A_1+\tilde{A_2},0)$&1\\
$_2(A_1 + \tilde{A_2})$&$F_4(a_3)$&40&40&$A_1$&$(\tilde{A_1}+A_2,0)$&1\\
$_3B_2$&$F_4(a_3)$&40&40&$2A_1$&$(C_2 \sim B_2, 0)$&2\\
$_4C_3(a_1)$&$F_4(a_3)$&40&30&$A_1$&$(B_3,[3,1^4])$&1\\
$_5F_4(a_3)$&$F_4(a_3)$&40&-&-&$(F_4,F_4(a_3))$&0\\ \hline 
$B_3$&$A_2$&30&30&$A_1$&$(C_3,0)$&1\\
$C_3$&$\tilde{A_2}$&30&30&$A_1$&$(B_3,0)$&1\\
$F_4(a_2)$&$A_1 + \tilde{A_1}$&28&-&-&$(F_4,A_1+\tilde{A_1})$&0\\
$F_4(a_1)$&$\tilde{A_1}$&22&-&-&$(F_4,\tilde{A_1})$&0\\
$F_4$&$0$&0&-&-&$(F_4,0)$&0\\ \toprule[0.75mm]
\end{tabular}
\end{center}

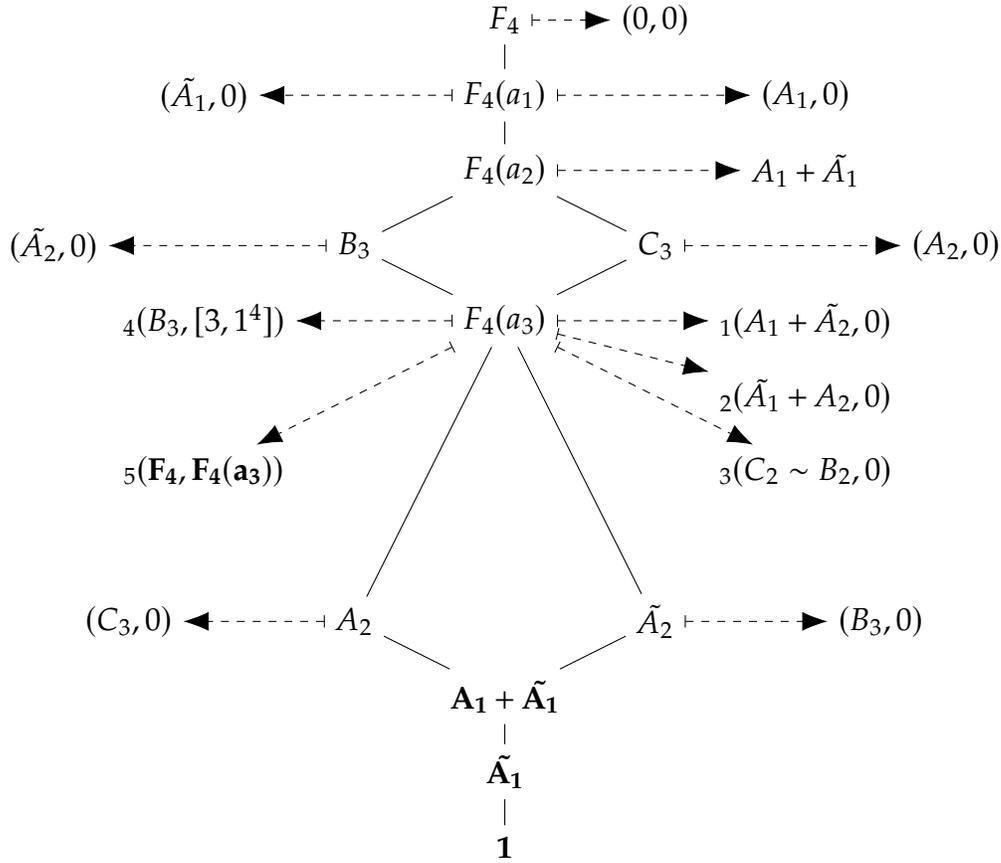
\begin{figure} [!h]
\begin{center}
\begin{tikzpicture}
 \node (a) at (0,0) {$F_4$};
 \node (a+) at (2,0) {$(0,0)$}; 
 
 \node (b) at (0,-1) {$F_4(a_1)$};
  \node (b+) at (4,-1) {$(A_1,0)$};
   \node (b++) at (-4,-1) {$(\tilde{A_1},0)$};
 
 \node (c) at (0,-2) {$F_4(a_2)$};
  \node (c+) at (4,-2) {$A_1+\tilde{A_1}$};
  
  \node (d) at (-2,-3) {$B_3$};
  \node (d+) at (-6,-3) {$(\tilde{A_2},0)$};
  
  \node (dd) at (2,-3) {$C_3$};
  \node (dd+) at (6,-3) {$(A_2,0)$};
  
   \node (e) at (0,-4) {$F_4(a_3)$};
    \node (e1) at (4,-4) {$_1(A_1+\tilde{A_2},0)$};
      \node (e2) at (4,-5) {$_2(\tilde{A_1}+A_2,0)$};
      \node (e3) at (4,-6) {$_3(C_2\sim B_2,0)$};
      \node (e4) at (-4,-4) {$_4(B_3,[3,1^4])$};
      \node (e5) at (-4,-6) {$_5\mathbf{(F_4,F_4(a_3))}$};
   
    \node (f) at (-2,-8) {$A_2$};
     \node (f+) at (-5,-8) {$(C_3,0)$};
    
    \node (ff) at (2,-8) {$\tilde{A_2}$};
     \node (ff+) at (5,-8) {$(B_3,0)$};
     
    \node (g) at (0,-9) {$\mathbf{A_1+\tilde{A_1}}$};
    
  \node (h) at (0,-10) {$\mathbf{\tilde{A_1}}$};  
  
  \node (i) at (0,-11) {$\mathbf{1}$}; 
  
  \draw[preaction={draw=white, -,line width=10pt}] (a)--(b)--(c)--(d)--(e)--(f)--(g)--(h)--(i);
   \draw[preaction={draw=white, -,line width=10pt}](c)--(dd)--(e)--(ff)--(g);
  
     \draw[|-{Latex[black,scale=2]}, dashed](a) -- (a+);
     \draw[|-{Latex[black,scale=2]}, dashed] (b) -- (b+);
     \draw[|-{Latex[black,scale=2]}, dashed] (b) -- (b++);
   \draw[|-{Latex[black,scale=2]}, dashed](c) -- (c+);
      \draw[|-{Latex[black,scale=2]}, dashed](dd) -- (dd+);
      \draw[|-{Latex[black,scale=2]}, dashed](d) -- (d+);
        \draw[|-{Latex[black,scale=2]}, dashed](e) -- (e1);
      \draw[|-{Latex[black,scale=2]}, dashed](e) -- (e2);
       \draw[|-{Latex[black,scale=2]}, dashed](e) -- (e3);
        \draw[|-{Latex[black,scale=2]}, dashed](e) -- (e4);
         \draw[|-{Latex[black,scale=2]}, dashed](e) -- (e5);
        \draw[|-{Latex[black,scale=2]}, dashed](ff) -- (ff+);
      \draw[|-{Latex[black,scale=2]}, dashed] (f) -- (f+);
     
 \end{tikzpicture}
\end{center}
\caption{This diagram shows the special sheets for the Lie algebra $F_4$.}
\label{hasseF4}
\end{figure}

\newpage
\bibliography{refs}
\bibliographystyle{utphys}

\end{document}